%% file: main.tex
\documentclass[useAMS,usenatbib,letterpaper]{mn2e}
\usepackage{journal_abbreviations, custom_commands}
\usepackage{lastpage}
\usepackage[totalwidth=500pt,totalheight=650pt,centering,layoutvoffset=-0.3cm]{geometry}

\def\wa{0.32\textwidth}

\defcitealias{turner14}{T14}
\def\obspaper{\citetalias{turner14}}

\title[Evidence for inflows around \textit{z}$\, \approx\,$2 star-forming galaxies]
{A comparison of observed and simulated absorption from HI, CIV, and SiIV 
around \textit{z}$\, \approx\,$2 star-forming galaxies suggests 
redshift-space distortions are due to inflows}
\author[Turner et al.]{Monica L. Turner,$^{1,2}$\thanks{E-mail: turnerm@mit.edu}
Joop Schaye,$^{2}$
Robert A. Crain,$^{3}$
Gwen Rudie,$^{4}$\newauthor
Charles C. Steidel,$^{5}$
Allison Strom,$^{5}$
and
Tom Theuns$^{6}$\\
$^{1}$MIT-Kavli Center for Astrophysics and Space Research, 
   Massachusetts Institute of Technology, \\
  \,  77 Massachusetts Ave., Cambridge, MA 02139, USA\\
$^{2}$Leiden Observatory, Leiden University, PO Box 9513, 2300 RA Leiden, The Netherlands\\
$^{3}$Astrophysics Research Institute, Liverpool John Moores University, 146 Brownlow Hill, Liverpool, L3 5RF, UK\\
$^{4}$Carnegie Observatories, 813 Santa Barbara Street, Pasadena, CA 91101, USA\\
$^{5}$California Institute of Technology, MS 249-17, Pasadena, CA 91125, USA\\
$^{6}$Institute for Computational Cosmology, Department of Physics, University of Durham, South Road, Durham, DH1 3LE, UK\\
}

\begin{document}	

\date{\today}

\pagerange{\pageref{firstpage}--\pageref{LastPage}} \pubyear{2017}

\maketitle

\label{firstpage}

\begin{abstract}
We study \hone\ and metal-line absorption around $z\approx2$ star-forming galaxies by
comparing an analysis of data  from the Keck
Baryonic Structure Survey to mock spectra generated from the EAGLE cosmological,
hydrodynamical simulations. We extract sightlines from the simulations
and compare the properties of the absorption by \hone, \cfour\ and \sifour\ around
simulated and observed galaxies using pixel optical depths. We
mimic the resolution, pixel size, and signal-to-noise
ratio of the observations, as well as the distributions of impact
parameters and galaxy redshift errors. We find that 
the EAGLE reference model is in excellent agreement with the observations. 
In particular, the simulation reproduces the high
metal-line optical depths found at small galactocentric distances, the
optical depth enhancements out to impact parameters of
2 proper Mpc, and the prominent redshift-space distortions which we
find are due to peculiar velocities rather than redshift errors. The
agreement is best for halo masses $\sim10^{12.0}$~\msol, for which the 
observed and simulated stellar masses also agree most closely. We
examine the median ion mass-weighted radial gas velocities around the galaxies,
and find that most of the gas is infalling, with the infall velocity depending 
on halo rather than stellar mass.  From this we
conclude that the observed redshift-space distortions are predominantly caused by
infall rather than outflows.
\end{abstract}



\section{Introduction}
\label{sec:intro}

Galaxy formation theory tells us that the cycling of gas 
through galaxies, from accretion streams that feed star formation
to the expulsion of gas by feedback from star formation and active galactic nuclei (AGN),
is a key process that largely determines a galaxy's properties. 
However, the
cosmological simulations that are typically used to explore this process are 
limited by their inability to directly resolve
the mechanisms responsible for expelling gas from galaxies,
requiring them to resort to subgrid models. As
these subgrid feedback prescriptions are typically calibrated
to reproduce properties of present-day galaxies, such as 
their stellar masses, it is important
to test the results against other observables.

In particular, the reservoir of gas around galaxies, known as
the circumgalactic medium (CGM), lies in the region where 
inflowing and outflowing gas meets, making it an ideal test bed for simulations. 
Observations of intergalactic metals,
which span wide ranges of ionization energies, can constrain
the composition, kinematics and physical state of the CGM.
By directly comparing simulations and observations, 
we can gain insights into the physics of the gas flows in and around galaxies.
Such studies can also help interpret the observations 
of the CGM.

Recent studies of simulations have encountered difficulties in
reproducing the observed \hone\ covering fractions
around massive galaxies at $z\approx2$ \citep{fumagalli14, fauchergiguere15, meiksin15}.
\citet{rahmati15} demonstrated that this tension
can be alleviated by better matching the simulated galaxy redshifts and stellar masses
to those of the observations, by measuring the absorption within the same velocity intervals,
and by using simulations with efficient stellar and AGN feedback. 
\citet{fauchergiguere16, meiksin17} also concluded that simulations employing strong feedback
can reproduce the observations.

For metal line absorption,  $z<1$ \osix\ observations from \citet{tumlinson11}
and \citet{prochaska11} have proven challenging for simulations to reproduce
\citep{suresh16, oppenheimer16}, though the correspondence 
is better for low ions, perhaps due to a paucity of hot gas in the models
\citep{hummels13, ford16}. In general, efficient
feedback is required to obtain agreement with
observations at large impact parameters
\citep{stinson12, hummels13}.
At higher redshift, \citet{shen13} were able to reproduce the metal-line
equivalent widths (EWs) reported by \citet{steidel10}. Their simulations indicate that the extended
absorption is mostly due to inflowing, rather than outflowing material,
contrary to the model presented by \citet{steidel10}. 
More recently, \citet{suresh15} compared simulated \cfour\ equivalent widths
to observations from \citet{turner14}, and found that even the most energetic wind
model was unable to match the observations at 200--300 proper kpc (pkpc) scales. 

Here we perform a comparison between
the observations of absorption by \hone, \cfour, and \sifour\ around $z\approx2$ galaxies
of \citet[][hereafter \obspaper]{turner14}, and the
Evolution and Assembly of Galaxies
and their Environments (EAGLE) simulations \citep{schaye15, crain15}.
The relatively high resolution ($\epsilon \sim 1$~kpc comoving) and large cosmological volume 
(100 cMpc)$^3$ of EAGLE permits the study of massive haloes 
while still marginally resolving the Jeans scale in the warm ($T\sim10^4$~K) interstellar medium.
EAGLE has been found to broadly reproduce a number of observables,
including the present day galaxy stellar mass function, 
galaxy sizes and the Tully-Fisher relation \citep{schaye15}, 
galaxy colours \citep{trayford15} and neutral gas scaling relations \citep{bahe16, crain17},
the evolution of galaxy stellar masses \citep{furlong15} and sizes \citep{furlong16},
and properties of \hone\ absorption at $z\approx2$--$3$ \citep{rahmati15}.
\citet{schaye15}, \citet{rahmati16} and \citet{turner16} found broad
agreement with observations of metal-line absorption along random
lines-of-sight for a variety of ions and redshifts. However, the careful like-for-like
comparison by \citet{turner16} revealed that EAGLE produces insufficient 
\cfour\ and \sifour\ associated with intergalactic \hone\ absorption at $z\approx3.5$. 

The analysis of \obspaper\ is based on data from 
the Keck Baryonic Structure Survey \citep[][KBSS]{rudie12, steidel14},
a galaxy redshift survey of 15 fields centred on hyper-luminous QSOs.
\obspaper\ employed an approach known as the pixel optical depth method
\citep{cowie98, ellison00, schaye00a, aguirre02, schaye03}
to measure absorption in the QSO spectra for \hone\ and five metal ions, 
accounting for saturation and contamination. 
A public version of the code used for this work can be found at 
\url{http://github.com/turnerm/podpy}.
The redshifts and impact parameters from the KBSS 
were then used
to characterize the statistical properties of the absorption
around 854  galaxies with $\langle z \rangle \approx 2.3$. 
The pixel optical depth
approach is particularly well-suited for comparison with simulations
as, unlike fits to absorption lines, it can be applied quickly, automatically and uniformly
to large numbers of observed and mock spectra alike. 

This work builds upon a previous study by \citet{rakic12, rakic13}, 
who also applied the pixel optical depth technique to 
the KBSS data in order to measure the median \hone\ absorption 
around the galaxies, and compared the results with the OWLS simulations
\citep{schaye10}. The authors measured a best-fit minimum halo mass of 
$\log_{10}\mhminm/\msolm = 11.6\pm0.2$
which agrees with independent estimates from the clustering analysis 
of \citet[$\log_{10}\mhminm/\msolm = 11.7$]{trainor12}. 
They also established that galaxy redshift errors and 
gas peculiar velocities had to be taken into account to reproduce 
the observed redshift-space distortions.
We expand upon the work of \citet{rakic13} by examining 
metal-line absorption due to \cfour\ and \sifour\ in addition to \hone,
and by comparing with the state-of-the-art EAGLE simulations. The improvements
provided by EAGLE compared to the OWLS simulations include
a much better match to observations of galaxies, 
a substantially larger volume (100~comoving~Mpc instead of 25~comoving~Mpc~h$^{-1}$) at a similar resolution,
a cosmology consistent with current constraints, and an improved hydrodynamics solver.

The structure of this paper is as follows. In \S~\ref{sec:method} we 
describe the simulations and our approach for generating mock spectra. 
In \S~\ref{sec:obs_compare} we present a direct comparison between the results
from \obspaper\ and the simulations, while in \S~\ref{sec:variations} we 
use the simulations to pinpoint the origin of the observed redshift-space distortions. 
Our discussion and conclusions can be found in \S~\ref{sec:conclusion}. Finally, in 
Appendix~\ref{app:test} we present resolution and box size tests.
Throughout this work, we denote proper and comoving distances as 
pMpc and cMpc, respectively. 
Both simulations and observations use cosmological parameters
determined from the Planck mission \citep{planck13}, i.e.
$\Omega_{\rm m} = 0.307$,  
$\Omega_{\Lambda} = 0.693$,
$\Omega_{\rm b} = 0.04825$,
$h = 0.6777$,
$Y_p = 0.24775$, and 
$\sigma_8=0.8288$.


\section{Method}
\label{sec:method}

\subsection{Simulations}

\begin{table*}
\caption{Characteristics of the EAGLE model variations used in this work. From left to right, the columns list the simulation identifiers, 
the box size, number of particles,
 initial baryonic particle mass, dark matter particle mass, comoving (Plummer-equivalent) gravitational softening,
   maximum physical softening, and differences with respect to the reference run.}
\begin{tabular}{lrrccccl}
\hline
Simulation & $L$       & $N$ & $m_{\rm b}$ & $m_{\rm dm}$ & $\epsilon_{\rm com}$ & $\epsilon_{\rm prop}$ & Deviations from Ref.\\  
                & [cMpc] &       & [\msol]    & [\msol]        & [ckpc]          &    [pkpc]                             &  \\
\hline 
\hline
{Ref-L100N1504} &   100 & $2\times1504^3$ & $1.81 \times 10^6$ & $9.70 \times 10^6$ & 2.66 & 0.70 &  Reference model\\
{Ref-L050N0752} &    50  & $2\times752^3$ & $1.81 \times 10^6$ & $ 9.70 \times 10^6$ & 2.66 & 0.70 &  Smaller volume \\
{Ref-L025N0376} &    25  & $2\times376^3$ & $1.81 \times 10^6$ & $ 9.70 \times 10^6$ & 2.66 & 0.70 &  Smaller volume \\
{Ref-L025N0752} &    25  & $2\times752^3$ & $2.26 \times 10^5$ & $ 1.21 \times 10^6$ & 1.33 & 0.35 & Higher resolution \\
{Recal-L025N0752} &    25  & $2\times752^3$ & $2.26 \times 10^5$ & $ 1.21 \times 10^6$ & 1.33 & 0.35 & Higher resolution, \\
 & & & & & & & recalibrated feedback\\
{NoAGN} &    50  & $2\times752^3$ & $1.81 \times 10^6$ & $ 9.70 \times 10^6$ & 2.66 & 0.70 & No AGN feedback \\
{WeakFB} &    25  & $2\times376^3$ & $1.81 \times 10^6$ & $ 9.70 \times 10^6$ & 2.66 & 0.70 & Weaker stellar feedback \\
{StrongFB} &    25  & $2\times376^3$ & $1.81 \times 10^6$ & $ 9.70 \times 10^6$ & 2.66 & 0.70 & Stronger stellar feedback \\
\hline
\end{tabular}
\label{tab:sims}
\end{table*}

We will compare the predictions of EAGLE,
a suite of cosmological, hydrodynamical simulations, 
to the results of \obspaper.
The EAGLE suite was run with a substantially modified version of the $N$-body TreePM 
smoothed particle hydrodynamics (SPH) code \texttt{GADGET}~3,
last described in \citet{springel05a}. In particular, the simulations employ the updated hydrodynamics solver
``Anarchy''  (Dalla Vecchia, in prep.; see also Appendix~A of \citealt{schaye15} and \citealt{schaller15}) which uses,
among other improvements, the pressure-entropy formulation
of SPH from \citet{hopkins13} and the time-step limiter from \citet{durier12}. 

The simulations include subgrid models for the following physical processes:
photo-heating and radiative cooling via eleven elements (hydrogen, helium, carbon, nitrogen, oxygen,
neon, magnesium, silicon, sulphur, calcium and iron) from \citet{wiersma09a} using a
\citet{haardt01} UV and X-ray background;
star formation from \citet{schaye08} with the metallicity-dependent 
gas density threshold of \citet{schaye04};
stellar evolution and enrichment from \citet{wiersma09b};
black-hole seeding and growth from \citet{springel05b, rosas13} and \citet{schaye15}.
Stochastic, thermal stellar feedback from star formation 
is implemented as described by \citet{dallavecchia12}, where the injected energy 
depends on both the local metallicity and density.
AGN feedback is also realized thermally as per \citet{booth09},
but implemented stochastically \citep{schaye15}.
 The feedback in EAGLE has been calibrated to match observations of the
$z=0$ galaxy stellar mass function, the galaxy--black hole mass relation,
and to give reasonable sizes of disk galaxies, as explained in detail by \citet{crain15}. 

The EAGLE suite includes simulations with variations in box size,
resolution, and subgrid physics. The largest run is the intermediate-resolution fiducial or reference (``Ref'') model, 
which uses a 100~cMpc periodic box with $1504^3$ particles of both dark matter and baryons
(denoted L100N1504). Information about the resolution and subgrid physics for each simulation 
used here can be found in Table~\ref{tab:sims}. In \S~\ref{sec:obs_compare}
we will focus on models with the reference subgrid physics, 
and present convergence testing for our results in Appendix \ref{app:test}.
We note that the high-resolution simulation (Ref-L025N0752) has also been run with 
subgrid physics recalibrated to better match the galaxy mass function (Recal-L025N0752). 
The alternative models that we consider in \S~\ref{sec:variations} and Appendix~\ref{app:feedback}
are NoAGN, in which AGN feedback has been disabled;
and WeakFB and StrongFB, which employ half and twice as strong stellar feedback as the reference
model, respectively. These subgrid variations are described in detail and compared with observations
by \citet{crain15}.

\subsection{Generating mock spectra}
\label{sec:gms}

Mock spectra were generated using the package
\texttt{SPECWIZARD} written by Schaye, Booth, and Theuns, which is implemented 
as described in  Appendix A4 of \citet{theuns1998}. 
The spectra were given the properties of the observed Keck/HIRES spectra:
a resolution of $\text{FWHM}\approx8.5$~\kmps\ and 
pixels of 2.8~\kmps\ in size. We then added Gaussian noise with 
an S/N ratio equal to that measured in \obspaper\ (see their Table~4),
which is about $\sim70$ for \hone\ and $\sim80$ for \cfour\ and \sifour. 
EAGLE imposes a minimum pressure as a function of density which 
most commonly applies to dense particles characteristic
of the multiphase ISM. 
 Before generating mock spectra,
we set the temperature of these particle to $10^4$~K, but we found that this
does not have a significant effect on our results because the cross section
of such dense absorbers is very small.

 \begin{table*}
 \caption{The intensity of the UV background ($\Gamma_{\honem}$) required to match
 observed \taurndhone, 
 and the median \hone, \cfour, and \sifour\ optical depths for random regions
 from the observations (top row) and various EAGLE model variations (subsequent rows). 
 The 1-$\sigma$ errors on the median observed optical depths
 were calculated by bootstrap resampling (with replacement) the QSO spectra 1000 times.
 For the simulations, we present the median \hone\ optical depths \textit{before} scaling 
  the UV background, while for \cfour\ and \sifour\ the results are derived after this 
  adjustment had been performed.  We note that in \citet{haardt01}, the interpolated 
  values of $\Gamma_{\honem}$ for $z=(2.01, 2.24, 2.48)$ 
  are $(1.32,1.31,1.24)$~$10^{-12}$~s$^{-1}$. }
\input{tables/med_od.tab}
\label{tab:medod}
\end{table*}

There is significant uncertainty in the normalization and shape of the 
ionizing background radiation. Therefore, we follow standard practice \citep[e.g.,][]{rakic13} 
and scale the strength of the ionizing background to match the observations of \hone.
More specifically, taking the \citet{haardt01} amplitude of the UV background
($\Gamma_{\honem}=1.31$~$10^{-12}$~s$^{-1}$ at $z=2.24$), 
we generate a grid of $64\times64$ spectra covering the full 100~cMpc box,
and measure the median optical depth of \hone\ \lya\ after convolving the spectra
with a Gaussian to achieve Keck/HIRES resolution ($\text{FWHM}\approx8.5$~\kmps) and adding noise.
For each EAGLE simulation, the resulting median \hone\ optical depth, which is representative of
random regions and denoted \taurndhone, 
is given in the third column of Table~\ref{tab:medod}. For subsequent runs,
the spectra are synthesized with $\Gamma_{\honem}$
scaled such that \taurndhone\ matches the observed value
for the sample of \obspaper\ ($\log_{10} \taurndhonem =-1.29$).

In this analysis, we do not consider radiation from 
nearby stars or AGN. In \citet{turner15}, we investigated the impact
of stellar ionizing photons from the KBSS galaxies 
on the gas located at small impact parameters.  According to their eq.~8,
at a transverse distance of 118~pkpc (corresponding to the median impact 
parameter of galaxies in the two innermost bins), 
a high escape fraction of 10\% would result in a reduction of \hone\ pixel optical depths
by a factor of two. While the higher ionization energy metal absorption lines are likely
not impacted by the stellar radiation because it drops steeply in intensity at wavelengths below 912~\AA, 
fluctuating AGN could be an issue if the time between outbursts is smaller than the metal ion 
recombination timescales \citep{oppenheimer13b, segers17}. However, the inclusion
of non-equilibrium ionization is beyond the scope of this work. 

The median \cfour\ and \sifour\ optical depths \textit{after} scaling
the ionization background are given in the fourth and fifth columns 
of Table~\ref{tab:medod}. 
These are lower than the 
observed values:
$\log_{10} \taurndcfourm \approx -3.6$ whereas $-3.0$ is observed,
and  $\log_{10} \taurndsifourm \approx -4.4$ whereas $-3.2$ is observed.
 Although the relative difference is large, these values are all very small compared with the noise.
We have compared histograms of the optical depth distributions for 
the simulations and observations, and found that the dominant causes of this 
discrepancy is that the simulated spectra have more pixels with 
negative values (due to noise). It is therefore likely that the observations
contain (small) contributions
from contamination due to the presence of other ions,
atmospheric lines, and are affected by continuum fitting errors, which suppress 
negative optical depth pixels and are not present in the simulations.
Indeed, it would be surprising if we could mimic the 
observed noise properties to such a high precision as contamination and
continuum fitting errors are not modelled at all. 

To account for this, we calculate the difference between
the observed and simulated median optical depths,  $\Delta \taurndm$,
and linearly add this difference to the simulated optical depths. 
In Appendix~\ref{app:contam} we verify that in most cases this addition does not change the securely
detected absorption. Rather, it serves to facilitate a
comparison between observations and simulations for detections close to the noise level. 
However, we note that in the case of the lowest minimum halo masses for \sifour, 
all of the optical depths are very close to the detection limit, 
and thus the addition of $\Delta \taurndm$ does result in a significant 
increase in all bins.

 \begin{table*}
 \caption{The median halo mass and number of galaxies for 
   the different \mhmin,
  for each box size, resolution, subgrid variation and redshift.}
\input{tables/gal_props_b.tab}
\label{tab:galpropb}
\end{table*}

We use the $z=2.24$ EAGLE snapshot, as it is closest to the median observed 
galaxy redshift of  $z = 2.34$. To investigate if our results our sensitive to 
the exact redshift used, we have repeated our study for runs 
at $z=2.0$ and $2.5$. In each case, the intensity of the UV background 
is fixed such that the median neutral
hydrogen optical depth in random regions agrees with the observations. 
We find that the differences in median optical depths for the runs
with different redshifts is negligible for both \hone\ and metal ions. 

Dark matter halos are identified by first linking neighbouring dark matter particles using a
friends-of-friends (FoF) algorithm with a linking length of 0.2 times the mean separation.
We then link each baryonic particle to its nearest dark matter particle,
such that baryons are connected to a FOF group if their nearest dark matter 
particle is, and invoke \texttt{SubFind} \citep{springel01, dolag09} to identify bound structures.
Here, we only consider central galaxies (the main progenitor of each FoF halo)
as we are selecting by halo mass,
and leave the consideration of satellite galaxies to a future work.\footnote{
The fraction of galaxies that are satellites in the fiducial Ref-L100N1504 simulation
with stellar mass above $\log_{10} M_{\ast} / \msolm = 10.2$ (the median 
value of the observations) is 16\%.}
The centre of each sub-halo is defined to be its
centre of mass, and the mass $M_{\rm halo}$ is defined as the total 
mass within a radius where the density is 
200 times the critical density of the Universe at the considered redshift. 
We note that although the measurement of $M_{\rm halo}$ is centred around the 
minimum gravitational potential of the most massive sub-halo, we find that our results are insensitive
to whether we define the sub-halo centres using the centre of mass or the minimum gravitational potential.

The observations are magnitude limited ($\mathcal{R} \leq 25.5$ mag,
\citealt{steidel10}) and will therefore to first order
probe a minimum stellar mass, and to second order a minimum halo mass. 
We follow \citet{rakic13} and realize halo samples by drawing randomly 
(with replacement) from all halos in the snapshot with fixed 
minimum halo mass (\mhmin) but without imposing a maximum halo mass. In practice,
due to the steepness of the halo mass function, our results will be dominated
by halos with masses close to \mhmin. 
In this work,
we present results for the range of minimum halo masses
$\log_{10}\mhminm/\msolm = (10.5, 11.0, 11.5, 12.0, 12.5)$.
This range is centred on the minimum halo mass measured by \citet{rakic13} 
($\log_{10}\mhminm/\msolm = 11.6\pm0.2$) and
 \citet{trainor12} ($\log_{10}\mhminm/\msolm = 11.7$). 
We give the median halo mass as well as the number of galaxies
that corresponds to a given minimum halo mass for each
simulation in Table~\ref{tab:galpropb}.

For each minimum halo mass, we generated 12,000 spectra with impact 
parameters ranging from 35~pkpc (the smallest in the KBSS sample)
to 5.64~pMpc. Although the KBSS data has little coverage beyond
2~pMpc, we show the simulations to larger distances
in order to explore some of the trends seen in the observations.
As in \obspaper,
the impact parameter distribution  is divided into twelve logarithmically 
spaced bins, and we generated 1000 
spectra in each bin. The innermost bin is 0.55 dex in width,
while the remaining eleven bins, starting at 0.13~pMpc, are all 0.15~dex wide.  	
For the exact bin edge values please see Table~5 in \obspaper.

 \begin{table}
 \caption{The fraction of galaxies per impact parameter bin (with edges $r_1$ and $r_2$)
   with redshifts measured from observations taken using 
   LRIS (LR, $\Delta v\approx150$~\kmps), 
   NIRSPEC (NS, $\Delta v\approx60$~\kmps), and 
   MOSFIRE (MF, $\Delta v\approx18$~\kmps). }
\input{tables/zerr.tab}
\label{tab:zerr}
\end{table}

Finally, as the observed galaxy redshifts are measured to a finite precision,
a fair comparison requires that we also add errors to the LOS positions
of the simulated galaxies. As discussed by \obspaper, the measurement 
uncertainty on the redshift of a galaxy depends on the manner in which 
it is measured and on the instrument used. \obspaper\ quote uncertainties of
$\Delta v\approx150$~\kmps\ for LRIS, $\approx60$~\kmps\ for NIRSPEC,
and $\approx18$~\kmps\ for MOSFIRE.  Because the observations have focused
on improving the redshift accuracy for the galaxies with smaller impact parameters,
the errors are not uniform  as a function of impact parameter. 
To capture this in the simulation, we have
tabulated the fraction of galaxies with redshifts measured from each instrument
as a function of impact parameter (given in Table~\ref{tab:zerr}). Then, in the simulations, 
we apply redshift errors drawn from a Gaussian distribution
with  $\sigma$ equal to the expected redshift error of the instrument,
to the same fraction of galaxies as a function of impact parameter.
For impact parameters larger than observed, we use the galaxy
fractions from the final observed impact parameter bin.


\def\mwa{0.243\textwidth}
\def\mwb{0.202\textwidth}
\def\mwc{0.333\textwidth}

\section{Comparison with observations}
\label{sec:obs_compare}

 \begin{table*}
 \caption{Typical galaxy properties for the different \mhmin\ considered,
  for each box size, resolution, subgrid variation and redshift. The quantities
   shown here are the median stellar mass and median SFR. 
   The observed galaxies from \obspaper\ have a median stellar mass of
  $\log_{10} M_{\ast} / \msolm = 10.2$ and median SFRs of $30$~\msol~yr$^{-1}$.}
\input{tables/gal_props_a.tab}
\label{tab:galpropa}
\end{table*}

\begin{figure*}
    \includegraphics[width=0.49 \textwidth]{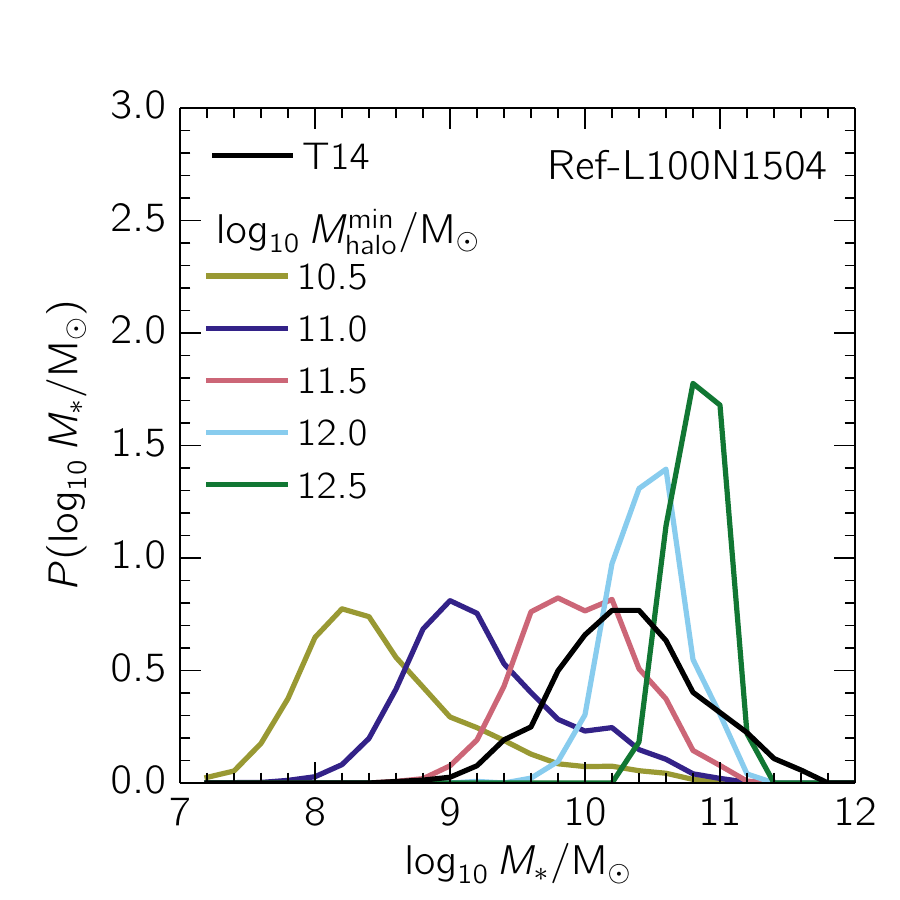} 
    \includegraphics[width=0.49\textwidth]{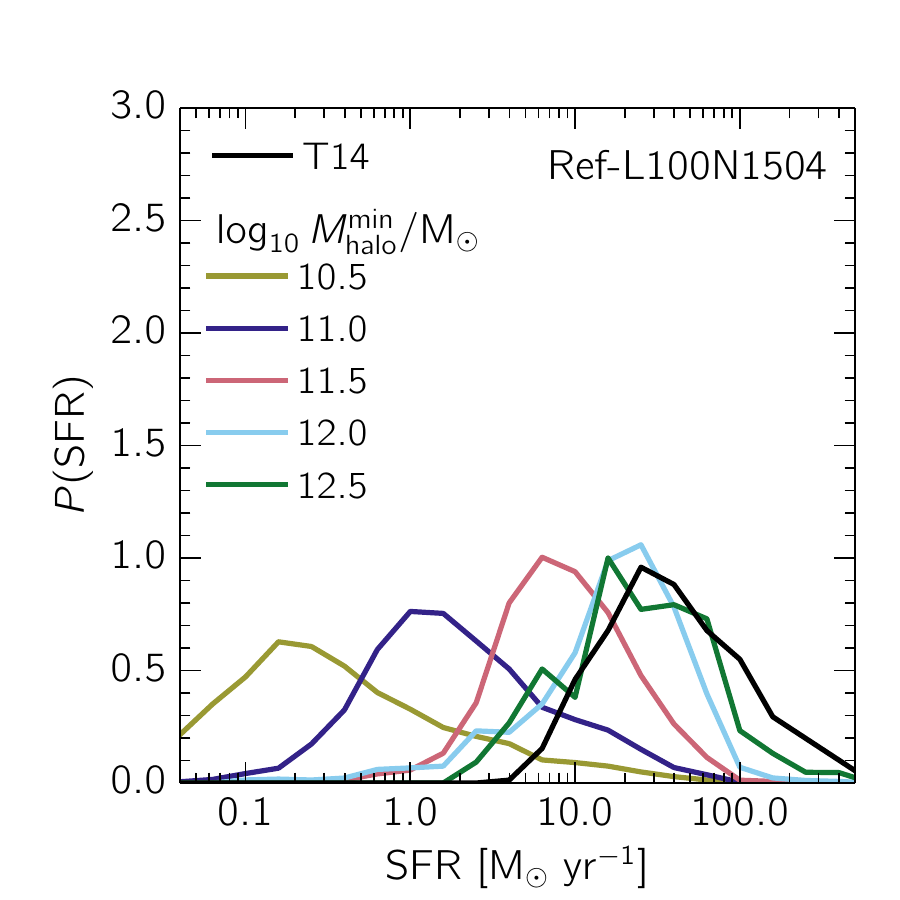} 
 \caption{PDFs of galaxy stellar masses (left panel) 
  and SFRs (right panel) for the KBSS galaxies in \obspaper\ (black lines), 
  as well as for different \mhmin\ in the Ref-L100N1504 simulation (coloured lines). Previous measurements
  of the observed galaxy halo masses \citep{trainor12, rakic13} estimate \mhmin\ to be between 
  $10^{11.5}$ and $10^{12.0}$~\msol\ (represented by the red and cyan lines, respectively). 
  The observed stellar masses agree very well with the simulations for these halo masses,
   while the observed SFRs are systematically somewhat higher than those of EAGLE by a factor of $\approx2$. }
\label{fig:galprop}
\end{figure*}

\subsection{The galaxy sample}

We present the median stellar masses ($ M_{\ast}$) and median star formation rates (SFRs)
for each \mhmin\ sample in Table~\ref{tab:galpropa}. We note that these are the true
values measured directly from the simulations, not from virtual observations. 
For the observed sample, the median stellar mass is
$\log_{10} M_{\ast} / \msolm = 10.2$ and the median SFR\footnote{
The observational quantities are estimated 
using spectral energy distribution (SED) fits
to 782 out of the 854 galaxies, while 180 of these 782 galaxies
have their SFRs measured from H$\alpha$. See \S~2.3 of \citet{steidel14}
and references therein for more details.} is $30$~\msol~yr$^{-1}$.
In both the observations and simulations, a \citet{chabrier03} initial 
mass function is assumed.
\citet{furlong15} found that observations of the evolution of the galaxy
stellar mass function are reproduced remarkably well in EAGLE.
However, they noted that the normalization of the specific SFR is 0.2--0.4~dex
below observations at all lookback times. 

To compare directly with our 
galaxy sample, in Fig.~\ref{fig:galprop} we present probability distribution functions
(PDFs) of the 
stellar masses and SFRs for different minimum halo masses taken
from the Ref-L100N1504 simulation, as well as from the observations. 
We will focus on the range of $\mhminm=10^{11.5}\text{--}10^{12.0}$~\msol,
which corresponds to median halo masses of $10^{11.8}$--$10^{12.2}$~\msol\ (see 
Table~\ref{tab:galpropb}) and agrees with the halo masses of the KBSS
\citep{trainor12, rakic13}. 
For this minimum halo mass range, we 
find very good agreement with the observed stellar masses,
while the SFRs from the observations are systematically higher than in 
EAGLE (consistent with the findings of \citealt{furlong15}), although
the medians only differ by a factor of $\approx2$ for $\mhminm=10^{12.0}$~\msol. 
We conclude that performing the galaxy selection by stellar mass or SFR
instead of halo mass would not significantly alter our results.\footnote{
While magnitudes across multiple bandpasses have been computed for the 
EAGLE galaxies \citep{trayford15, mcalpine16}, the extinction corrections due to dust have not. 
Nevertheless, we have compared the observed and simulated galaxy magnitude
histograms, and found that, as expected given the neglect of extinction, the observed values are consistent 
with a somewhat lower minimum halo mass (between $10^{11.0}$ and $10^{11.5}$~\msol) 
than predicted by the other metrics in this work as well as by independent
halo mass estimates.}

\begin{figure} 
  \includegraphics[width=0.5\textwidth]{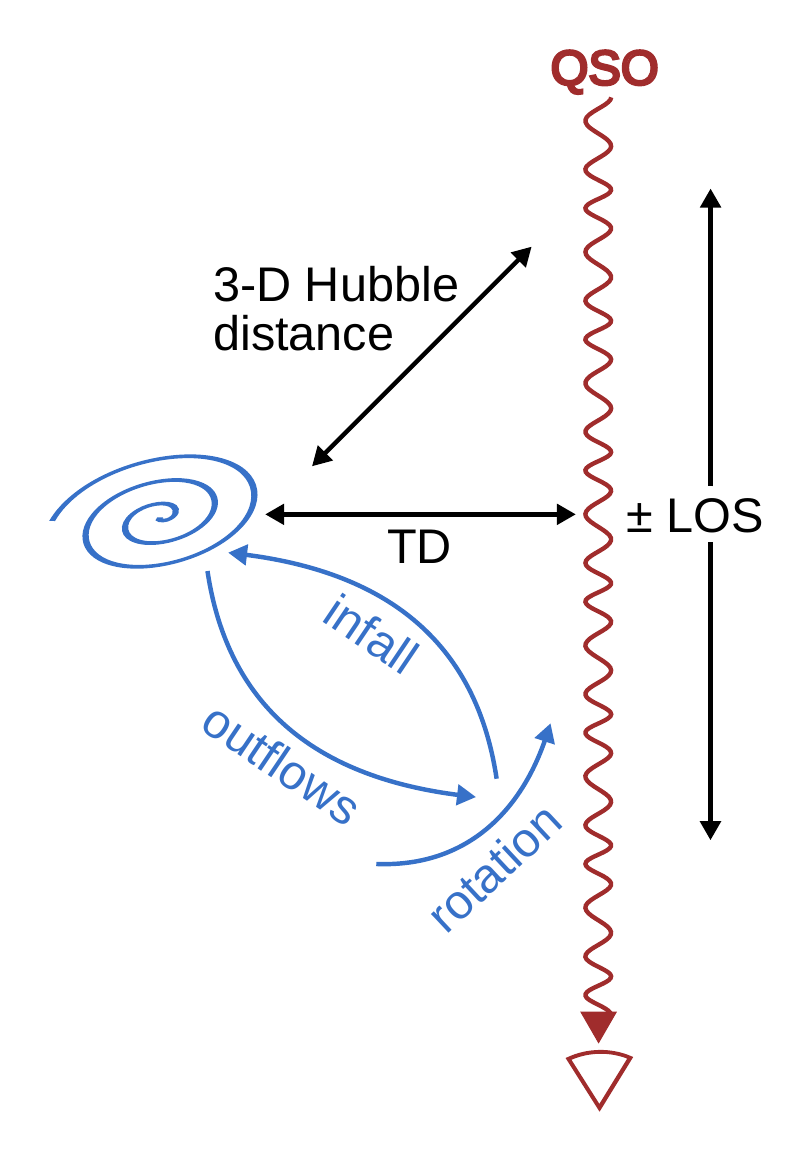} 
  \caption{A schematic outlining the geometry of the observations. The 
    blue spiral represents a galaxy, which has associated inflowing, outflowing
    and rotating gas, which is probed by the QSO sightline. The black lines
    denote the transverse distance (TD), LOS distance, and 3-D Hubble distance.}
 \label{fig:schematic}
\end{figure}

\subsection{Synthesis of galaxy and quasar data}
\label{sec:schematic}

The large number of galaxies in the KBSS has provided a unique opportunity to
analyze the absorption along the quasar LOS in a statistical fashion. To outline
the geometry required to do so, we present a schematic of a galaxy
next to a quasar sightline in Fig.~\ref{fig:schematic}. Note that the observational
sample from \obspaper\ has 854 of such galaxies. 

The galaxy is denoted by a blue spiral, and its redshift allows us to associate
it with a specific region along the QSO sightline. 
The gas from the galaxy, which could be static, infalling,
outflowing or rotating, is probed by the light from the background quasar. 
The velocity component of the gas tangent to the sightline will result in velocity- 
or redshift-space anisotropies in the absorption. 
    
The black lines mark some
 quantities frequently presented in \obspaper\ and this work. The impact parameter or
transverse distance (TD) is the 
projected distance between the galaxy and the sightline. The LOS distance, 
which also measures redshift and velocity, runs 
along the QSO sightline. Finally, the 3-D Hubble distance is simply the 
3-D distance assuming pure Hubble flow. 

\begin{figure*}
  \includegraphics[width=\mwa]{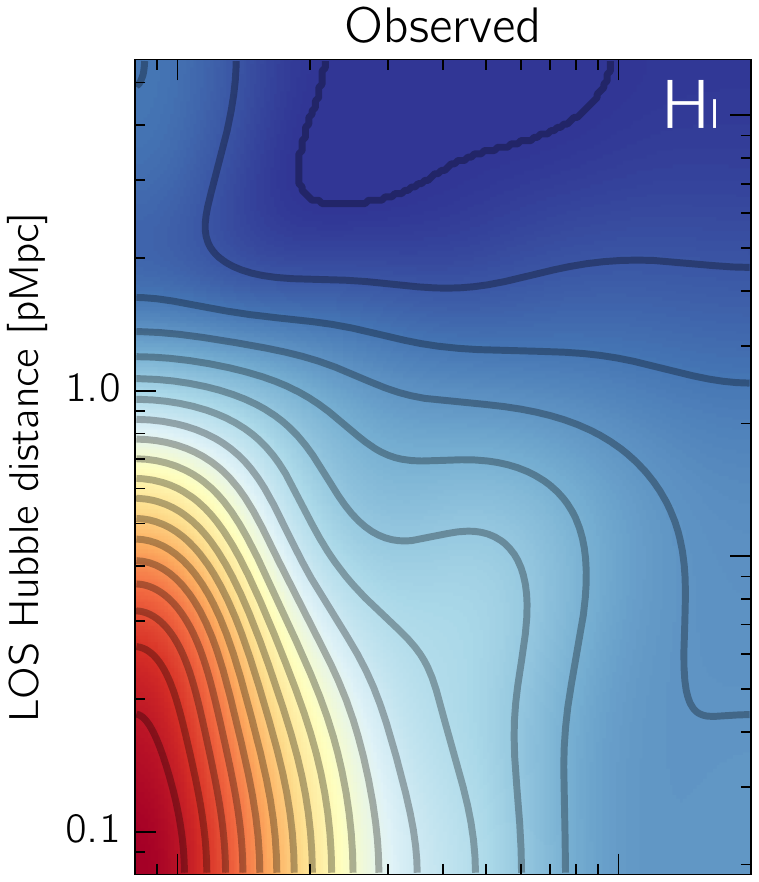} 
  \includegraphics[width=\mwb]{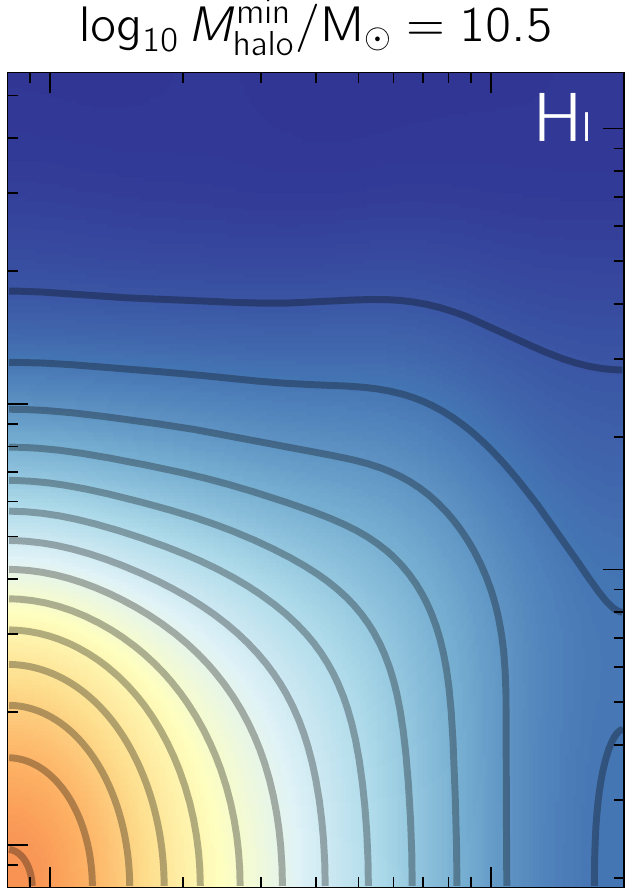} 
  \includegraphics[width=\mwb]{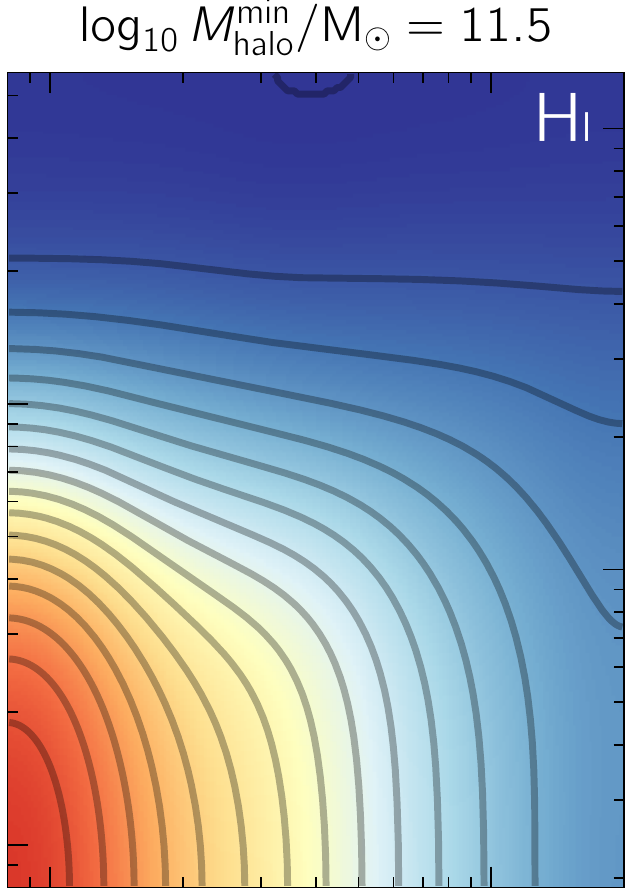} 
  \includegraphics[width=\mwc]{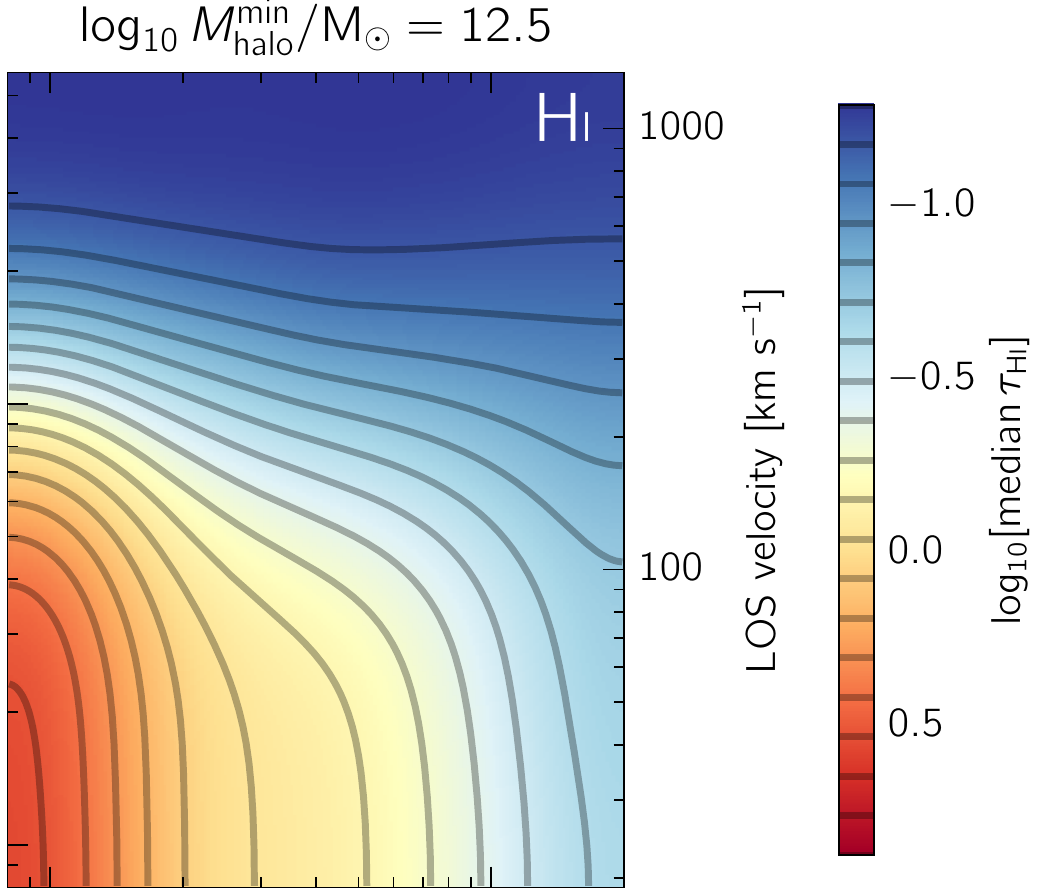} \\
  \includegraphics[width=\mwa]{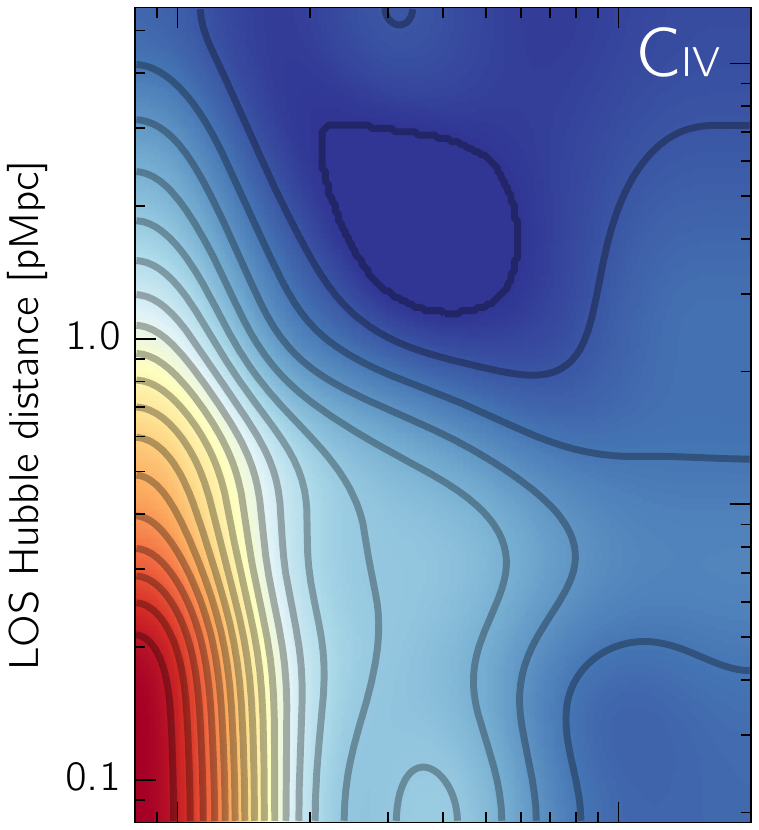} 
  \includegraphics[width=\mwb]{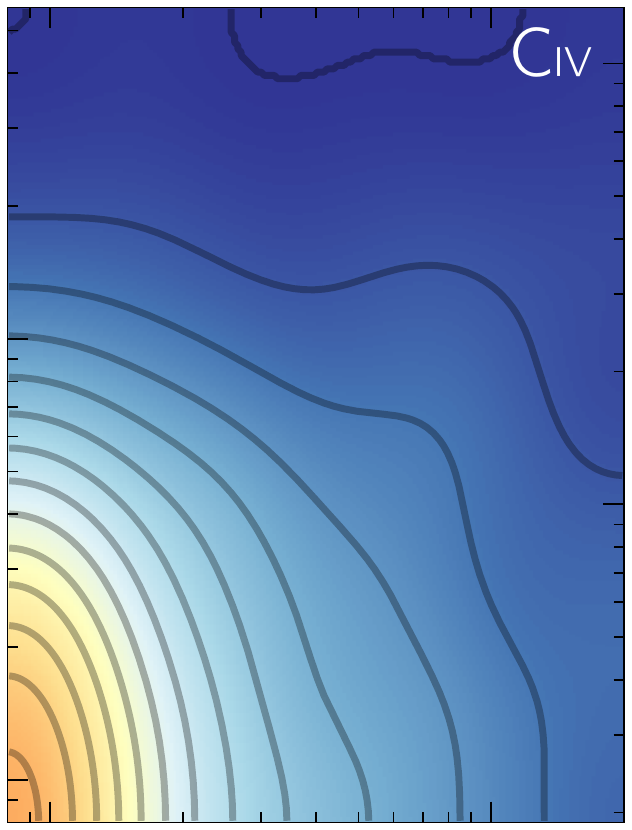} 
  \includegraphics[width=\mwb]{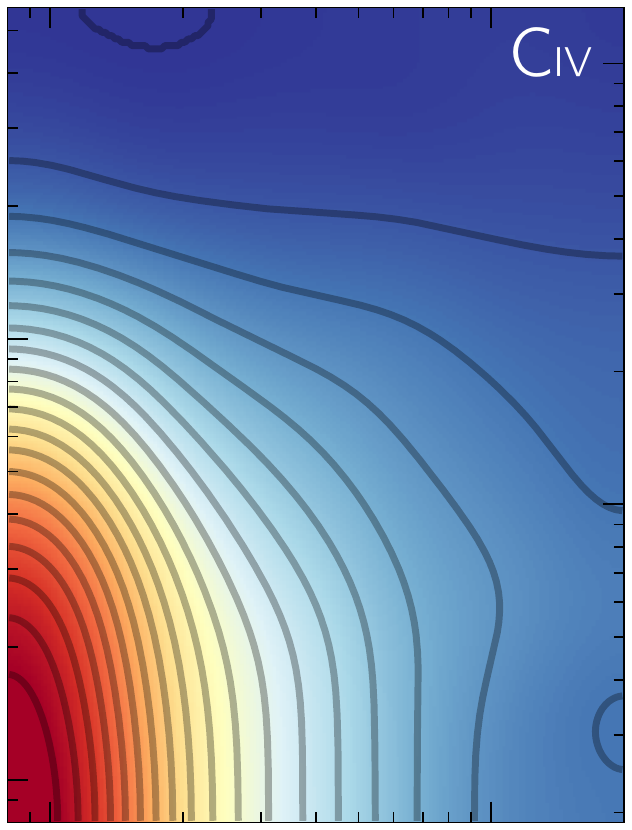} 
  \includegraphics[width=\mwc]{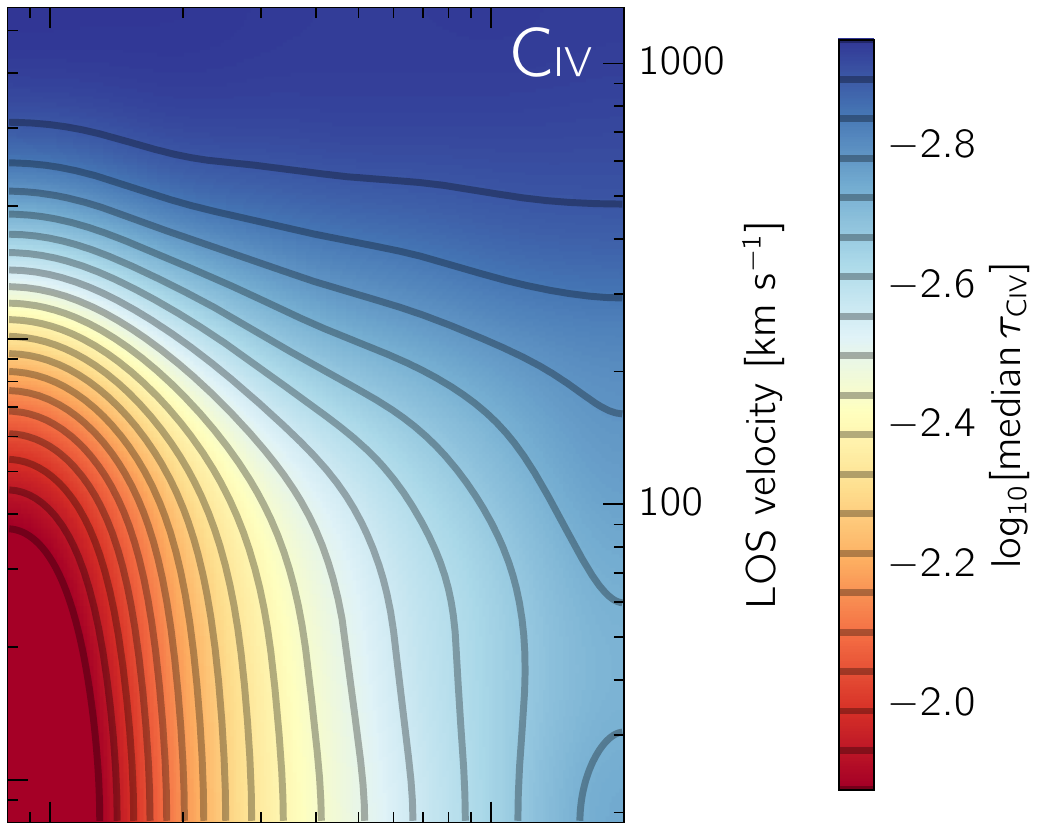} \\
    \includegraphics[width=\mwa]{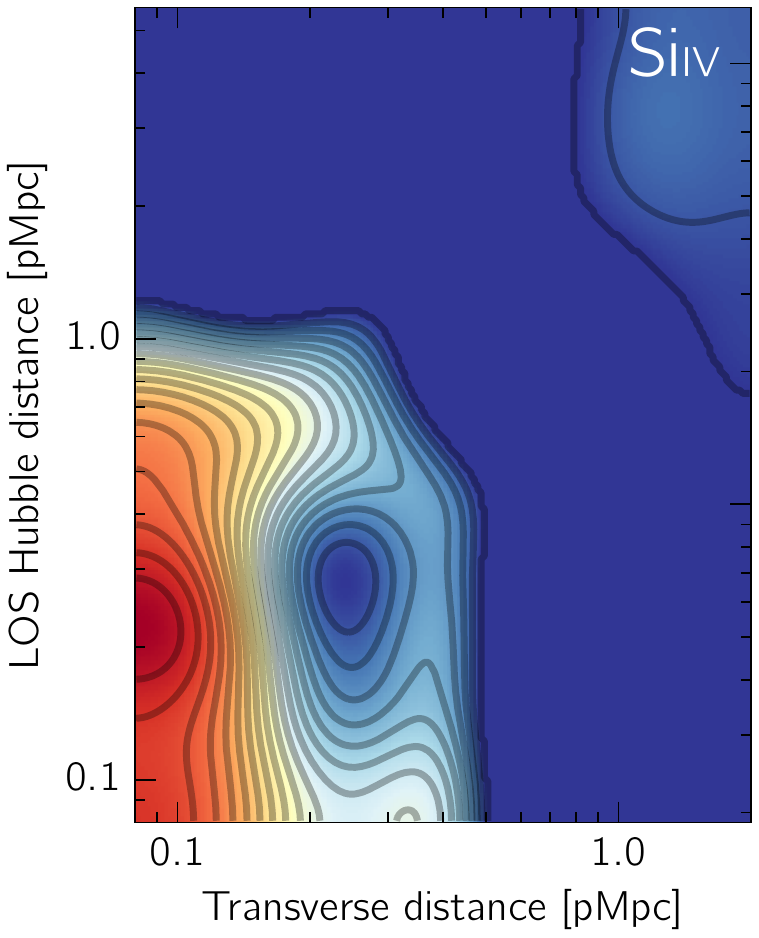} 
  \includegraphics[width=\mwb]{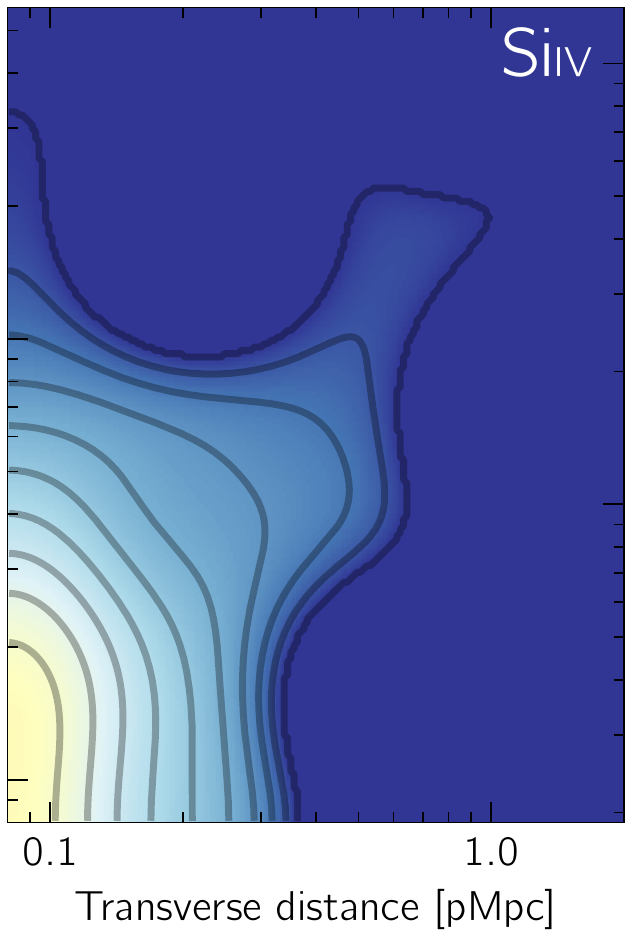} 
  \includegraphics[width=\mwb]{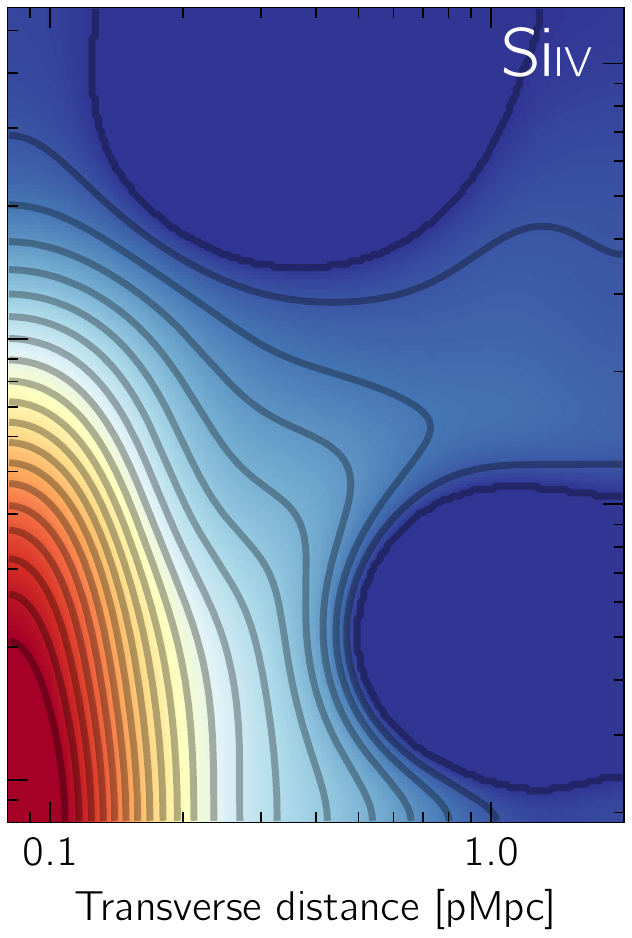} 
  \includegraphics[width=\mwc]{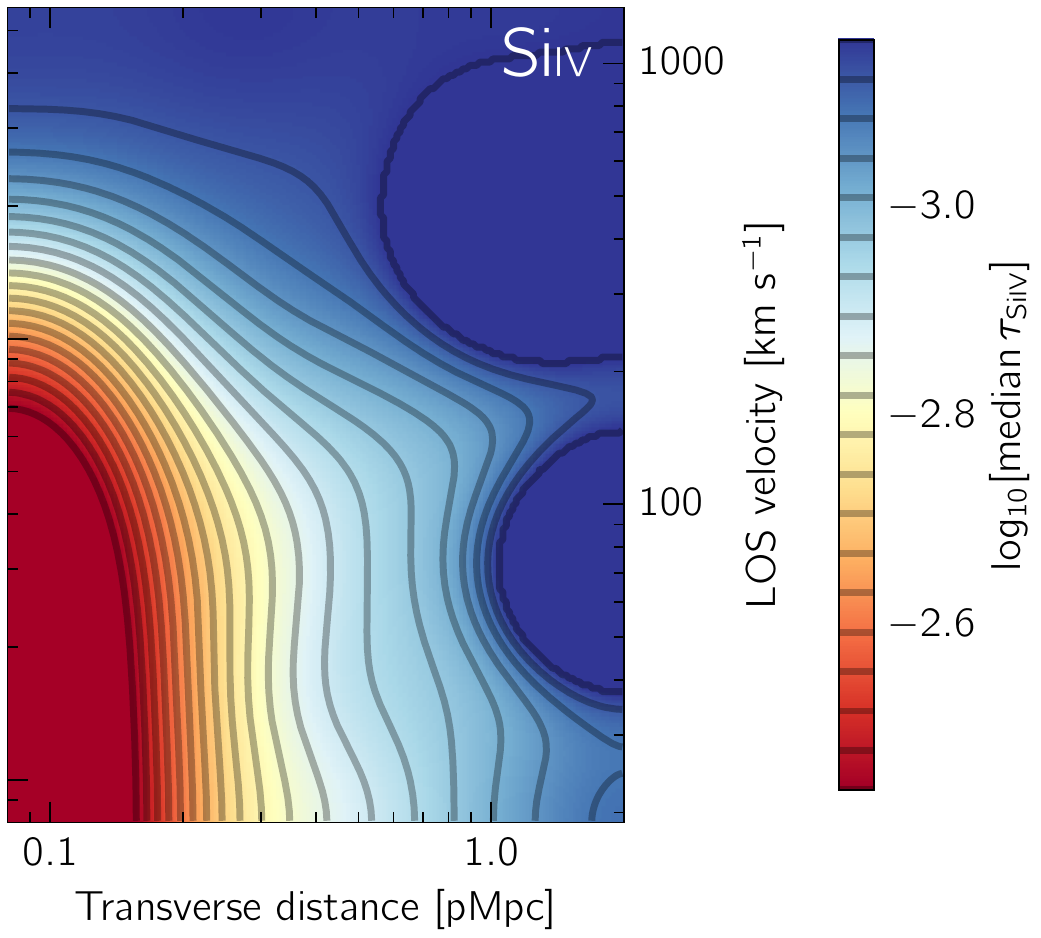} \\
  \caption{2-D maps of median metal-line optical depth around galaxies, for \hone\ (top row),
   \cfour\ (centre row) and \sifour\ (bottom row). The leftmost column shows the KBSS
    observations from \obspaper, while the following three
    columns display the results from the Ref-L100N1504 EAGLE simulations, with an increasing
    minimum halo mass of $\mhminm=10^{10.5}$, $10^{11.5}$, and $10^{12.5}$~\msol\ from left to right. 
    The observations have been smoothed with a Gaussian that has a $\sigma$ equal to the bin size (0.15~dex). 
    The minimum optical depth values used in the colour scale are set to the median optical depth for 
     random regions, which for each ion is by construction
    identical for the simulations and observations, while the maximum optical depth
    values are set by the maxima in the observations.
    The $p$-values, i.e. the probability that a model (with a given \mhmin) is consistent with 
    the observed unsmoothed data, are given in Table~\ref{tab:2dmap_chisq}.  }
 \label{fig:maps}
\end{figure*}

 \begin{table*}
 \caption{
   $p$-values (i.e. the probability that a model, with a given \mhmin, is consistent with 
    the observed unsmoothed data) from Fig.~\ref{fig:maps}. We measure $p$-values
    using data from the full maps, and for the case of \hone\ also for the same subset of bins used to measure 
     halo mass as in \citet{rakic13}. The data rule out all models with $\mhminm\geq10^{12.0}$~\msol,
    while all but the full map of \cfour\ rule out  $\mhminm=10^{10.5}$~\msol. 
}
\input{tables/2dmap_chisq.tab}
\label{tab:2dmap_chisq}
\end{table*}

\subsection{2-D optical depth maps}
\label{sec:maps}

The primary results of \obspaper\ can be summarized by their
Fig.~4, where they presented the first 2-dimensional (2-D)  maps of metal-line
absorption around galaxies. These maps
 were constructed by  using the galaxy redshifts to associate each galaxy with a 
wavelength region in the QSO spectrum, as illustrated in Fig.~\ref{fig:schematic}.
The recovered pixel
optical depths for every galaxy were then binned by the TD
and LOS distance to the galaxy, and the maps were made by taking the median
pixel optical depth in each bin. 

We reproduce the \obspaper\ maps of the median observed \hone, \cfour\ and \sifour\ optical depths in
the leftmost column of Fig.~\ref{fig:maps}. \obspaper\ noted
the presence of a strong enhancement of the absorption near galaxies with respect to random 
regions that extends $\sim180$~pkpc in the transverse direction and 
$\sim\pm240$~\kmps\ along the LOS.
A second result was the detection of weak excess absorption 
out to a transverse distance of 2~pMpc, the maximum 
impact parameter probed, for \hone\ and \cfour. 

The observations are compared with simulated samples with minimum
halo masses of $10^{10.5}$, $10^{11.5}$, and $10^{12.5}$~\msol, shown
respectively in columns 2--4 of Figure~\ref{fig:maps}.  
Although we will explore the comparison more quantitatively in the subsequent
figures, we can already infer some characteristic behaviours from the 2-D maps. 
Primarily, we find that the strength of absorption, characterized by the pixel optical depths, by metals and \hone\ in the vicinity
of galaxies increases with halo mass (from left to right), and that the effect is stronger for the metals.
For \cfour\ and \sifour, it appears that the $\mhminm=10^{11.0}$~\msol\
model produces too little absorption while the $\mhminm=10^{12.0}$~\msol\ produces too much, 
which suggests that the data agree most closely with a mass that lies within this range. 

We would like to assess the goodness-of-fit of each of the models, 
however the $\chi^2$ statistic cannot be used because the errors are correlated and non-Gaussian.
Instead, we compute $p$-values for each \mhmin\ considered
here in the following manner. For a given realization of the simulated data,
we use 10,000 bootstrap realizations of the observations (performed by bootstrapping
the galaxies in each transverse bin)
and determine, for each bin in the transverse and/or LOS direction, the percentile of the bootstrap
realization that agrees with the simulations,
where the percentiles are normalized to range from 0 to 50
(that is, if a percentile was $>50$, it was subtracted from $100$). 
Then, the percentiles corresponding to each LOS-impact parameter bin 
are divided by 100 and multiplied together to give a probability of the model.
This procedure is also applied to all of the bootstrap realizations of the data,
so that each realization has an associated probability. The $p$-values which we quote here
are then given by the fraction of the realizations of the data that have a probability lower
than that of the model. We note that if we pretend that the observations 
are the model, then this procedure yields $p$-values 
$\geqq0.988$. We reject models with $p$-values $<0.05$. 

In Table~\ref{tab:2dmap_chisq}, we present the $p$-values for each \mhmin,
for both the full (unsmoothed) map shown in Fig~\ref{fig:maps} as well as for distances along the LOS limited to the four bins within
170--670~\kmps\ (or 0.71--2.83~pMpc for pure Hubble flow), as these bins were used in \citet{rakic13} to measure
the halo mass from the \hone\ data.\footnote{The exact bin edge values differ slightly from \citet{rakic13},
as the ones used here were chosen to be consistent with \citet{rakic12} and \obspaper.} \citet{rakic13} chose this velocity range
because there they found the \hone\ absorption to be insensitive to variations in subgrid physics models.  
For \hone, only the $\mhminm=10^{11.0}$~\msol\ model is consistent with the full map,
but the $p$-value for $10^{11.5}$~\msol\  (0.033) is less than a a factor of two smaller than for
$\mhminm=10^{11.0}$ (0.059). However, when using the restricted bin range of \citet{rakic13}, 
the $\mhminm=10^{11.5}$~\msol\ model
is accepted with $p=0.34$, in good agreement with \citet{rakic13}. 

Fitting to the full range of data
points or the subset from \citet{rakic13} may, however, not be the best approach for \sifour, 
because most of the map is dominated by noise, as we do not detect significant 
absorption beyond $\sim500$~pkpc in the transverse
direction. Indeed, we find that 
all models are formally ruled out by the data, although the disagreement is only
marginal for $\mhminm=10^{11.5}$~\msol\ ($p=0.049$). 
On the other hand, \cfour\ fares much better, likely because the absorption above \taurndcfour\ extends
out to the full impact parameter range of 2~pMpc. For the full map, we find that models with
 $\mhminm=10^{10.5}\text{--}10^{11.5}$~\msol\ are consistent with the \cfour\ observations.

\begin{figure*}
  \includegraphics[width=\wa]{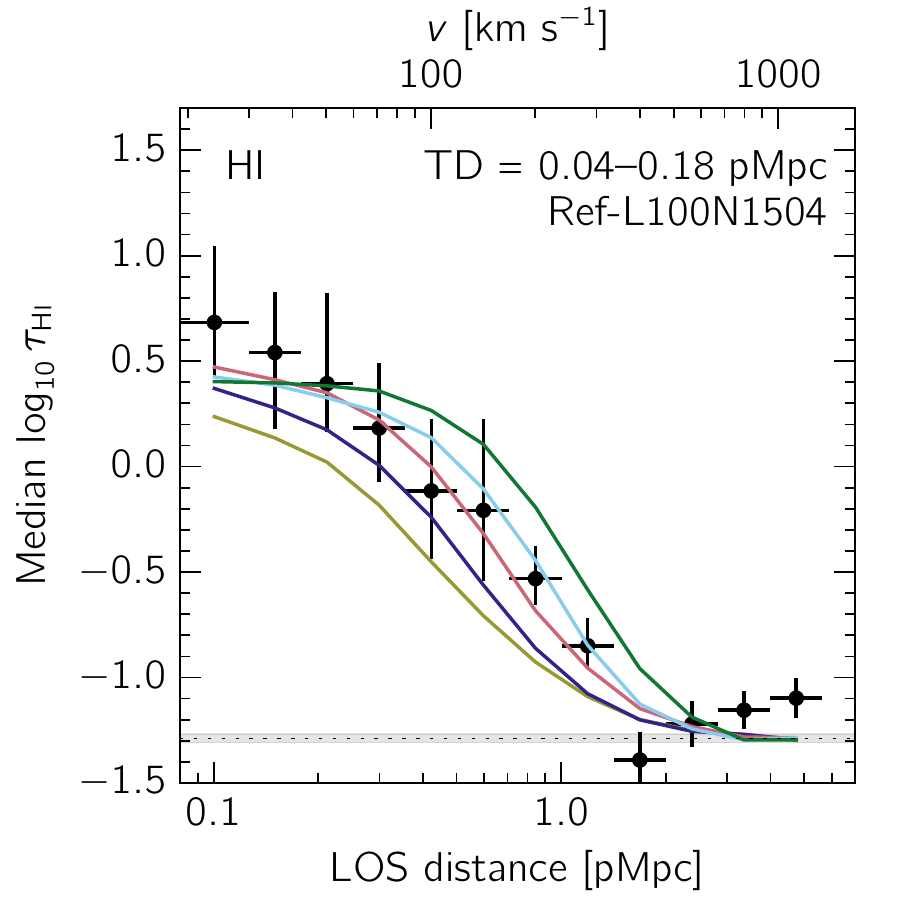} 
  \includegraphics[width=\wa]{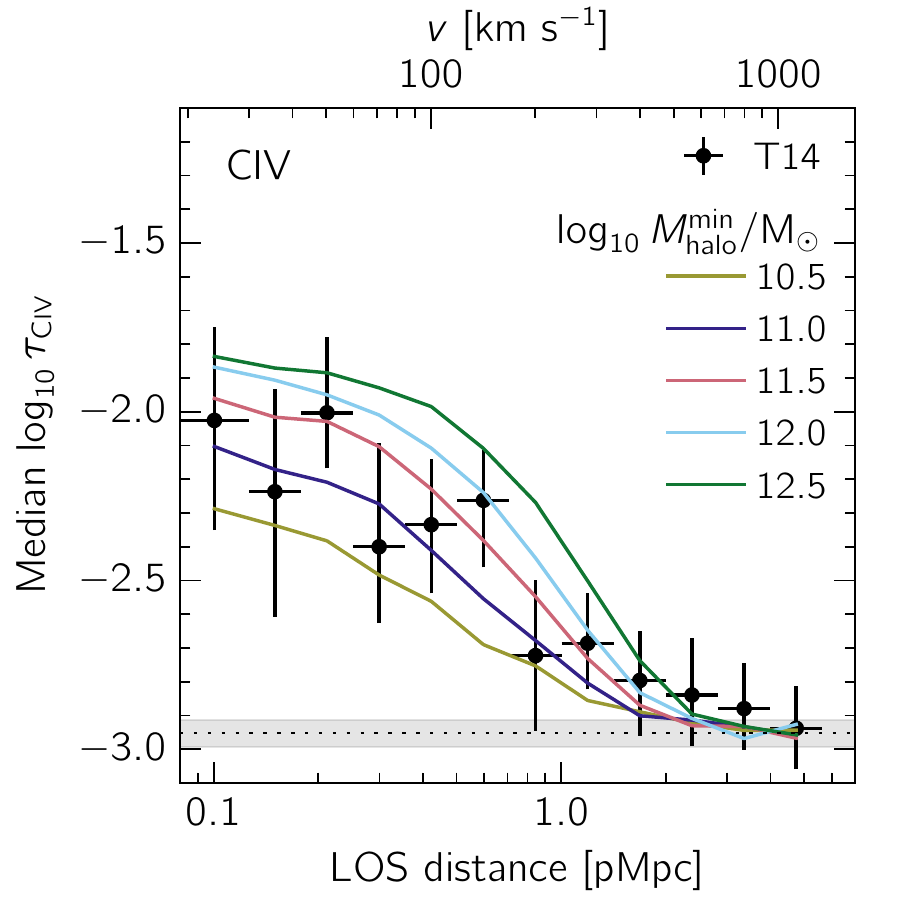} 
  \includegraphics[width=\wa]{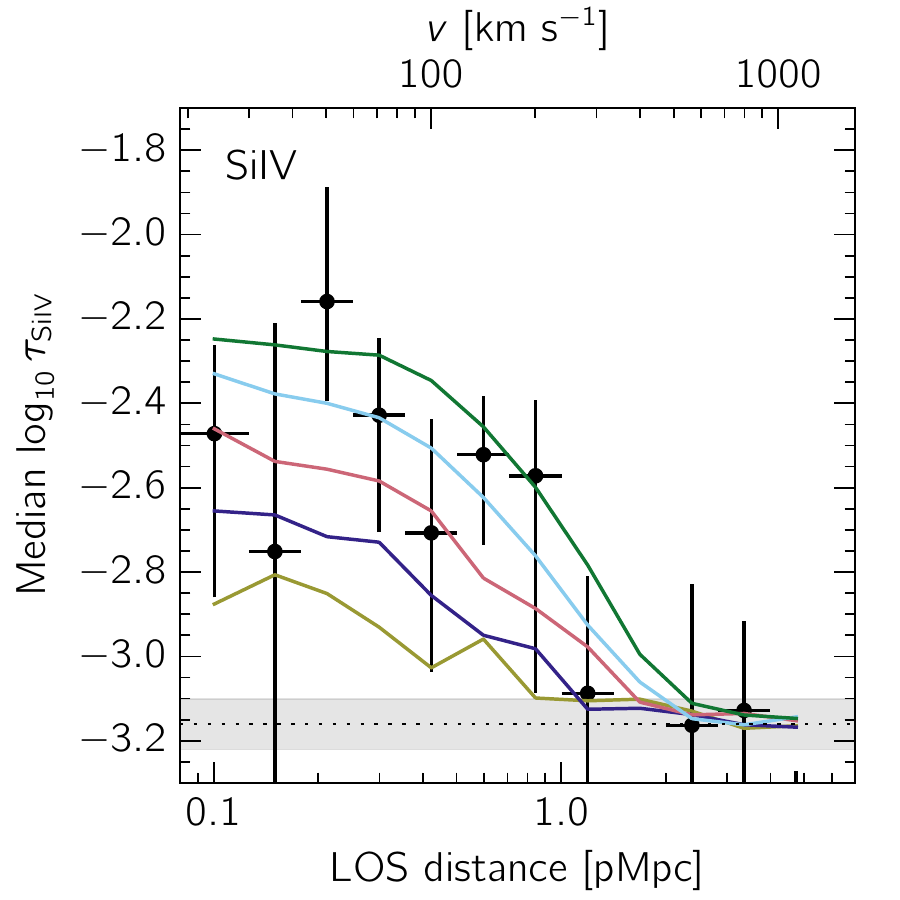} \\
 \caption{Cuts taken along the LOS through the maps from Fig.~\ref{fig:maps} 
 (right column of Fig.~6 from \obspaper). The upper horizontal axis indicates the velocity 
 along the LOS, while the lower horizontal axis gives the corresponding physical distance
 assuming pure Hubble flow.
 The size of the transverse bin, 0.04--0.18~pMpc, was chosen
 to include the strong absorption in both \hone\ and metals.
 The observations from \obspaper\ are shown as black circles with 
 1-$\sigma$ error bars, while the simulations
 are denoted by the lines where each colour represents a different minimum halo mass.
 The horizontal dotted lines indicate \taurnd, and the grey region shows the 1-$\sigma$ scatter.
 The $p$-values for each halo mass model, based on the comparison shown here and on the 
 cuts along the transverse direction shown in 
  Fig.~\ref{fig:cuts_trans}, are given in Table~\ref{tab:cuts_chisq}.
Only the model with $\log_{10} \mhminm / \msolm = 11.5$ is consistent with all three ions.}
 \label{fig:cuts_los}
\end{figure*}

\subsection{Cuts along the LOS and transverse directions}

To avoid comparing noise with noise, we confront the simulations with 
the regions of the 2-D maps that contain the most signal (Fig.~6 from \obspaper).
In Fig.~\ref{fig:cuts_los} we show
a cut that runs along the LOS, where the inner two impact parameter bins 
(ranging from 35 to 180~pkpc) were combined. From left to right, 
the panels show the median optical depth as a function of LOS distance
for \hone, \cfour\ and \sifour. Note that the y-axis range depends on the ion 
in question, and that the dynamic range is much larger for \hone\ than for the metal
ions.  The observations are shown with black circles, while the coloured curves show the simulation
results for halo masses ranging from 
$10^{10.5}$ to $10^{12.5}$~\msol, in intervals of 0.5~dex. 

Examining \hone\ in the left panel, we see that 
the \hone\ optical depths are most sensitive
to halo mass at intermediate velocities ($v\approx100$--$200$~\kmps). For all minimum halo masses considered,
the optical depths appear to be systematically lower than the $\sim2$
innermost measurements (out to 40~\kmps). However, we emphasize that the errors along the LOS
are correlated (to scales of $\sim10^2$~\kmps, see \citealt{rakic12}) and that the observations 
and model agree at the 1-$\sigma$ level for all but the lowest halo masses, 
so this apparent discrepancy is not significant. 

Turning next to the metals, we find that in contrast to \hone, the median optical
depths within 60~\kmps\ of the galaxy positions are sensitive to
halo mass. This halo mass dependence continues up to $\sim600$~\kmps\ where all curves asymptote to \taurnd.
Overall, we find that all the minimum halo masses considered here
are consistent with the qualitative behaviour of the observations, which is that
the metal-line absorption is strongly enhanced with respect to \taurnd\ out 
to $\sim\pm200$~\kmps\ along the LOS.

\begin{figure*}
  \includegraphics[width=\wa]{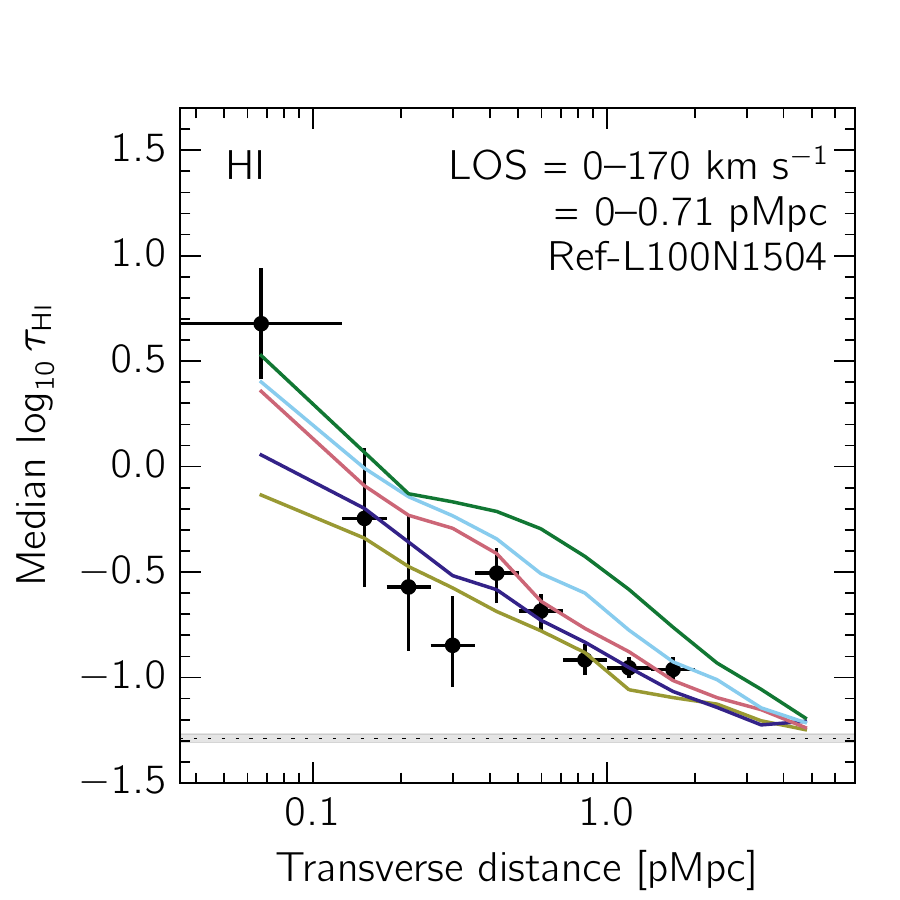} 
  \includegraphics[width=\wa]{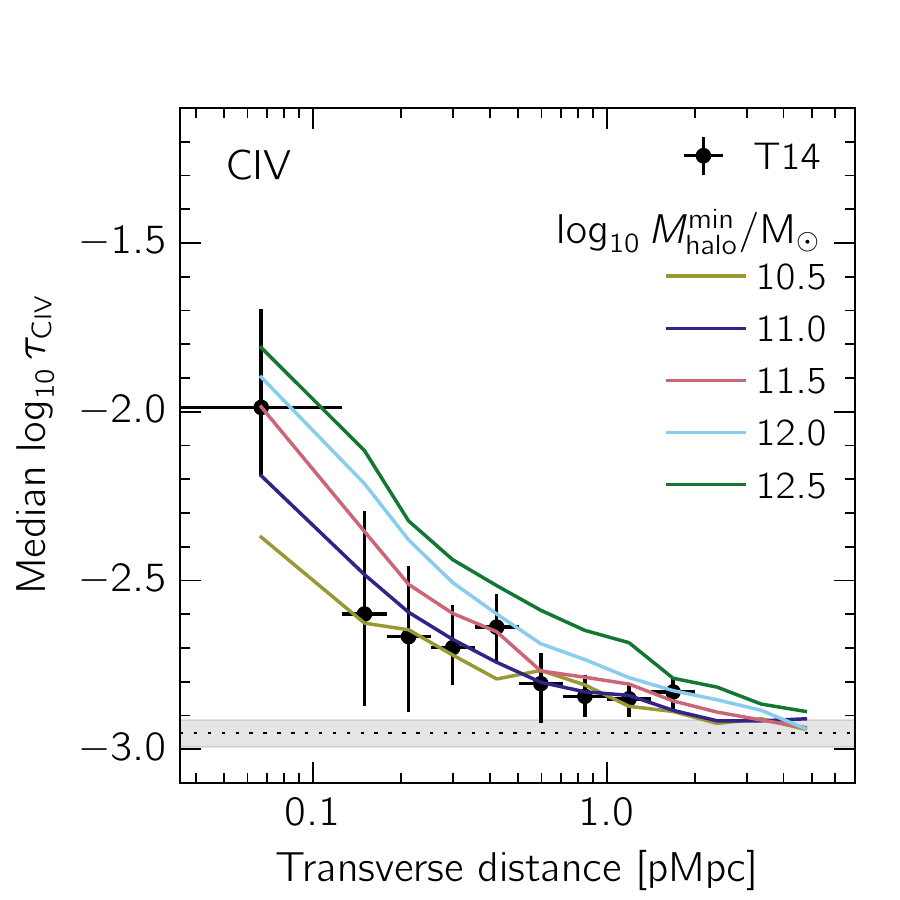} 
  \includegraphics[width=\wa]{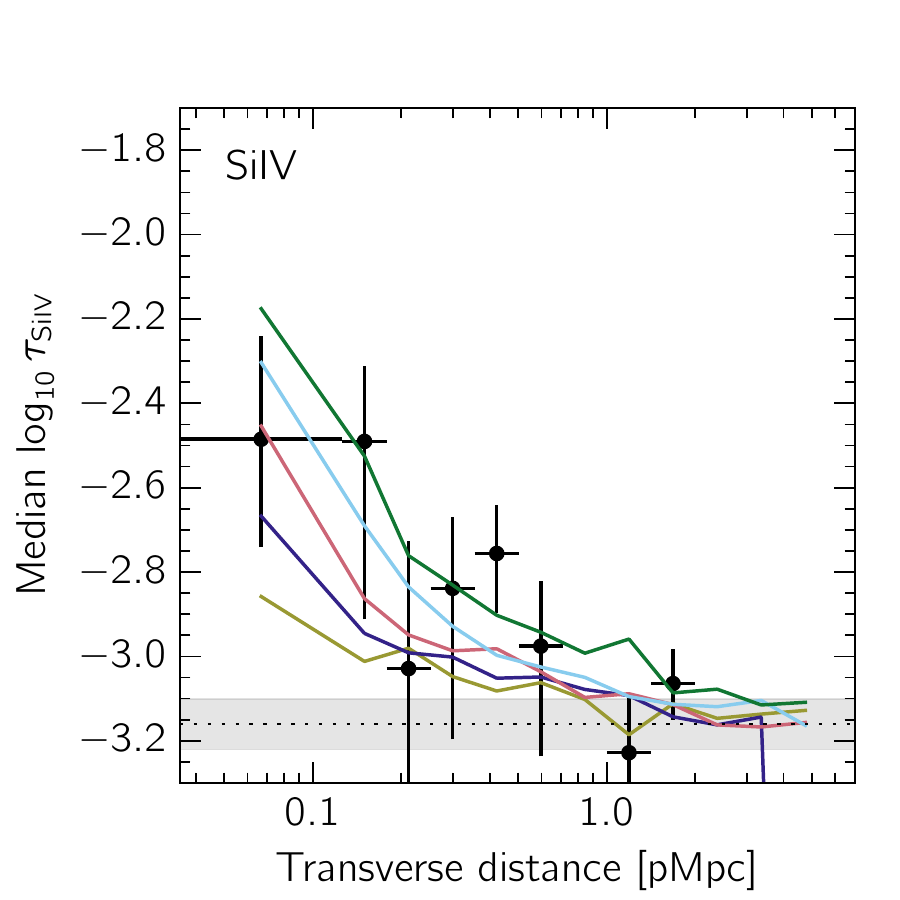} \\
 \caption{The same as Fig.~\ref{fig:cuts_los}, except taking a cut along the 
  transverse direction  of Fig.~\ref{fig:maps} with a width
   $\Delta v =\pm170$~\kmps\ of the LOS (left-hand column of Fig.~6 in \obspaper) 
  The enhancement of \hone\ and \cfour\ absorption out to the 
  maximum observed impact parameter of 2~pMpc is also seen in the 
  simulations, where it extends out to $\sim5$~pMpc. The $p$-values resulting from a simultaneous comparison to this
  data plus that in Fig.~\ref{fig:cuts_los} are given in Table~\ref{tab:cuts_chisq}. }
 \label{fig:cuts_trans}
\end{figure*}

 \begin{table}
 \caption{
 $p$-values for the models given the data shown in Figs.~\ref{fig:cuts_los} and \ref{fig:cuts_trans}.
 Only the model with $\mhminm=10^{11.5}$~\msol\ is consistent with all the data.
 }
\input{tables/mass_cuts_chisq.tab}
\label{tab:cuts_chisq}
\end{table}

In Fig.~\ref{fig:cuts_trans} we examine cuts along the transverse direction, 
with a velocity width of $\pm170$~\kmps. This velocity cut was chosen because
it corresponds to the scale on which $\tau_{\honem}$ is smooth, and because
it is larger than the correlation scale \citep{rakic11}.

The simulations again capture the qualitative behaviour of the observations. 
The enhancement in the absorption decreases with the impact parameter and this 
gradient decreases with the distance from the galaxy. 
For all halo masses considered, the \hone\ and \cfour\ median
optical depths show enhancement above \taurnd\ out to $\approx5$~pMpc in EAGLE.
\obspaper\ postulated that the significant excess absorption extending out to the maximum
impact parameter of 2~pMpc
in the observations (i.e., many virial radii away from the host galaxy) is likely due
to clustering effects. Support for this argument is given by Fig.~\ref{fig:box_test}, 
which shows that the median optical depths are not fully converged for 
box sizes $<50$~cMpc, where clustering on $\sim$Mpc scales is not properly captured.

Finally, in Table~\ref{tab:cuts_chisq} we present the $p$-values of simultaneous
comparisons to the data from Figs.~\ref{fig:cuts_los} and \ref{fig:cuts_trans}. 
The \hone\ data are consistent with \mhmin\ between 
$10^{11.0}$ and $10^{11.5}$~\msol. While the \cfour\ data only exclude
the highest minimum halo mass model, $10^{12.5}$~\msol, the \sifour\ data are consistent 
with $\mhminm=10^{11.5}\text{--}10^{12.0}$~\msol. Therefore, only 
the model with $\mhminm=10^{11.5}$ is consistent with all of the data. 

\begin{figure*}
  \includegraphics[width=\wa]{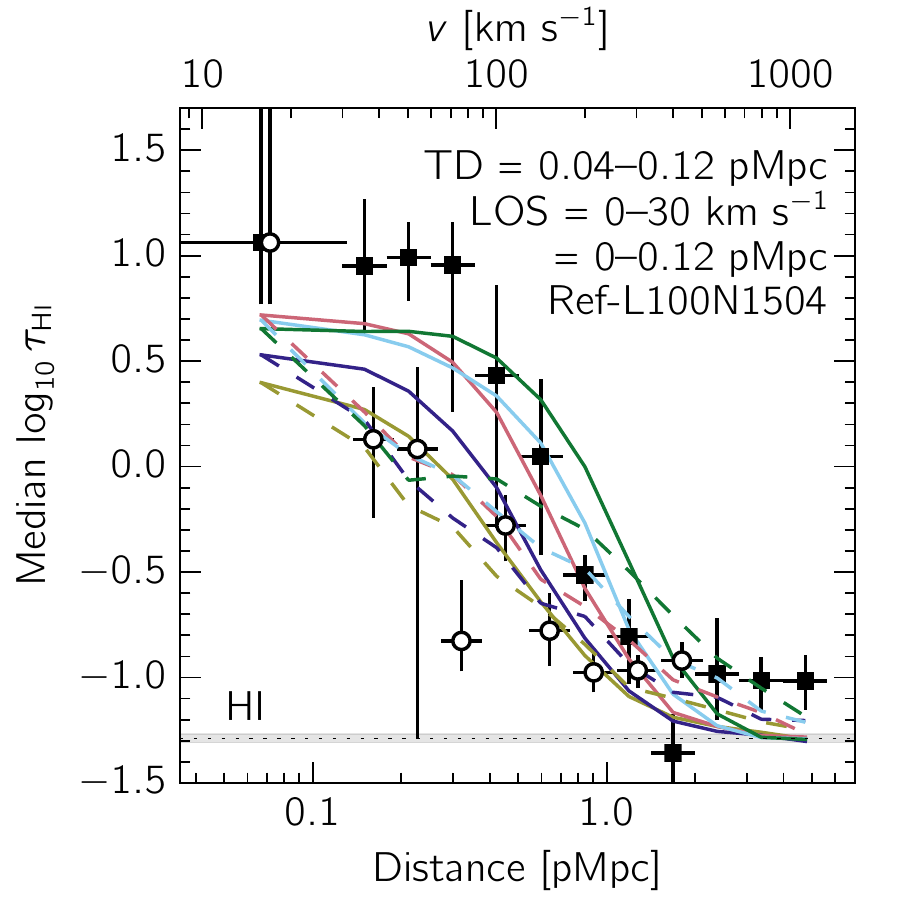} 
  \includegraphics[width=\wa]{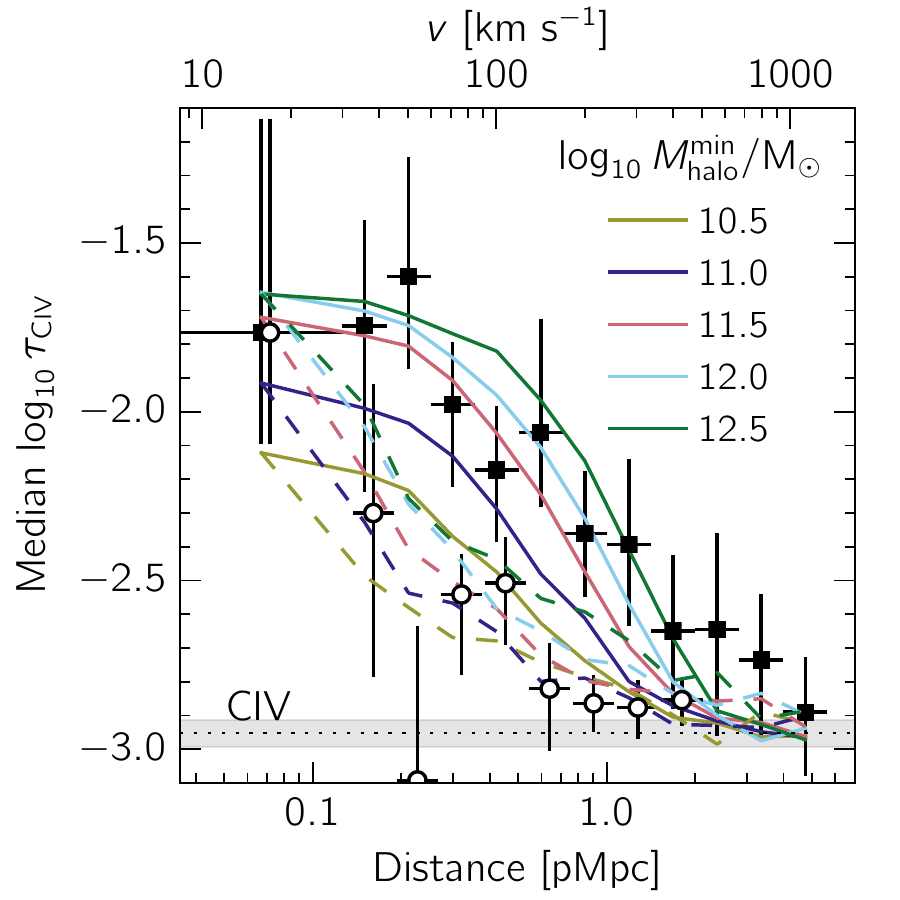} 
  \includegraphics[width=\wa]{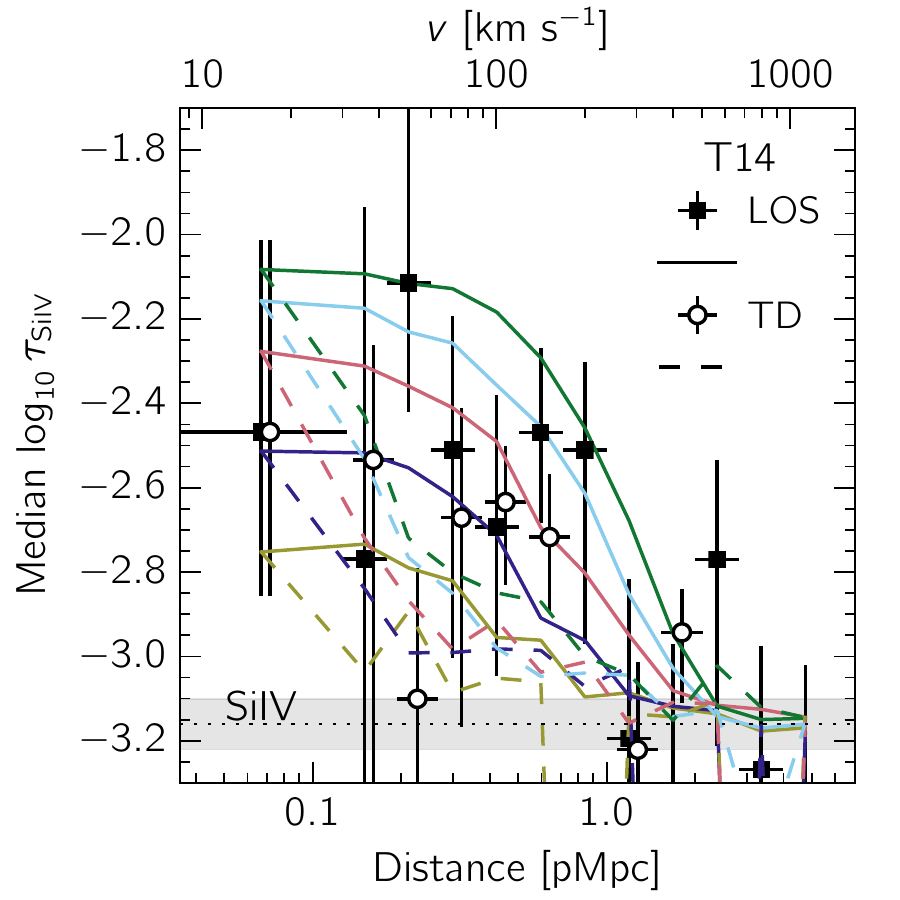} 
  \caption{The median optical depth from cuts of equal size taken in the
  transverse direction (0.04--0.12~pMpc) and along the LOS ($\Delta v=\pm30$~\kmps, 
  equivalent to 0.12~pMpc assuming pure Hubble flow), 
  in order to study the redshift-space distortions (Fig.~5 in \obspaper). 
  The observed medians and 1-$\sigma$ errors are indicated in black, while the coloured lines show 
  simulations using different \mhmin. The filled squares and solid lines denote the 
   median optical depths along the line-of-sight, denoted LOS, while the open circles and dashed lines indicate
   the transverse direction, denoted TD. The $p$-values for the models are given in Table~\ref{tab:zsd_chisq}.
  }
 \label{fig:zsd}
\end{figure*}

 \begin{table}
 \caption{$p$-values for the models given the data shown in Fig.~\ref{fig:zsd}. While the \hone\ data
   rule out all minimum halo masses except for $\mhminm=10^{11.5}$~\msol, the 
   metal-line absorption data are less constraining.
  }
\input{tables/mass_zspacedist_chisq.tab}
\label{tab:zsd_chisq}
\end{table}

\subsection{Redshift-space distortions}

In Fig.~\ref{fig:zsd} we examine redshift-space distortions,
 i.e., how the absorption along the LOS differs from that of
 the transverse direction (Fig.~5 from \obspaper). We have plotted the median optical depths from the innermost
bin (TD=0.04--0.13~pMpc, $\Delta v=\pm30$~\kmps) along the LOS (solid lines, filled squares for the observations) and the transverse
direction (dashed lines, open circles for the observations). 
Note that along the LOS, this bin spans a total of 60~\kmps, so its size is not significantly smaller
than the magnitude of most redshift errors. 

Fig.~\ref{fig:zsd} demonstrates that the data and models show strong and comparable redshift-space distortions.
The $p$-values for the comparisons
are given in Table~\ref{tab:zsd_chisq}. The \hone\ data rule out all models
except for $\mhminm=10^{11.5}$~\msol, but the metal data do not provide strong
constraints on the model halo mass. Although a simple ``chi-by-eye''
would eliminate the lowest-mass models, we reiterate that the errors are non-Gaussian and 
that along the LOS they are correlated. 

The results for \hone\ in Fig.~\ref{fig:zsd} can be directly contrasted with the top left panels 
of Fig.~7 (for points along the LOS) and Fig.~8 (for points in the transverse direction)
in \citet{rakic13}. In these figures, the authors examined the same 
optical depth profiles from the KBSS observations as are shown here, but using an 
older set of KBSS measurements with larger redshift errors, and compared the results to the OWLS
simulations. They found that in the smallest transverse distance
bin, the OWLS simulations significantly underestimated the highest observed \hone\ optical depths.
The relative success of the EAGLE simulations compared to those presented in \citet{rakic13}
 can be partly attributed to the smaller redshift errors in the latest
KBSS data, however even after removing redshift errors the \hone\ optical depths
in  \citet{rakic13} were still well below those from the observations. We conclude that the EAGLE
simulations provide much better agreement to observations of \hone\ in absorption
than was found for OWLS.

\begin{figure*}
  \includegraphics[width=\wa]{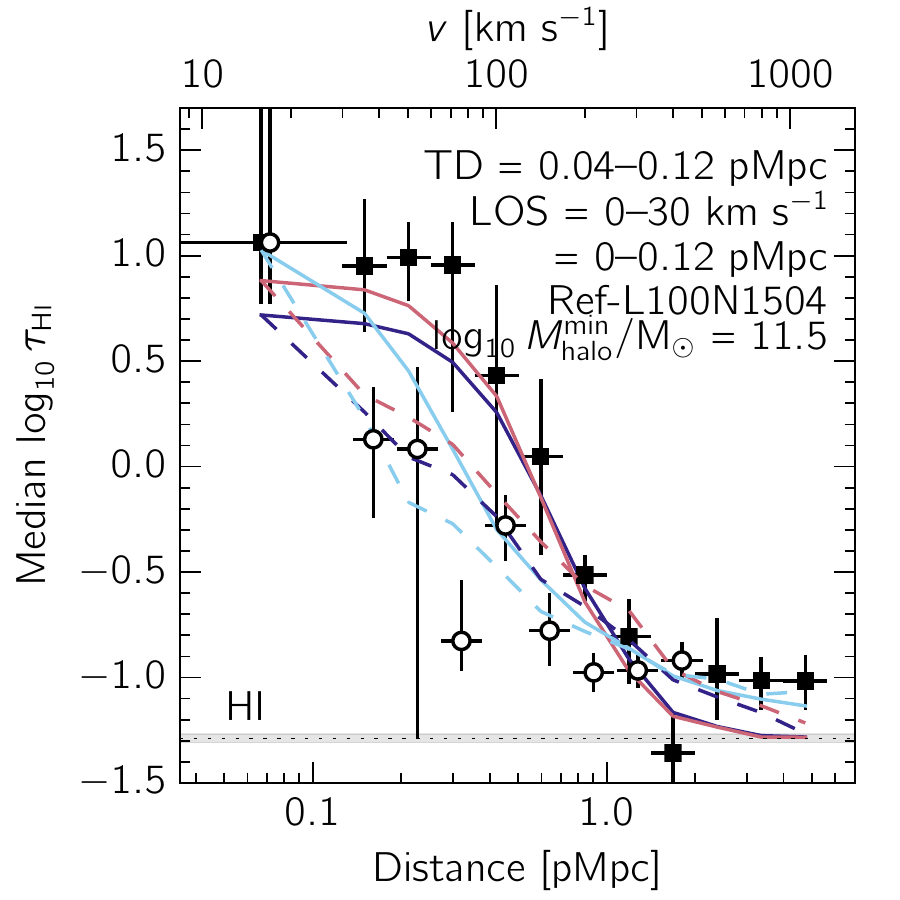} 
    \includegraphics[width=\wa]{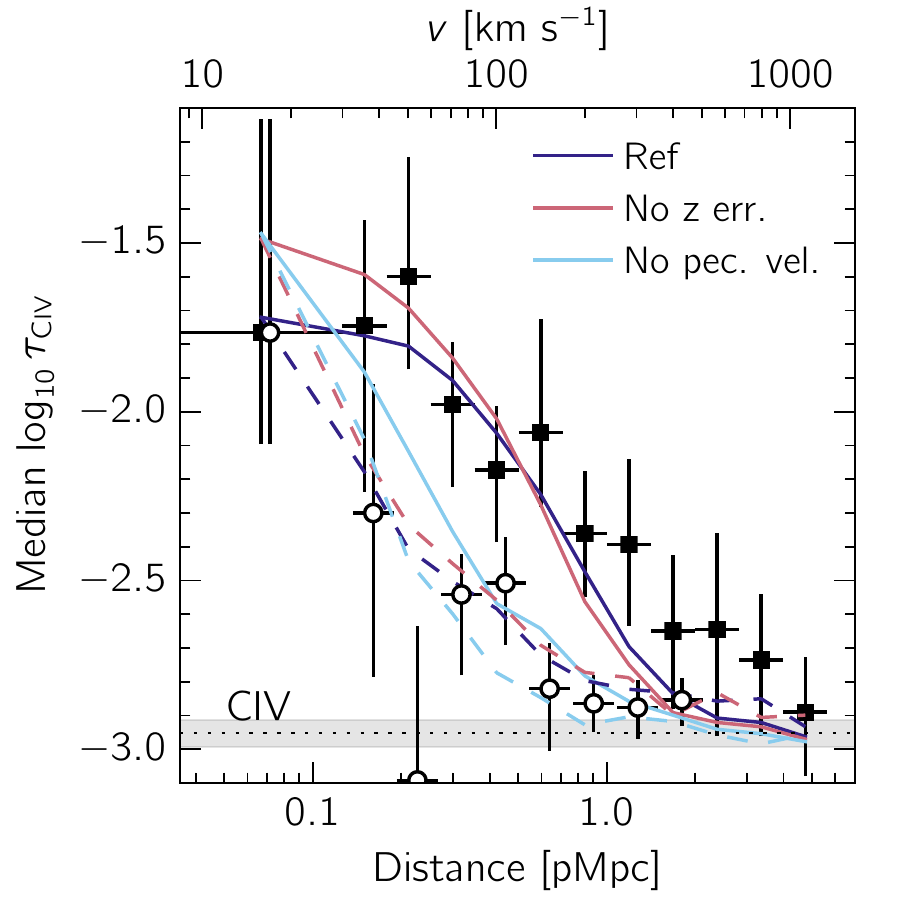} 
  \includegraphics[width=\wa]{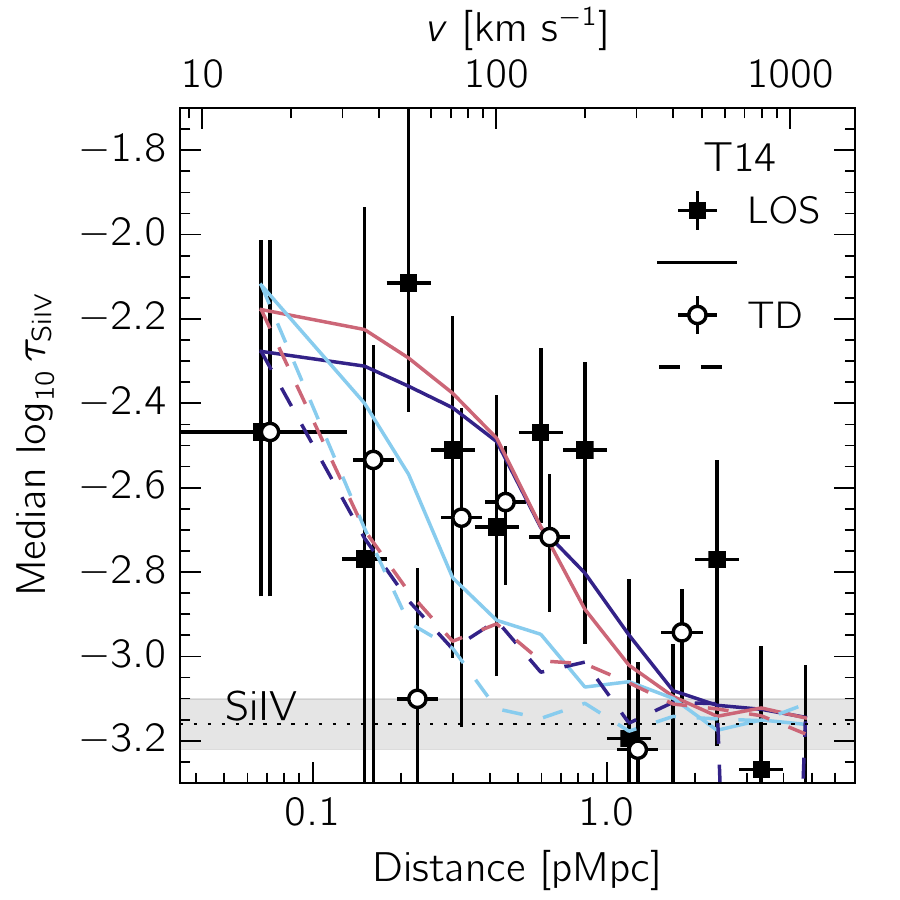} \\
   \caption{The same as Fig.~\ref{fig:zsd}, but only presenting $\mhminm=10^{11.5}$~\msol, and
    comparing to models without redshift errors (red line) and without peculiar velocities (cyan line).
    Without redshift errors, the median optical depths decrease more steeply along the LOS, and 
    there is also more absorption along the transverse direction for the velocity bin shown (0--30~\kmps).
    Without peculiar velocities,
    the median optical depths along the LOS and transverse directions are almost identical, with the 
   former still being slightly smoothed by the redshift errors. Thus, peculiar velocities are required
   to obtain the large observed redshift space distortions.}
 \label{fig:nopecvel}
\end{figure*}

 \begin{table*}
 \caption{$p$-values for the models given the data shown in Fig.~\ref{fig:nopecvel}. 
 In this case, rather than computing the likelihood of the individual pixel optical depth 
 values of a model given the data, we compute the likelihood of the difference between 
 the LOS and TD optical depths, i.e. the magnitude of the anisotropy. We find that models
 in which peculiar velocities have been turned off are ruled out by the \hone\ and 
\cfour\ data. 
  }
\input{tables/nopecvel_chisq.tab}
\label{tab:nopecvel_chisq}
\end{table*}

\section{Origin of redshift-space distortions}
\label{sec:variations}

In this section we will exploit the simulations to probe the origin of the observed redshift-space
distortions, which may be caused by gas inflows, outflows, virial motions, or galaxy redshift errors. 
We begin by attempting to disentangle the effects of redshift errors from gas peculiar velocities. 
In Fig.~\ref{fig:nopecvel}, we show
the same redshift-space distortions as in Fig.~\ref{fig:zsd}, but we
have now added a simulation  without any redshift errors. 
As demonstrated previously in \citeauthor{rakic13} (\citeyear{rakic13}, see also \citealt{tummuangpak14})
we find that the removal of redshift 
errors increases the median optical depths at small galactocentric distances,
both in the transverse direction and along the LOS. However, the effect is modest
because the KBSS redshift errors are small for most galaxies (see \S~\ref{sec:gms}).
Thus, it is likely that the elongation along the LOS is mostly due to differences between
the peculiar velocities of the galaxy and the absorbing gas.

To test this, we have
run \texttt{SPECWIZARD} with galaxy and gas peculiar velocities set to zero, and
plot the result as the cyan lines in Fig.~\ref{fig:nopecvel}. 
The redshift-space distortions for the cyan lines 
are strongly suppressed compared to the other cases, i.e., the solid and dashed
lines are much closer together, with the remaining difference due to redshift errors.
Turning off peculiar velocities
decreases the median optical depths along the LOS, except for the innermost bin. 
We note that if we exclude redshift errors and turn off peculiar velocities 
simultaneously, the optical depths along the LOS and in the transverse direction are 
nearly identical.

We know from the previous section that the models with $\mhminm=10^{11.5}$ are
not ruled out by the data. To check whether in the absence of redshift errors or
peculiar velocities the predicted redshift space distortions 
are consistent with the observations, instead of computing the $p$-values
using the individual pixel optical depth values, we use the difference in optical depths
between the LOS and transverse directions, i.e. the magnitude of the anisotropy. 
The results of this calculation are presented in Table~\ref{tab:nopecvel_chisq}.
While the data do not fully rule out the models without redshift errors, 
for \hone\ and \cfour\ we find that the models without peculiar velocities
do not produce redshift-space anisotropies consistent with the data.
We note that the \sifour\ data, which are much noisier than 
\hone\ and \sifour, are inconclusive on this point.
Overall, the simulations provide strong 
evidence that the redshift-space distortions arise due to the
motion of the gas, rather than galaxy redshift errors.  

Next, we can employ the simulations to explore the net \textit{direction} of the gas velocity, i.e.\ whether we are 
probing infalling, outflowing or rotating gas.
\obspaper\ measured a strong enhancement of the optical depths along the LOS to
about $\pm240$~\kmps, which is
close to the galaxies' circular velocities
($\approx217$~\kmps). Thus, it was not previously possible to differentiate
between the virial motion and outflow scenario. 
Other observations also do not offer conclusive evidence. On the one hand, 
\citet{steidel10} found ubiquitous outflows by 
using the KBSS galaxies as background sources to probe their own gas 
in absorption.
On the other hand, large scale infall of \hone\ has been observed not just around
galaxies in the KBSS \citep{rakic12}
but in other samples of Lyman break galaxies \citep[e.g.,][]{bielby16}. 
Indeed, simulations by \citet{kawata07} found that symmetric absorption features 
around a galaxy were produced by filamentary accretion. 

In Fig.~\ref{fig:velocity}, we present the
gas particle radial velocities as a function of radial distance from galaxy centres.
To calculate the velocities, we consider all galaxies in the simulation box above a specific mass threshold, 
and for each of these galaxies we calculate the radial velocity of each gas particle relative to 
its centre of mass. 
The median velocities are then weighted by, from left to right in 
Fig.~\ref{fig:velocity}, mass, volume, and ion mass for 
\hone, \cfour, and \sifour, respectively. 
We note that ion masses are calculated using the same interpolation 
tables that were used to generate the mock spectra.

While we have plotted the mass- and volume-weighted median
gas velocities to get a picture of the behaviour of the total gas content, 
in order to use the results from Fig.~\ref{fig:velocity} to interpret the observed redshift-space distortions, 
we argue that the ion mass-weighted quantities provide the most appropriate comparison. 
Along a single sightline that intersects various gas clouds, the pixels 
will be ion column density-weighted, given by   
the ion density multiplied by the kernel. However, a group 
of sightlines is also cloud cross section-weighted, since larger gas clouds 
are more likely to be intercepted. This effect contributes a kernel$^2$ term
to the weighting, and taken together, the product of the ion density and kernel$^3$
results in the ion mass.

\afterpage{
\begin{landscape}
\begin{figure}
\includegraphics[width=22cm]{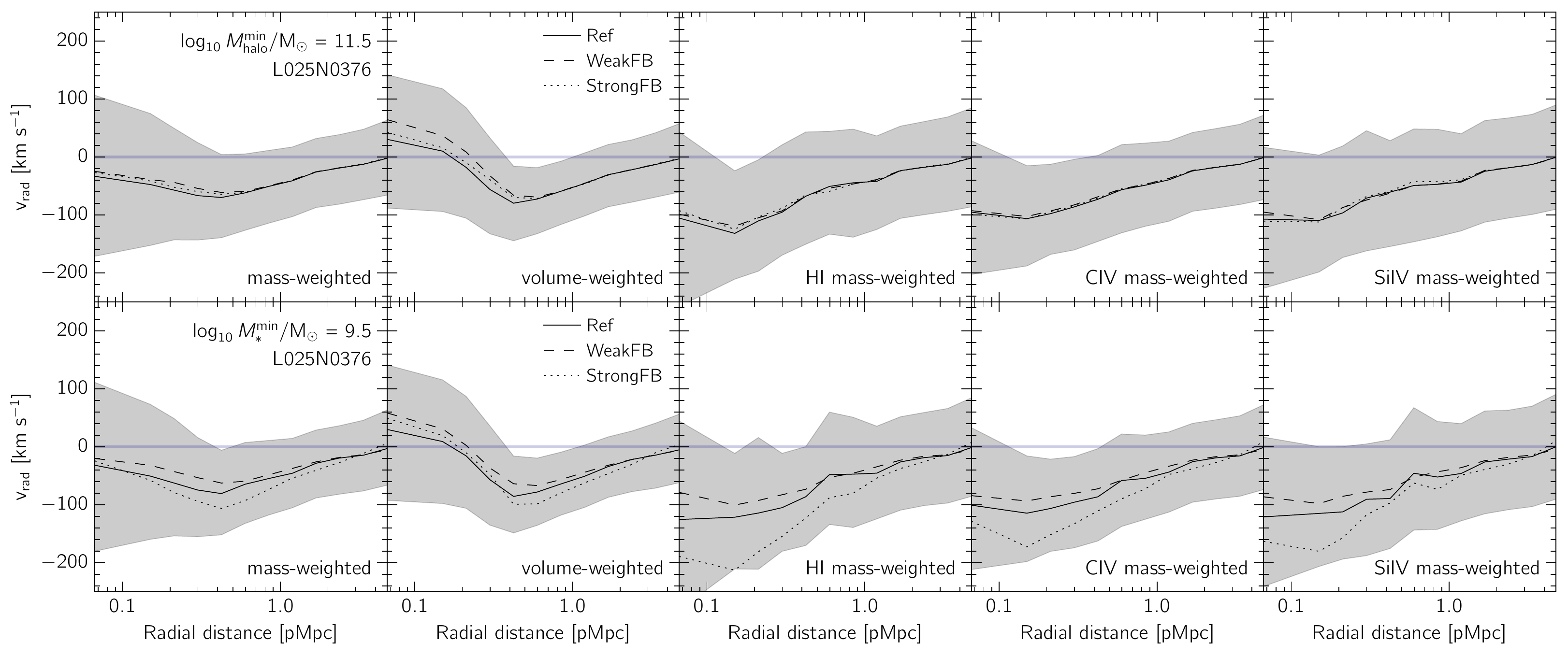} 
  \caption{Median gas particle radial 
  velocities as a function of radial distance from the galaxy centre,
  for the Reference (solid) model as well as the WeakFB (dashed) and StrongFB (dotted) variations.
  The grey regions show the 1-$\sigma$ scatter around the Reference model, and the horizontal blue line
  separates the positive (outflowing) and negative (infalling) directions. 
  From left to right the columns show mass-, volume-, \hone\ mass-, \cfour\ mass-, and \sifour\ mass-weighted medians, while the top and bottom rows 
  use a minimum fixed halo mass ($\mhminm=10^{11.5}$~\msol) and stellar mass ($\msminm=10^{9.5}$~\msol), 
  respectively. We find that the ion mass-weighted gas velocities are mostly negative (meaning that the gas is infalling), and that 
  the velocities depend on halo mass rather than stellar mass (since at fixed halo mass the 
  different models, which predict different stellar masses, have identical velocities). 
  This indicates that the ions examined here are primarily 
  tracing infalling, rather than outflowing, gas.}
 \label{fig:velocity}
\end{figure} 
\end{landscape}
}

In the top row of  Fig.~\ref{fig:velocity}, we show results for all galaxies 
with a minimum halo mass of $10^{11.5}$~\msol, while in the bottom row
we use a minimum fixed \textit{stellar} mass of $10^{9.5}$~\msol, and in each case we not only consider the 
reference model, but also the strong and weak feedback models (hence, the smaller L025N0376 box was used). 
We do so because we would like to know how the gas velocities vary with stellar mass at fixed halo mass,
and vice versa.  Table~\ref{tab:galpropa} demonstrates that for a fixed minimum halo mass, the StrongFB (WeakFB) model produces
galaxies with smaller (larger) stellar masses than the reference model. It follows that at a fixed minimum stellar mass,
the StrongFB (WeakFB) model galaxies reside in more (less) massive halos. 
Specifically for $\log_{10}\msminm/\msolm = 9.5$, we find median $\log_{10}\mhalom/\msolm = (11.5, 11.7, 12.1)$ for 
WeakFB, Ref, and StrongFB, respectively

The first point to note in Fig.~\ref{fig:velocity} is that 
the ion mass-weighted gas particle velocities, 
including the 1-$\sigma$ scatter, are mostly negative, which indicates that the net direction
of the gas probed by these ions is typically towards the host galaxy (infalling).  
In contrast, the mass- and volume-weighted velocities are more positive, and in the latter case
are mostly outflowing at the smallest radial distances. The volume-weighted velocities 
are most sensitive to outflows because outflows tend to be hot and diffuse with high 
volume-filling factors.

Next, focusing on the ion mass-weighted velocities in the top row, where we have 
used a fixed minimum halo mass, 
we find that varying the feedback model (and hence the stellar mass) has almost no effect on the gas 
velocities. Furthermore, in the bottom row where we have used a fixed minimum stellar mass, the gas velocities 
\textit{do} depend on the feedback model. In particular, at fixed stellar mass the velocities decrease
(become more infalling) with increasing feedback strength (and thus increasing halo mass). 
Since the galaxy gas accretion velocity is expected to increase with halo mass, in addition to the fact that we 
are seeing primarily negative velocities, this suggests
that \hone, \cfour\ and \sifour\ primarily trace infalling gas.

Since outflows are thought to be driven by energetic 
feedback from star formation and AGN, if the observed redshift-space distortions
were dominated by outflowing gas, one would expect a significant difference
between models with substantially different feedback physics. 
In Appendix~\ref{app:feedback}, we study how
the optical depth profiles are affected by varying the feedback model, and 
we find only modest differences in the metal-line 
absorption properties. This supports the above finding,
that the observed redshift-space anisotropies are dominated by 
virial motions and/or inflows, rather than by outflowing gas.


\section{Discussion and conclusions}
\label{sec:conclusion}

We have compared the circumgalactic \hone\ and metal-line absorption
in the EAGLE cosmological hydrodynamical simulations with observations.
EAGLE includes subgrid prescriptions for stellar and AGN feedback
that have been calibrated to match observations of the $z\sim0$ galaxy stellar
mass function, the black hole-galaxy mass relation, and galaxy disk sizes.
The fiducial EAGLE model has been run 
with a relatively high resolution 
($\epsilon \sim 1$~pkpc at $z\approx2$)
in a cosmologically representative box size (100~cMpc).

In this work we compare the results from EAGLE to observations from \obspaper.
\obspaper\ used data from the KBSS, a spectroscopic galaxy survey in 15 QSO fields,
to study the properties of the gas around 854 star-forming galaxies at $z\approx2$.
They applied the pixel optical depth technique to the QSO spectra
and combined the absorption information with the positions and redshifts of 
the galaxies in the field to construct the first 2-D maps of metal-line absorption
around galaxies. 

We have used the simulations to generate mock spectra that were designed
to mimic the properties of the QSO spectra from the KBSS.
The galaxy impact parameter distribution and redshift errors were also matched
to those of the observations. We compared 
the simulated and observed optical depths of \hone, \cfour\ and \sifour\ as 
a function of transverse and LOS separation from galaxies. 
 Our main conclusions are:

\begin{itemize}
  \item The galaxy stellar masses from EAGLE are in very good agreement with the 
  observations for $\mhminm=10^{11.5}$--$10^{12.0}$~\msol (corresponding to median halo
    masses of $10^{11.8}$--$10^{12.2}$), which is 
    the halo mass range inferred from observations \citep{trainor12, rakic13}. 
  The SFRs also show broad agreement, especially for 
   $\mhminm=10^{12.0}$, which matches the observed SFR PDF closely and 
  has a median SFR only a factor
  of two below that of the observations (Fig.~\ref{fig:galprop}).
  \item The EAGLE simulations reproduce the observed \hone, \cfour\ and
   \sifour\ absorption around galaxies in detail, including the redshift-space distortions.
  The observations are most consistent with a minimum halo mass of $10^{11.5}$~\msol, 
  in agreement with \citet{rakic13}
  (Figs.~\ref{fig:maps},  \ref{fig:cuts_los}, \ref{fig:cuts_trans}). 
  \item \obspaper\ detected enhanced median optical depths for
   \hone\ and \cfour\ in the transverse direction out to impact parameters of 
   2~physical Mpc (pMpc); the maximum in their sample). The simulations are consistent with this
   result, and predict that this enhancement extends 
   to $\approx5$~pMpc  (Fig.~\ref{fig:cuts_trans}). This large-scale enhancement 
    is not seen in the 25~cMpc simulation box, and we conclude that it is due to
     clustering (Fig.~\ref{fig:box_test}). 
  \item Although redshift errors smooth optical depths along the LOS 
    and reduce them slightly in the transverse direction, 
    the very small redshift errors for galaxies observed with 
    MOSFIRE ($\approx18$~\kmps) have a negligible  
   effect on the results (Fig.~\ref{fig:nopecvel}).
  \item Gas peculiar velocities generate the significantly higher
  median optical depths along the LOS compared to the transverse direction 
   (Fig.~\ref{fig:nopecvel}).
  \item The median ion mass-weighted radial gas velocities 
  indicate that the bulk of the absorbing gas is flowing inward towards the galaxies, 
  and that inflow rates increase with halo mass, but are insensitive to the 
  strength of feedback at fixed halo mass (Fig.~\ref{fig:velocity}).  
  We also find that the optical depths do not change significantly when the stellar
  feedback strength is varied by a factor of two (Fig.~\ref{fig:wind}), or when AGN feedback 
  is neglected (Fig.~\ref{fig:agn}). 
  It is therefore likely that the observed enhancement
   of optical depths along the LOS compared to the transverse direction is not caused by 
   outflows, but rather by infalling gas. 
\end{itemize}

The results from this work are seemingly in tension with those of \citet{steidel10}, 
who used  the same sample of galaxies as background sources to probe their own gas
in absorption , and measured ubiquitous 
outflows in \cfour\ and \sifour.  It is plausible that this discrepancy is due
to geometrical effects, for example such ``down-the-barrel'' observations may be sensitive to the gas 
at much smaller galactocentric distances.
As a caveat we mention that our 
conclusion that the gas is accreting is based entirely on the comparison with simulations.

It is also informative to compare these findings with those of \citet{turner15},
who used the same KBSS dataset and found that at small galactocentric distances, 
\osix\ traces hot, collisionally ionized, outflowing gas, in contrast to what 
we see in this study for \cfour\ and \sifour. However, \osix\ has
a higher ionization potential than \cfour\ and \sifour. 
Furthermore, \citet{turner15} found that the hot outflowing material can only
be identified when there is no chance superposition of pixels with 
high \hone\ optical depths, while
conversely, in this work we are considering all pixels close to galaxies.
Taken together, the results from \citet{turner15} and 
this work imply that the two distinct gas phases inhabit the same 
regions around galaxies, while having different kinematics. 
The hot phase may be significant, but it is much more difficult to detect.

As a next step, we would like to perform an analysis similar to \citet{turner15} 
using the simulations. 
However, \osix\ is not as straightforward to 
analyze in the simulations as the ions studied in this work, as it 
is located in the \lyb\ forest and therefore suffers
from strong contamination from \hone\ Lyman series lines. 
This contamination must be modelled, which we leave for a future study.

\section*{Acknowledgements}

We would like to thank the anonymous
referee for a constructive report that improved this article. 
This work used the DiRAC Data Centric system at Durham University, 
operated by the Institute for Computational Cosmology on behalf of 
the STFC DiRAC HPC Facility (www.dirac.ac.uk). This equipment was 
funded by BIS National E-infrastructure capital grant ST/K00042X/1, 
STFC capital grants ST/H008519/1 and ST/K00087X/1, STFC DiRAC Operations 
grant ST/K003267/1 and Durham University. DiRAC is part of the National
E-Infrastructure. This work was supported by the Netherlands
Organisation for Scientific Research (NWO), through
VICI grant 639.043.409 and by the European Research Council
under the European Union's Seventh Framework Programme
(FP7/2007- 2013) / ERC Grant agreement 278594-GasAroundGalaxies.
RAC is a Royal Society University Research Fellow.


\bibliographystyle{apj} 
\bibliography{bibliography}


\appendix

\section{Resolution and box size tests}
\label{app:test}

\begin{figure*}
  \includegraphics[width=\wa]{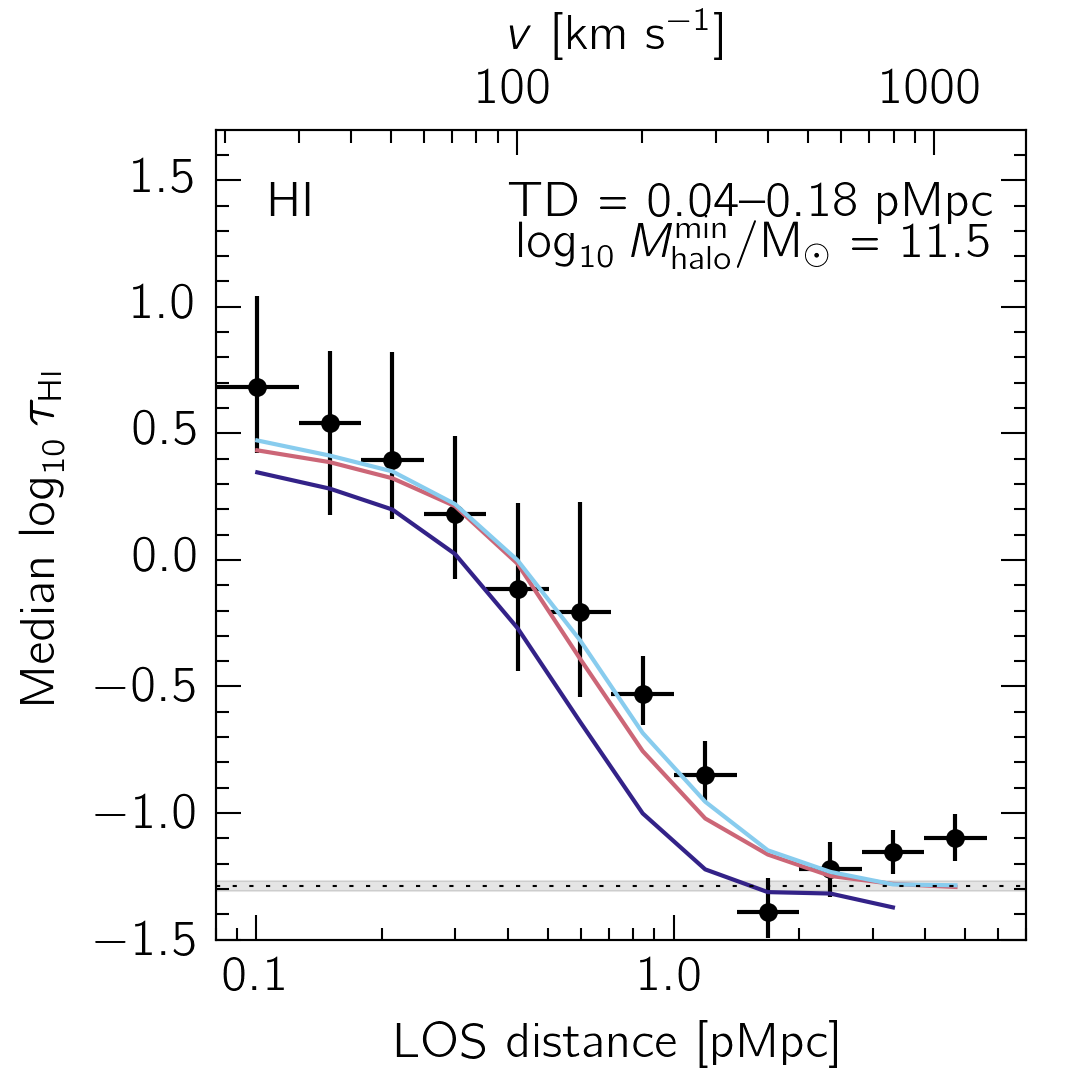} 
  \includegraphics[width=\wa]{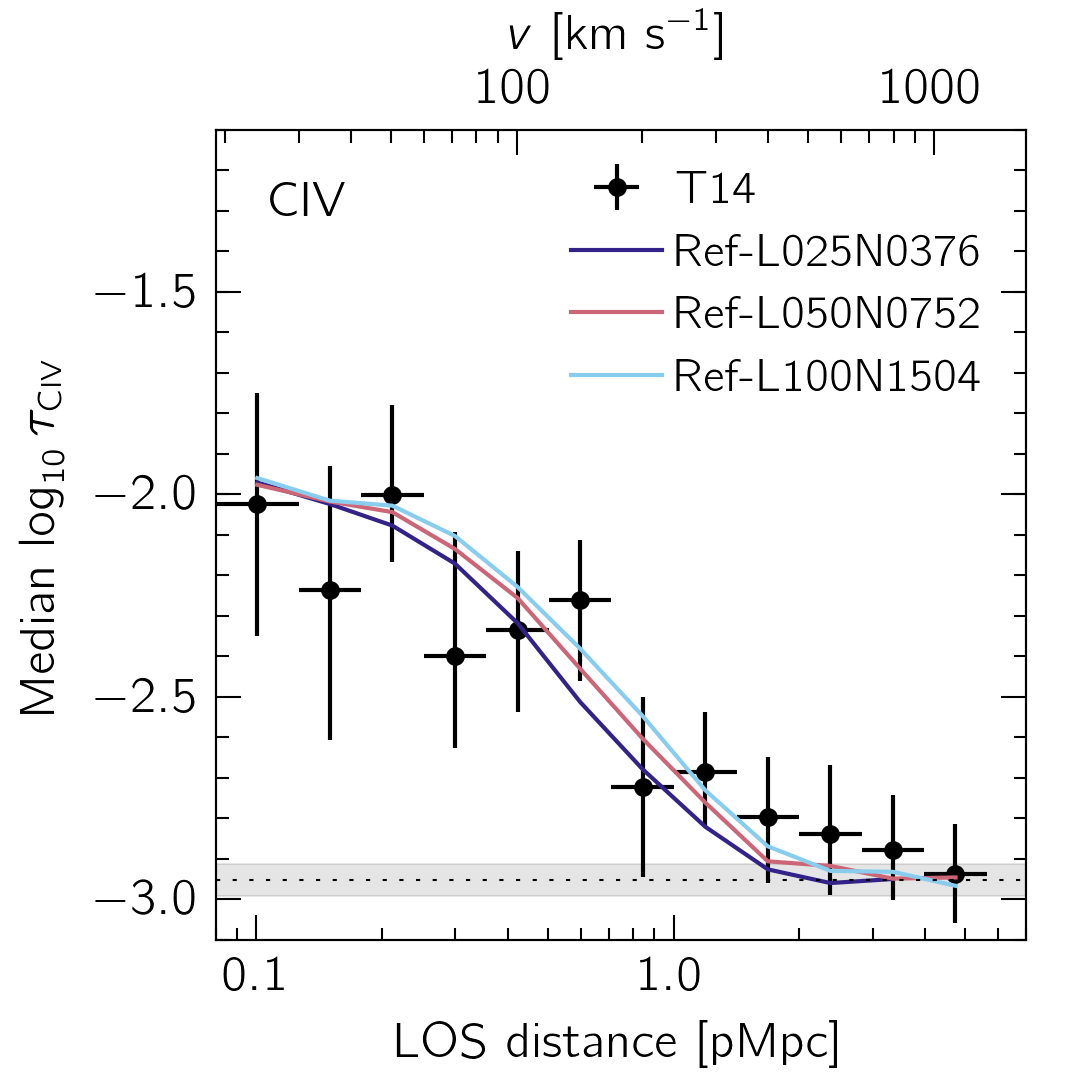} 
  \includegraphics[width=\wa]{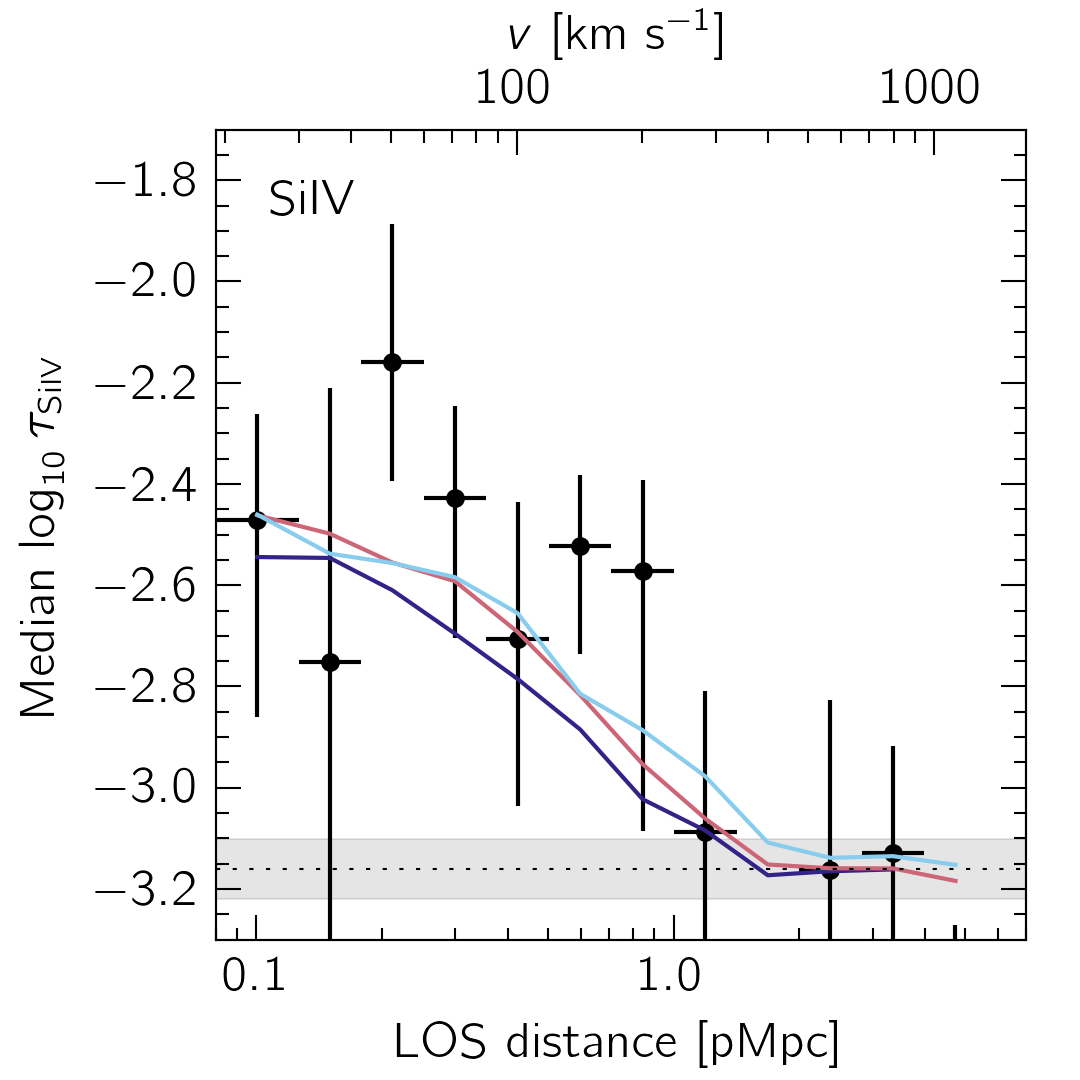} \\
  \includegraphics[width=\wa]{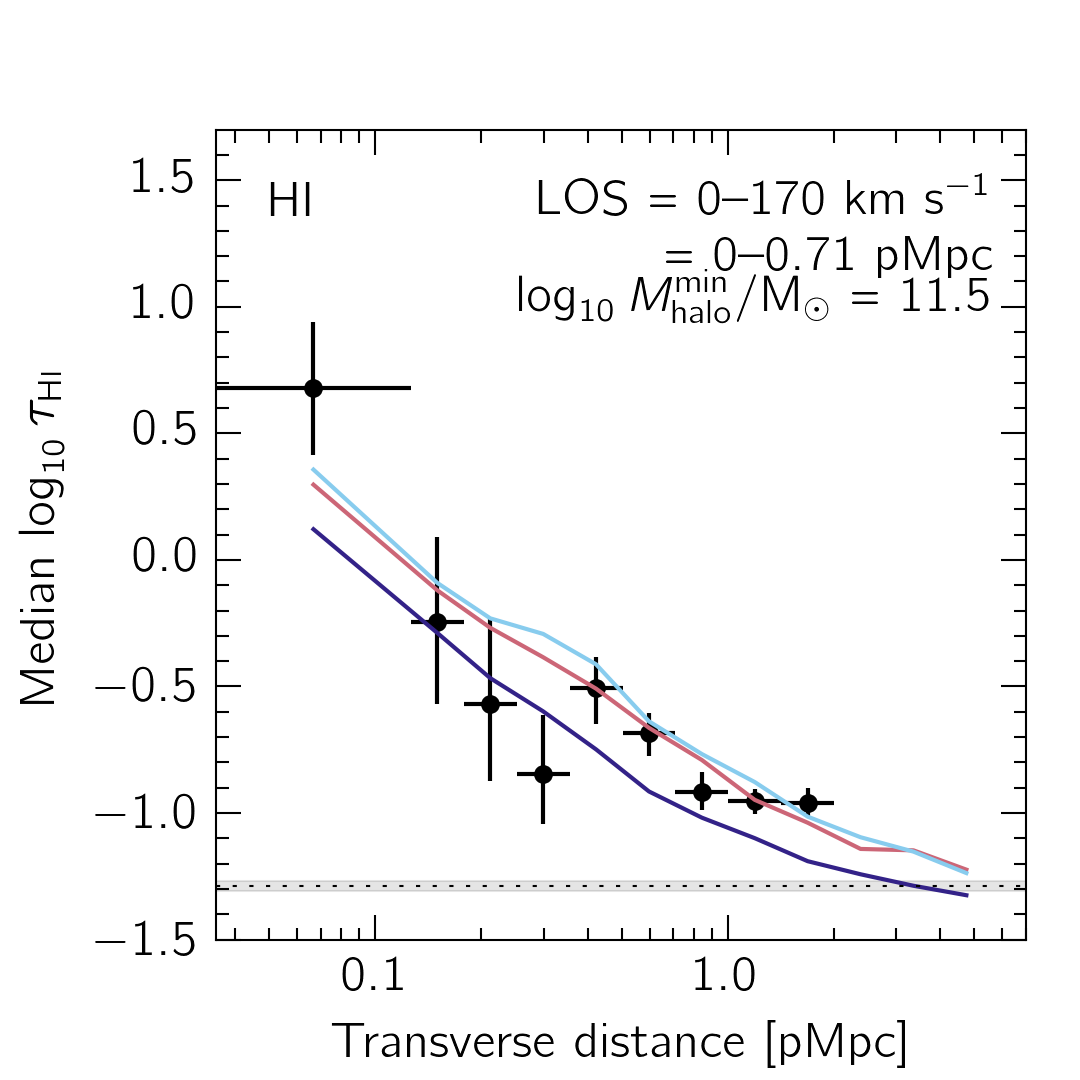} 
  \includegraphics[width=\wa]{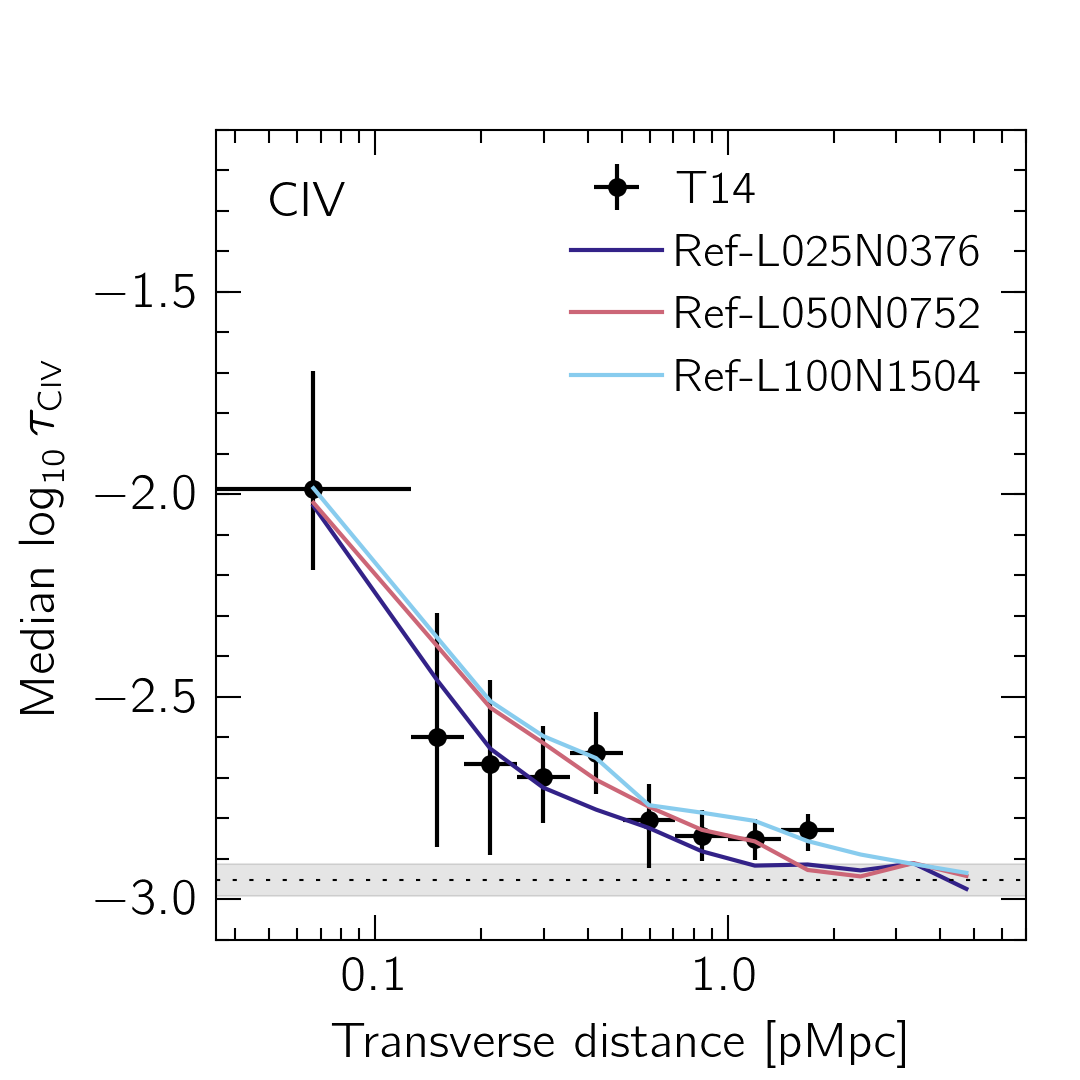} 
  \includegraphics[width=\wa]{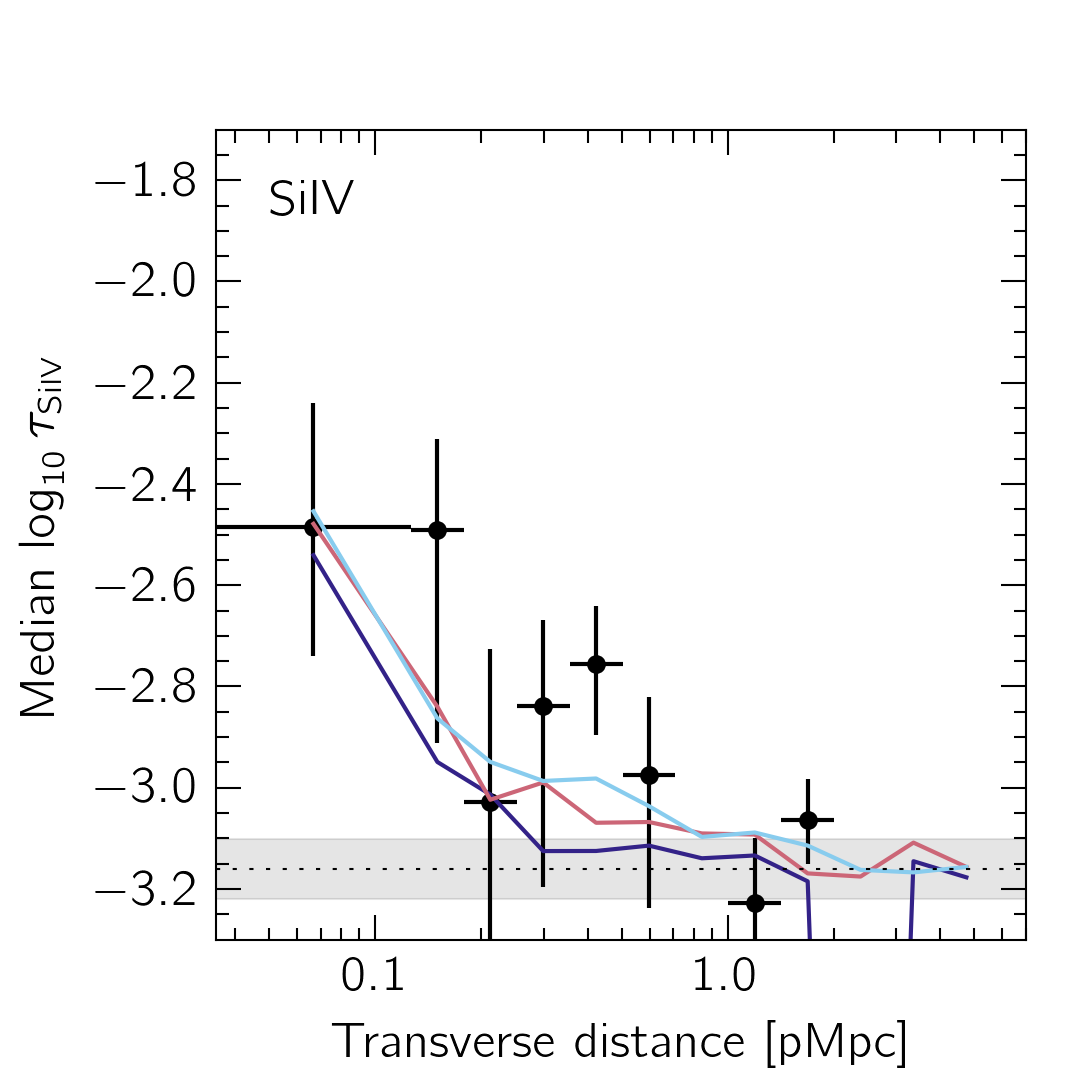} \\
 \caption{ 
Convergence with respect to simulation box size, where 
  we plot the median optical depths as a function of
   distance from galaxies along the LOS (top row) 
   and in the transverse direction (bottom row).
  From left to right we show \hone, \cfour, and \sifour\ for 
   $\mhminm=10^{11.5}$~\msol. 
  While the optical depths from the 25~cMpc box 
   are systematically lower than for the other runs,
our fiducial simulation is converged.
 }
 \label{fig:box_test}
\end{figure*}

\begin{figure*}
  \includegraphics[width=\wa]{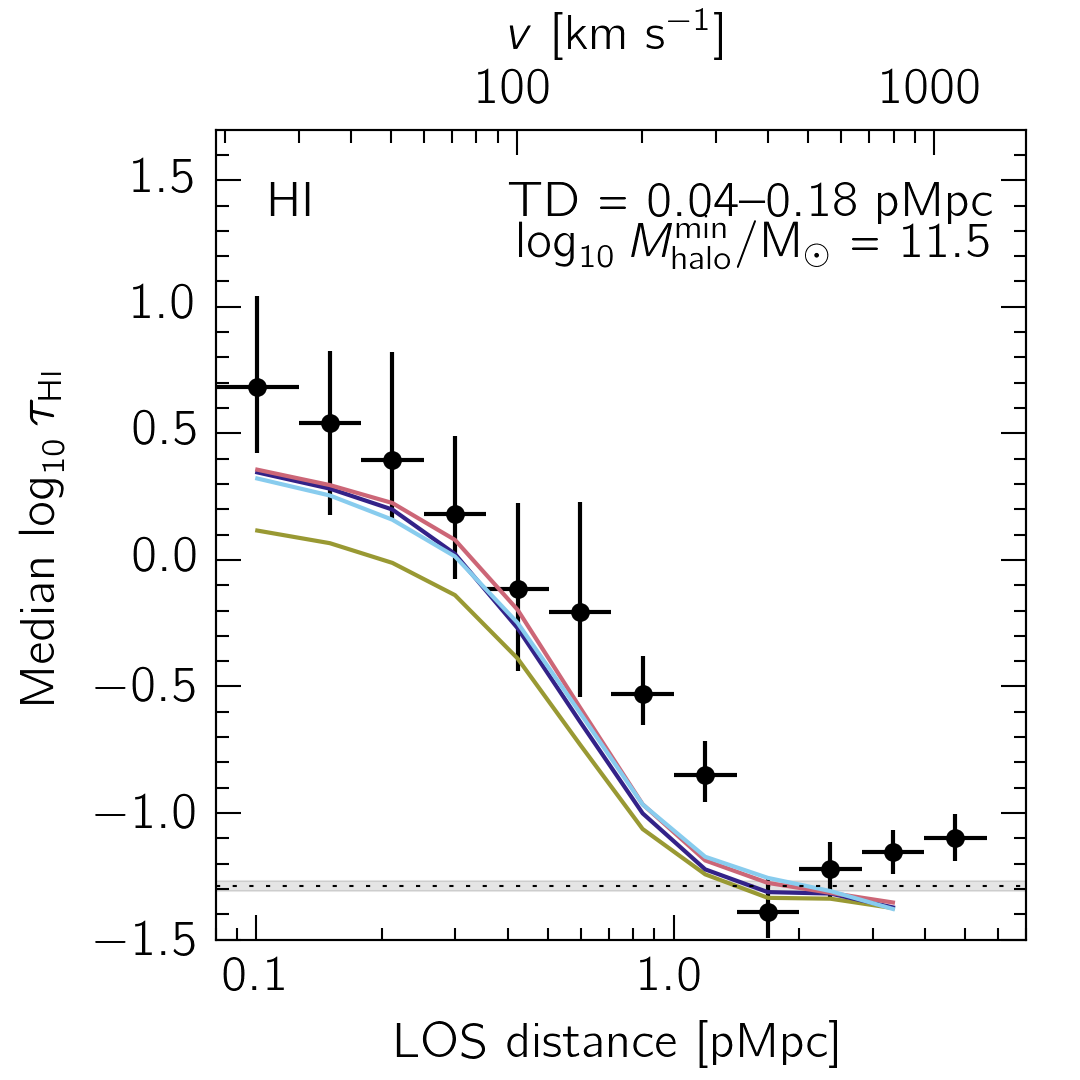} 
  \includegraphics[width=\wa]{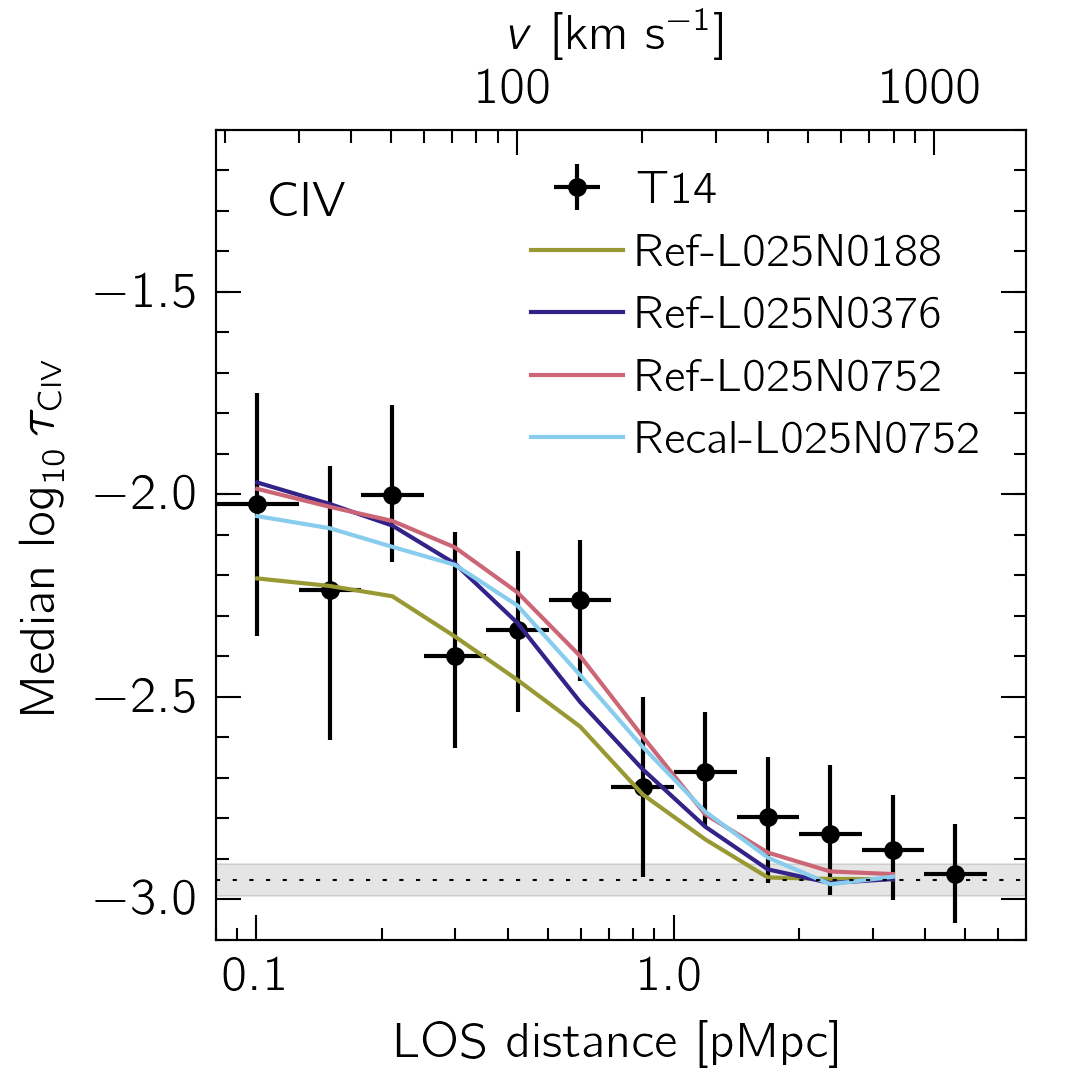} 
  \includegraphics[width=\wa]{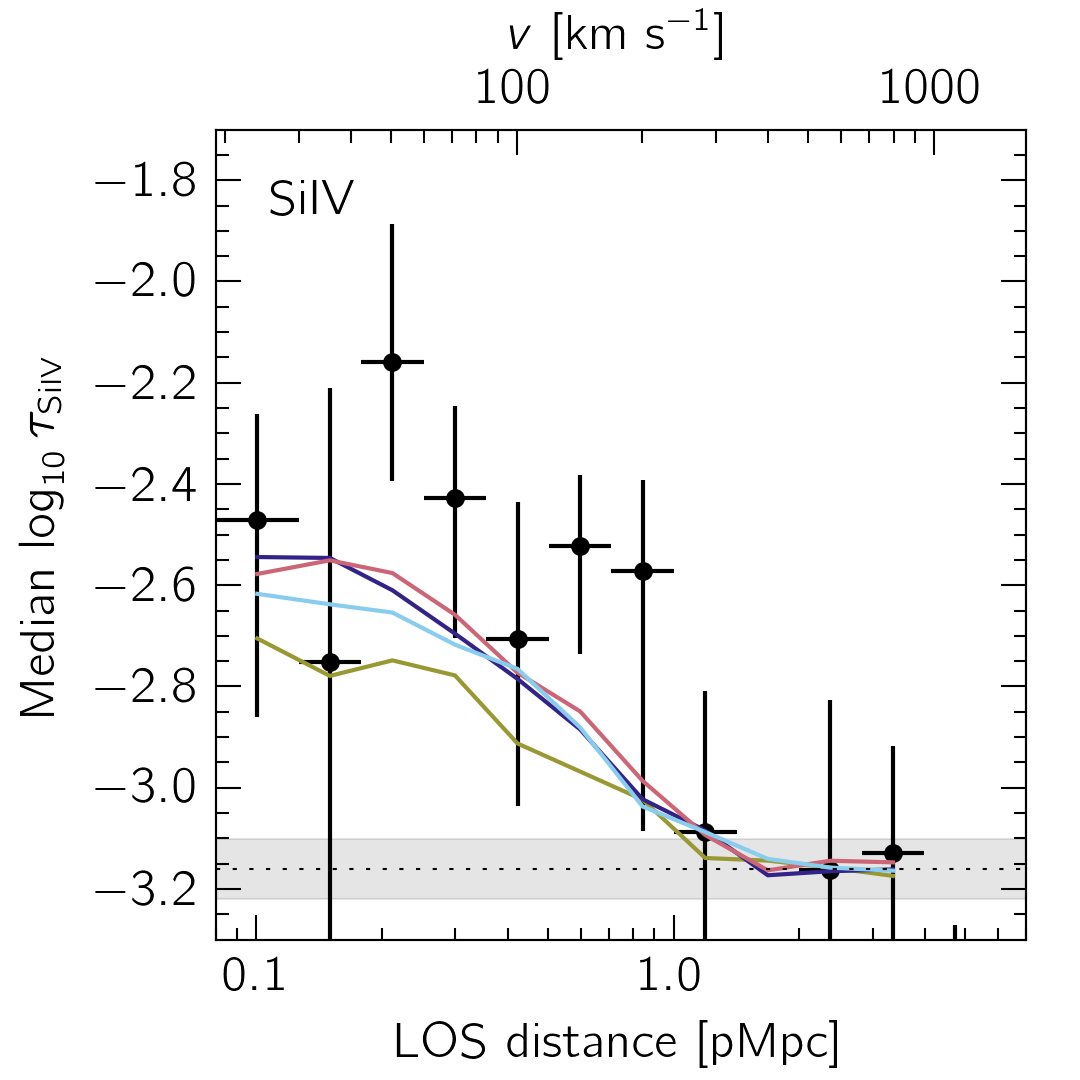} \\
  \includegraphics[width=\wa]{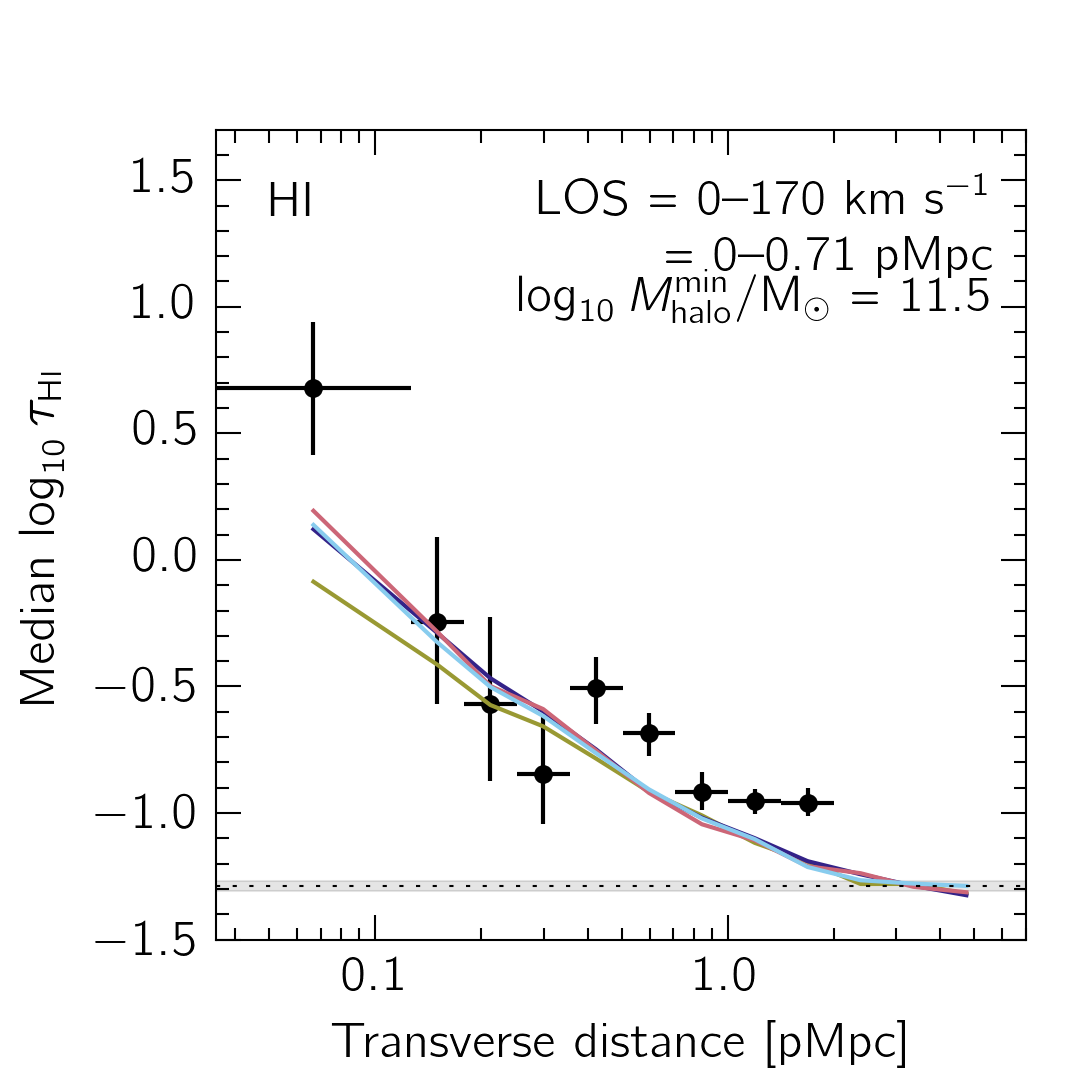} 
  \includegraphics[width=\wa]{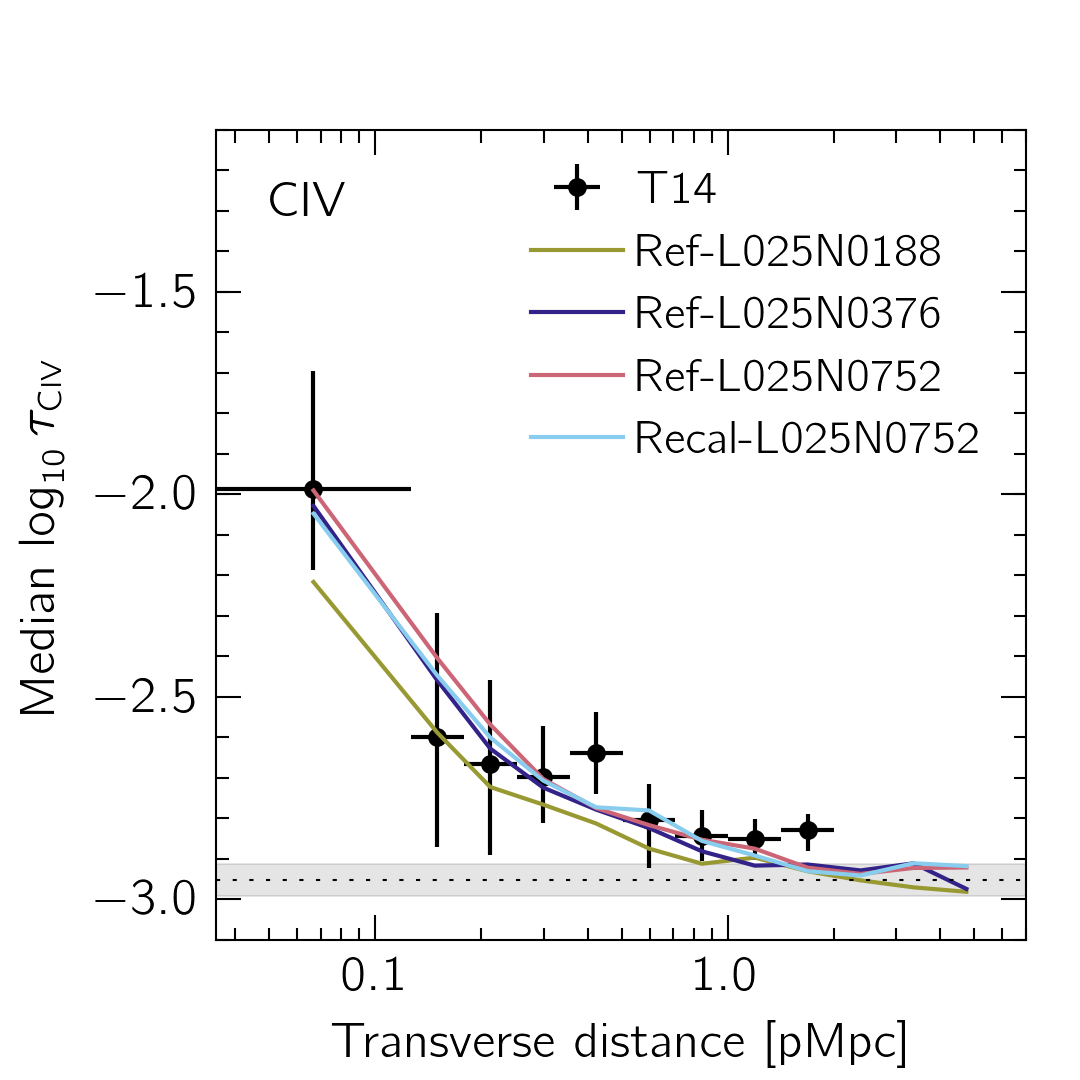} 
  \includegraphics[width=\wa]{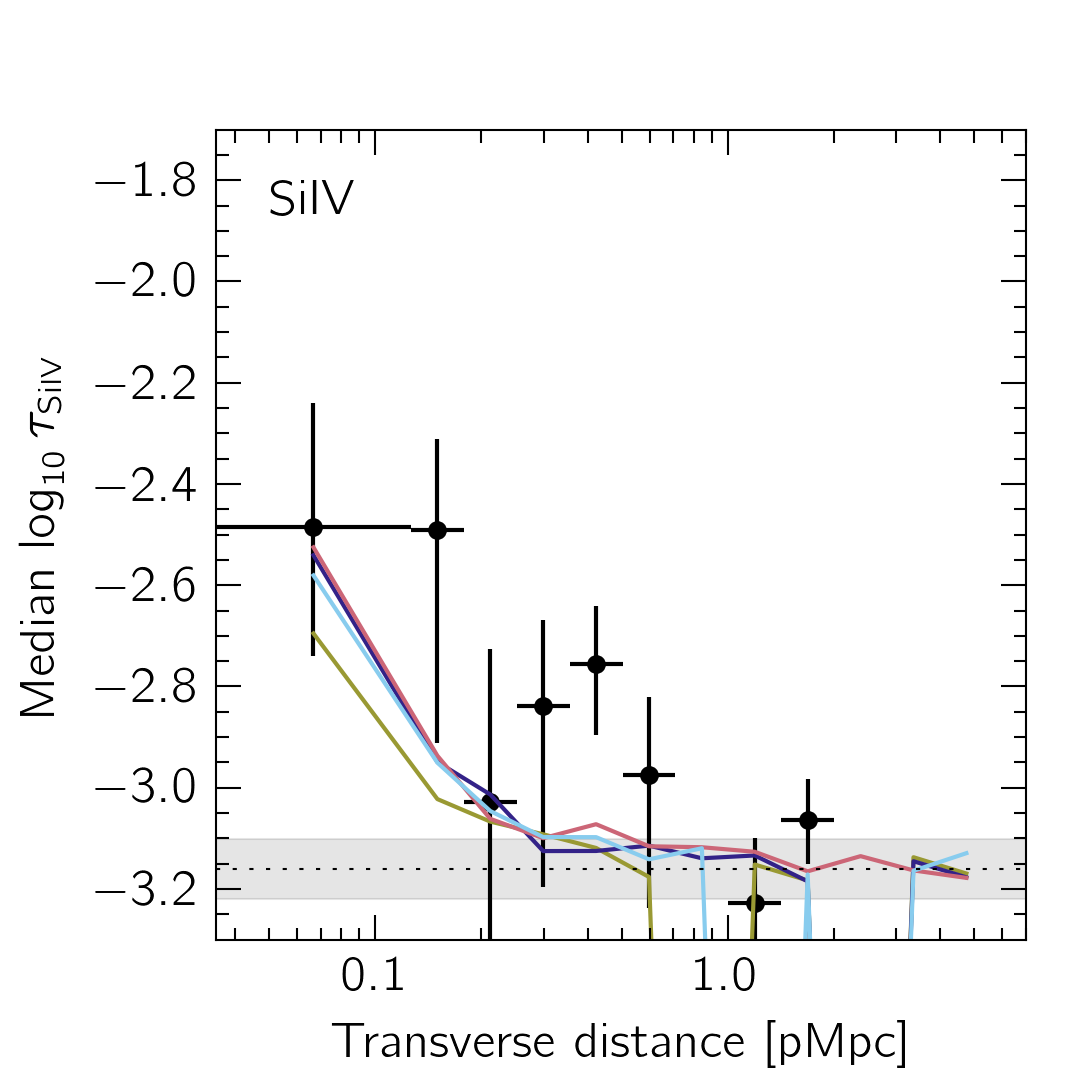} \\
 \caption{The same as Fig.~\ref{fig:box_test}, but instead presenting 
  convergence with respect to the numerical resolution. We use the   25~cMpc 
  box simulations showing the reference model for $188^3$, $376^3$ and $752^3$
 particles, in addition to the recalibrated L025N0752 run.
The lowest-resolution run, Ref-L025N0188, has lower median optical depths 
 in the innermost LOS and transverse distance bins compared to realizations with
 higher resolutions,but the fiducial intermediate- and high- resolution runs agree, indicating
that the fiducial simulation is converged. }
 \label{fig:res_test_small}
\end{figure*}

We examine the effects of varying the simulation box size in 
Fig.~\ref{fig:box_test}, where we plot the same cuts
along the LOS (top row) and transverse distance (bottom row)
as were shown in  Figs.~\ref{fig:cuts_los} and \ref{fig:cuts_trans}. 
We show results from the fiducial model Ref-L100N1504,
as well as from the reference run in the 50 and 25~cMpc boxes 
with the same resolution. The median optical
depth profiles for the 50 and 100~cMpc runs are converged, while for the 
25~cMpc box the optical depths tend to be lower, likely because
the median halo masses are smaller in the 25~cMpc box (see Table~\ref{tab:galpropb}). 

To test the effects of varying resolution, we turn to the 25~cMpc box
which has been realized with resolutions higher than the
fiducial one used in this work. The L025N0752 simulation
includes a version that has been run using the subgrid physics of the reference
model (Ref-) and one that has been recalibrated to better match the $z\approx0$ galaxy stellar mass function
(Recal-). In Fig.~\ref{fig:res_test_small} we 
plot the median optical depths along the LOS (top row) and transverse distance (bottom row)
for these high-resolution runs as well as for our fiducial resolution
($376^3$ particles in the 25~cMpc box) and finally a lower resolution 
of $188^3$ particles. 

We find that for all ions and halo masses, the optical depth profiles  from the fiducial
and high-resolution runs are in agreement, while those from the low-resolution runs
are systematically lower. We can therefore conclude that our fiducial simulation
is converged.

\section{Addition of minimum contamination level}

\label{app:contam}
\begin{figure*}
   \includegraphics[width=\wa]{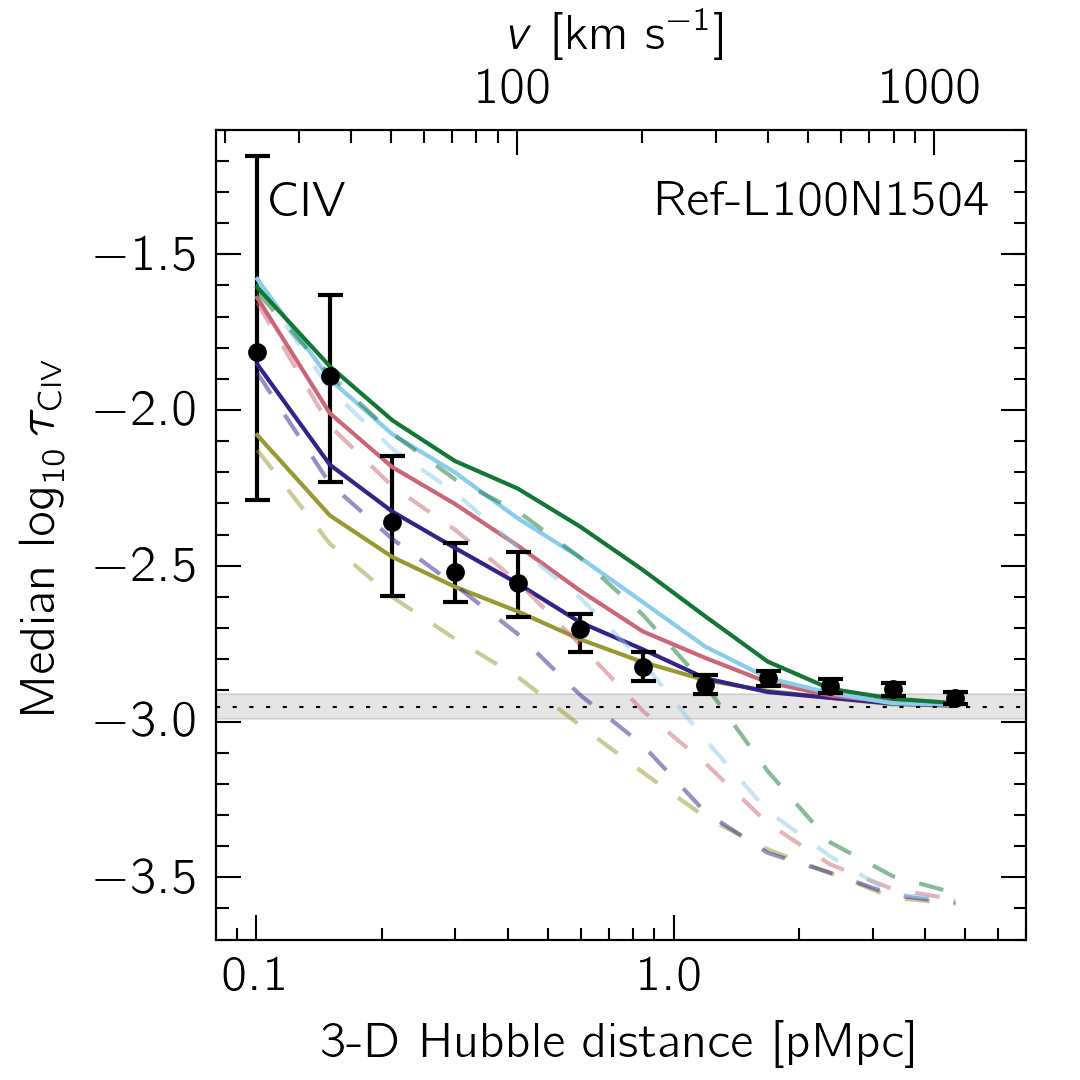} 
   \includegraphics[width=\wa]{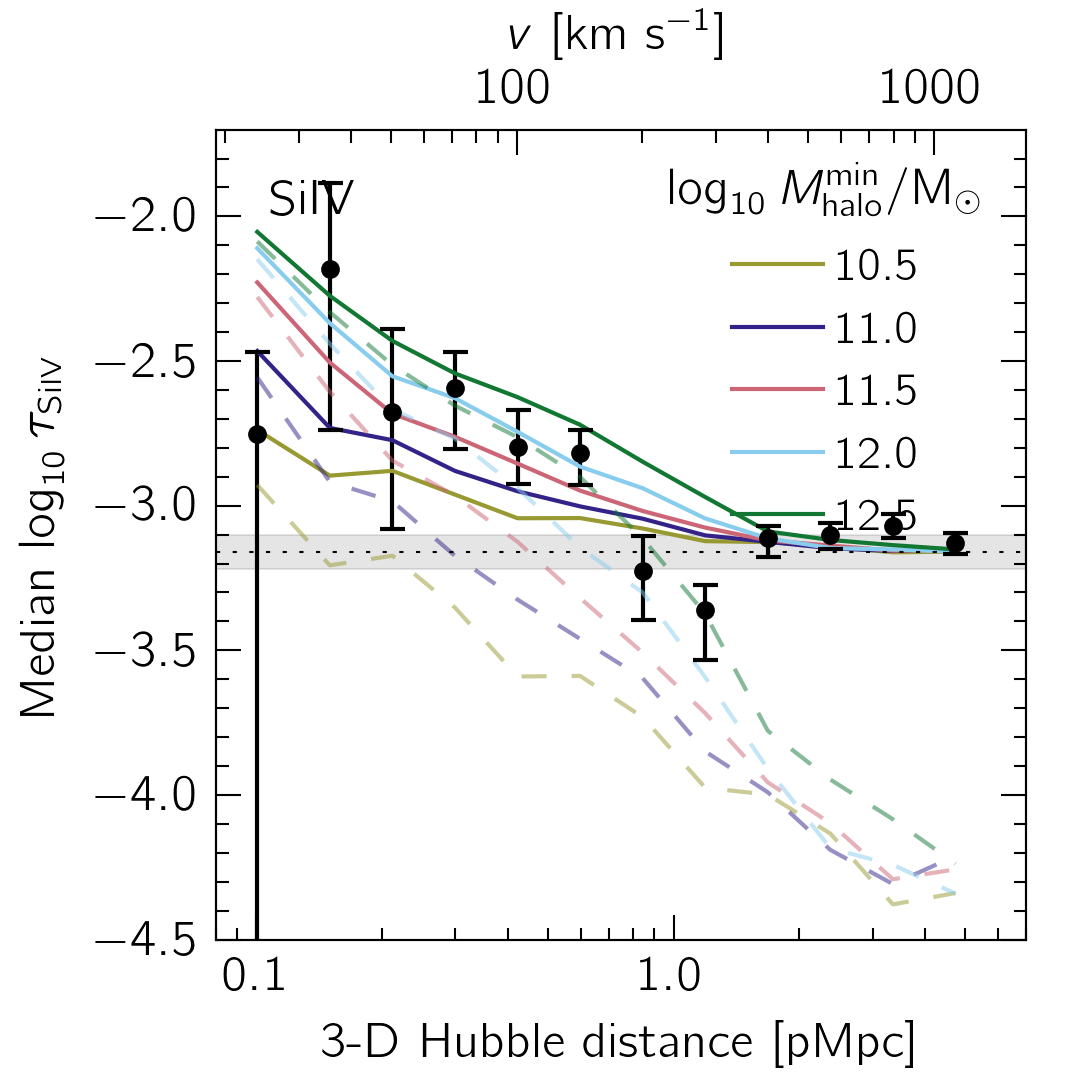} \\
   \caption{Median optical depths against 3-D Hubble distance, for \cfour\ (left) and \sifour (right).
     As in previous figures, the filled black points denote the observations, while the simulations 
      are represented by the lines. Here, the dashed (solid) lines show the optical depth profiles
      before (after) the addition of $\Delta \taurndm$. While equalizing \taurnd\ between 
       the observations and simulations is important for comparing the optical depths at larger 
       distances, it does not substantially alter the well-detected signal in the innermost bins ($v\lesssim100$~\kmps).}
 \label{fig:hubble}
\end{figure*}

In \S~\ref{sec:method}, we describe how \taurnd, the median pixel optical depth
in random regions, is lower in the simulations than in 
the observations, probably due to the absence of contamination in the mock spectra. For 
the figures in the main text, we simply add $\Delta \taurndm$, defined as the 
difference between simulated and observed \taurnd, to the simulated optical depths,
primarily to facilitate comparison
at larger galactocentric distances. 

To quantify the impact of this procedure, in Fig.~\ref{fig:hubble} 
we have plotted the median pixel optical depths for \cfour\ and \sifour\ as a function of 3-D 
Hubble distance, defined as $\sqrt{b^2 + \left(\Delta v / H\left(z\right) \right)^2}$,
a metric that allows us to combine together both the transverse and LOS direction 
information. The dashed lines in Fig.~\ref{fig:hubble} show the simulated median pixel optical 
depths before adding $\Delta \taurndm$, while the solid lines demonstrate the result of
the addition. 

The dashed lines demonstrate that the addition of $\Delta \taurndm$ does
not significantly impact the innermost bins
where the optical depths are detected
with high confidence above \taurnd. This gives us assurance that the selected 
simulated galaxies are producing enough metals to reproduce the observations. 
Rather, the main impact of $\Delta \taurndm$ is on the slope 
of the optical depth profiles. With or without the addition of $\Delta \taurndm$, 
it is clear that the observed and simulated median optical depths exhibit the 
same qualitative behaviour, with values peaking in the innermost bins and 
decreasing with distance until they reach \taurnd. 

\section{Variation of feedback models}
\label{app:feedback}

\begin{figure*}
     \includegraphics[width=\wa]{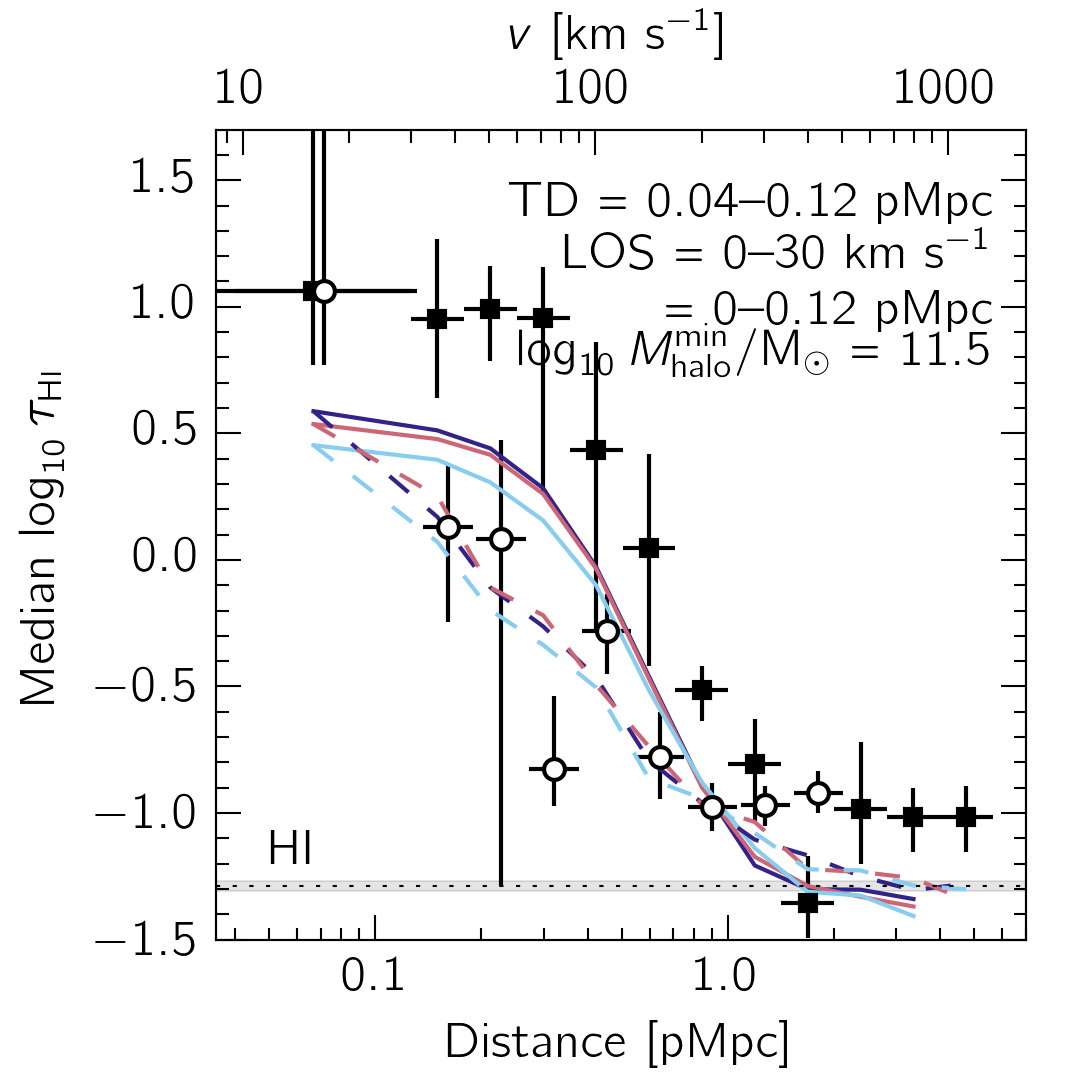} 
   \includegraphics[width=\wa]{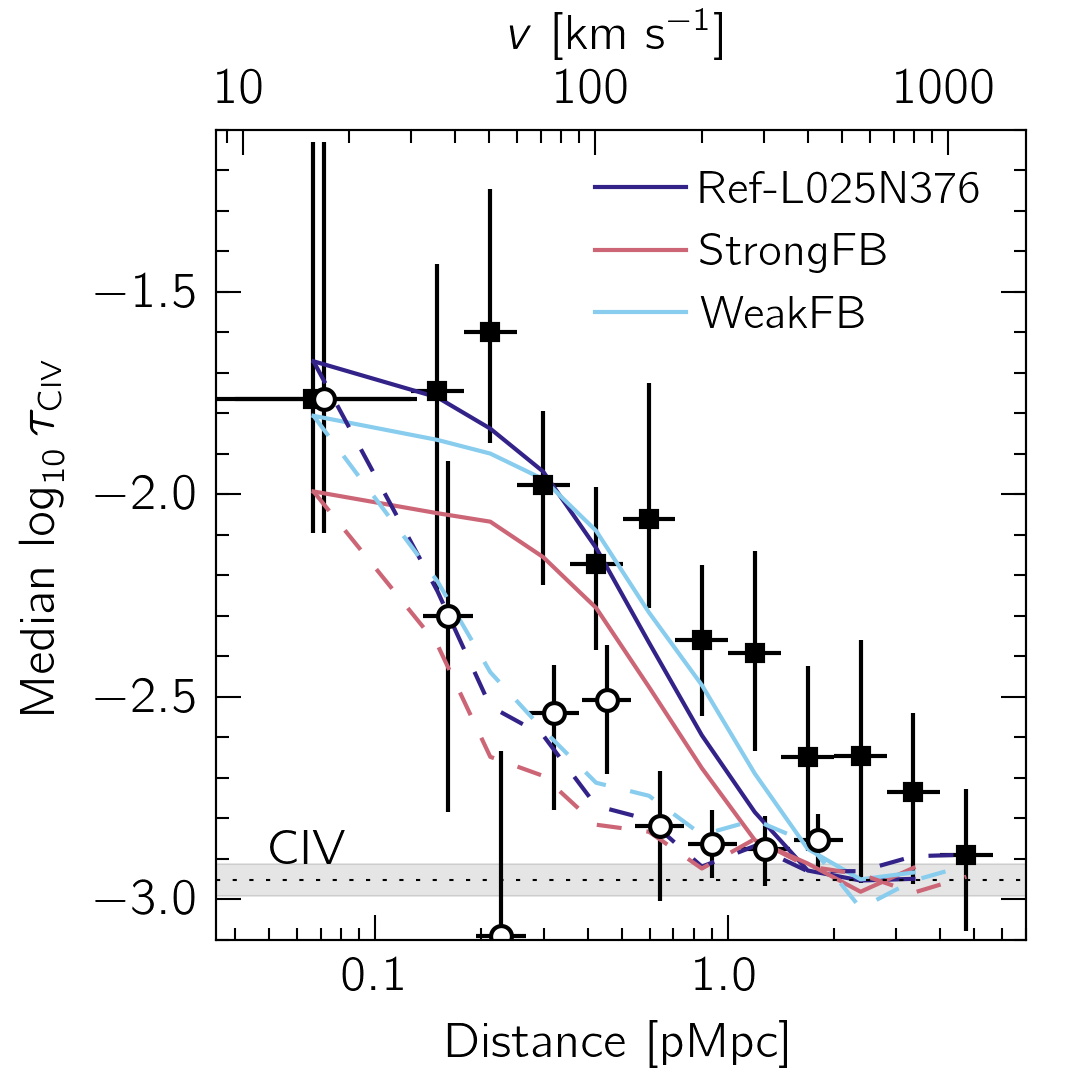} 
   \includegraphics[width=\wa]{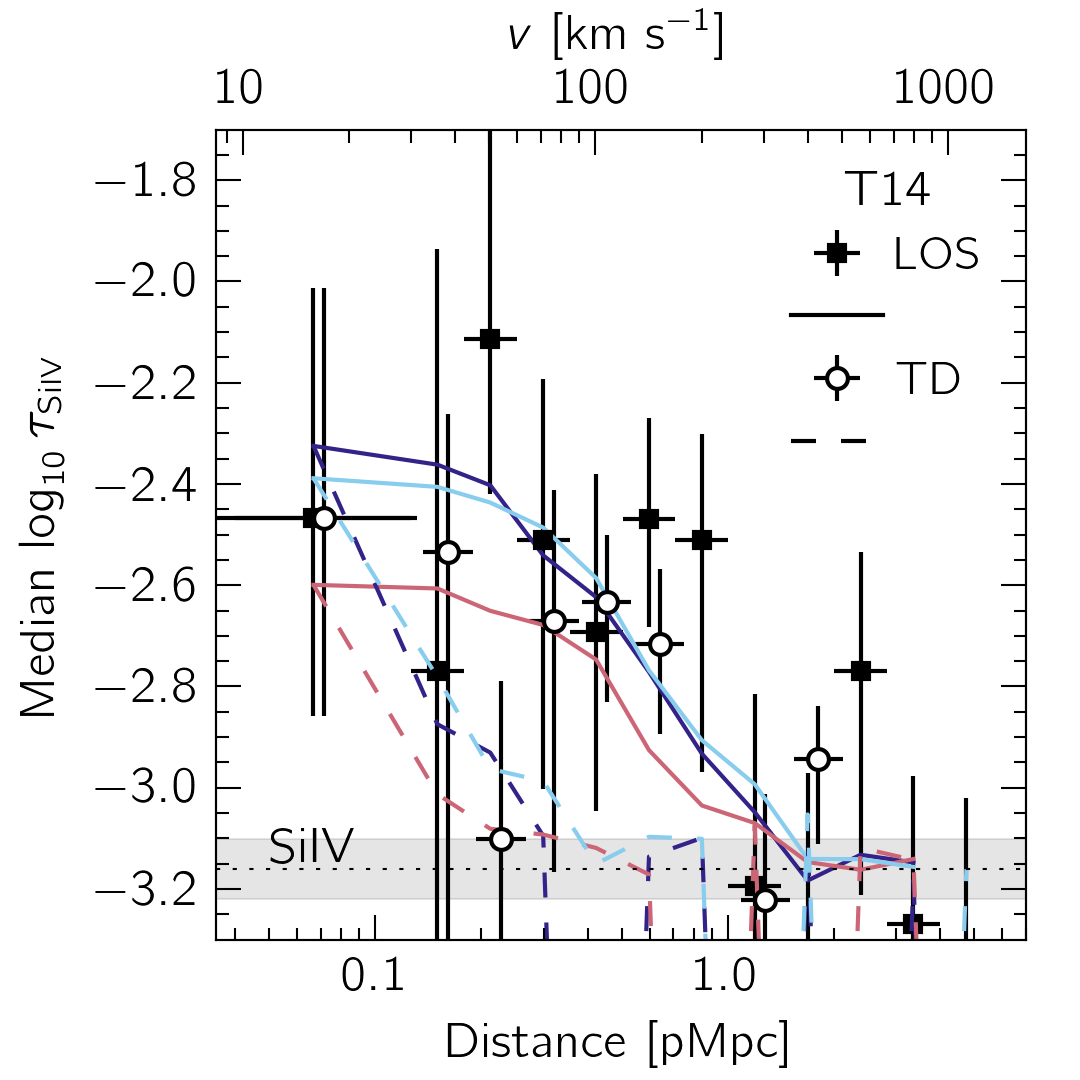} \\
        \includegraphics[width=\wa]{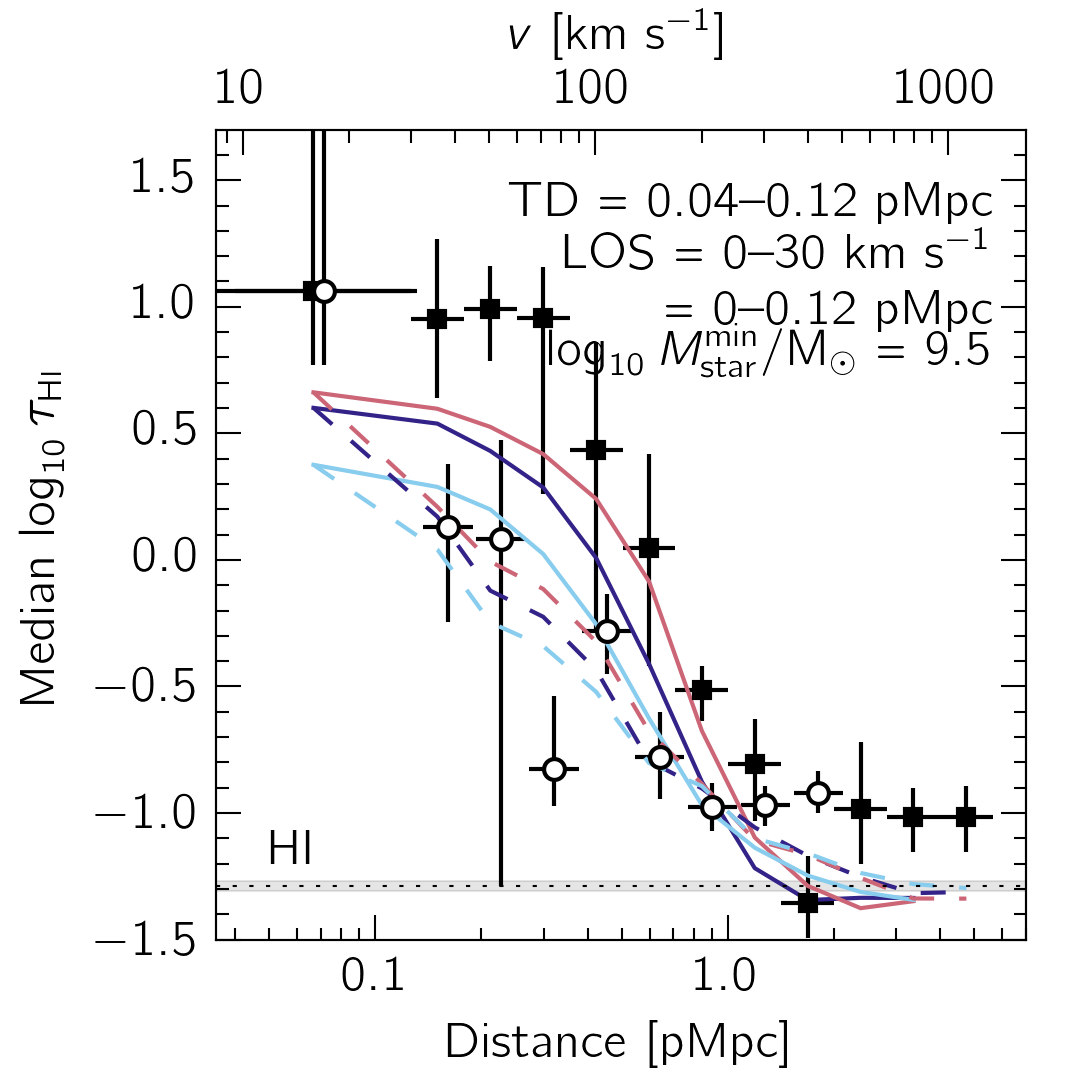} 
        \includegraphics[width=\wa]{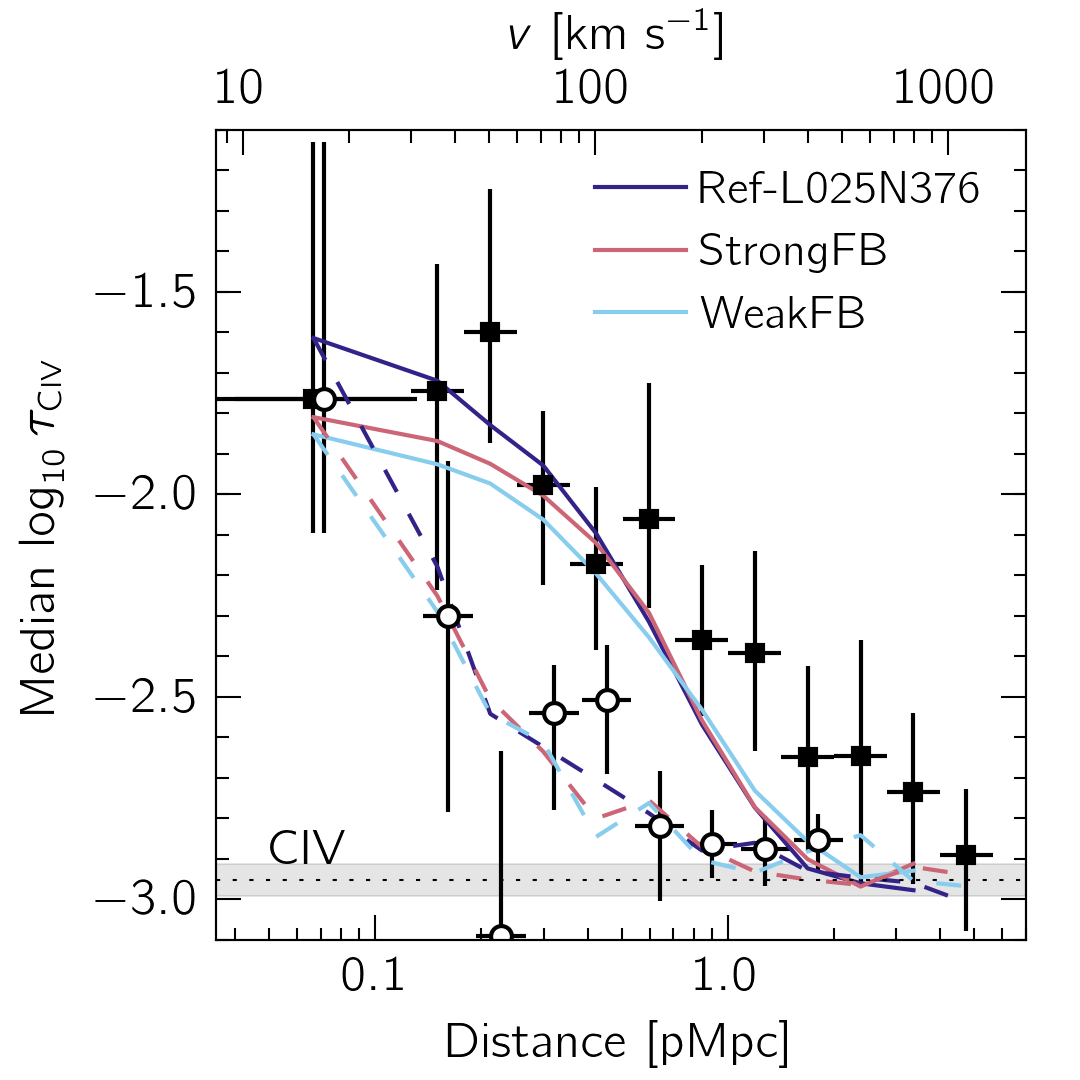} 
        \includegraphics[width=\wa]{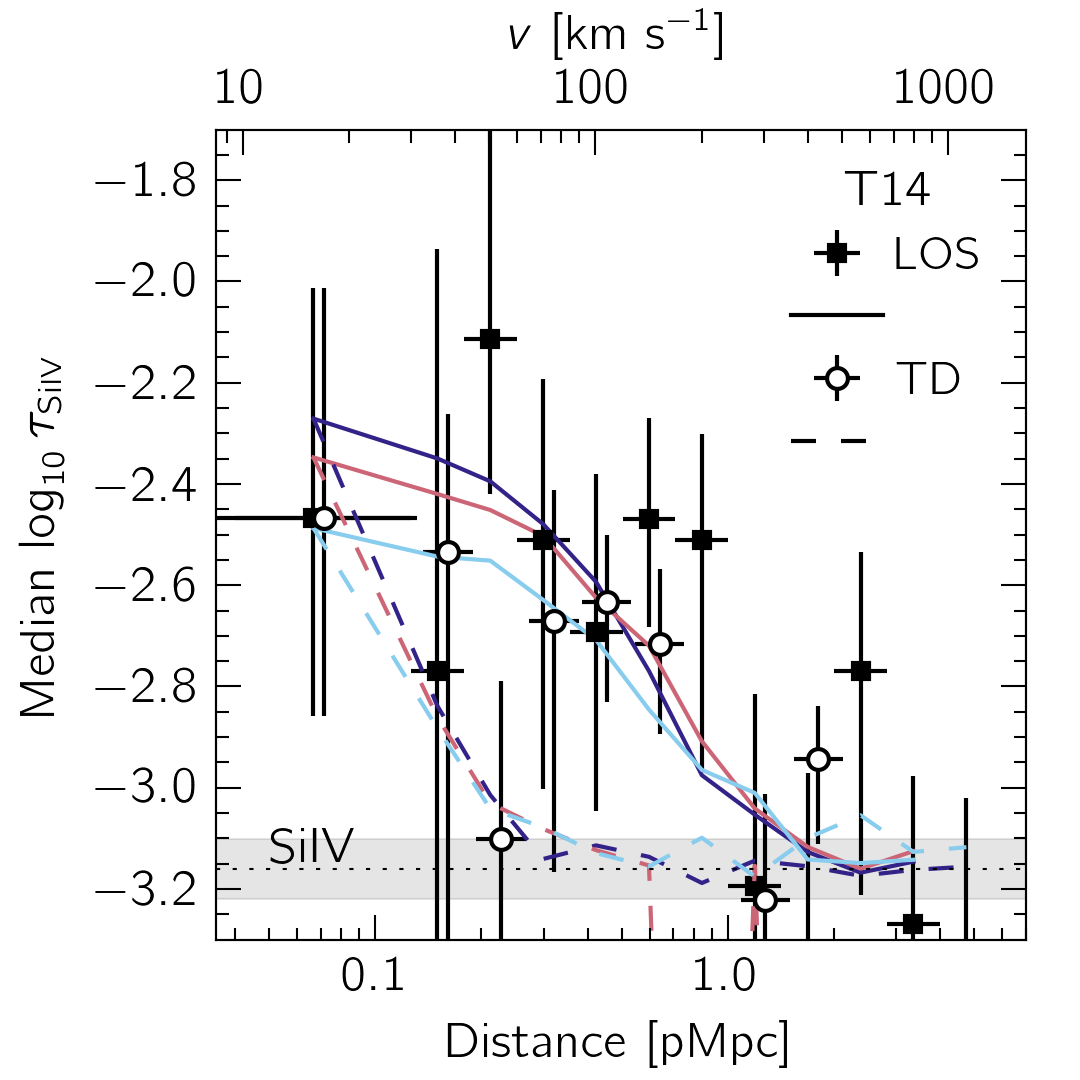} 
   \caption{
   As Fig.~\ref{fig:zsd}, but for different EAGLE stellar subgrid prescriptions.
   In the top row we have used a fixed minimum halo mass of $\mhminm=10^{11.5}$~\msol, 
   while in the bottom row a fixed \textit{stellar} mass of $\msminm=10^{9.5}$~\msol is considered.	
   We show the reference model,
   as well as results from the WeakFB and StrongFB runs, all using a 25~cMpc box. 
   At fixed halo mass, the metal-line absorption around our galaxies does not depend strongly on
   feedback models. This indicates that the observed redshift-space distortions
    are likely not dominated by outflows, but rather by infall or virial motions
    set by the mass of the halo.}
 \label{fig:wind}
\end{figure*} 

\begin{figure*}
    \includegraphics[width=\wa]{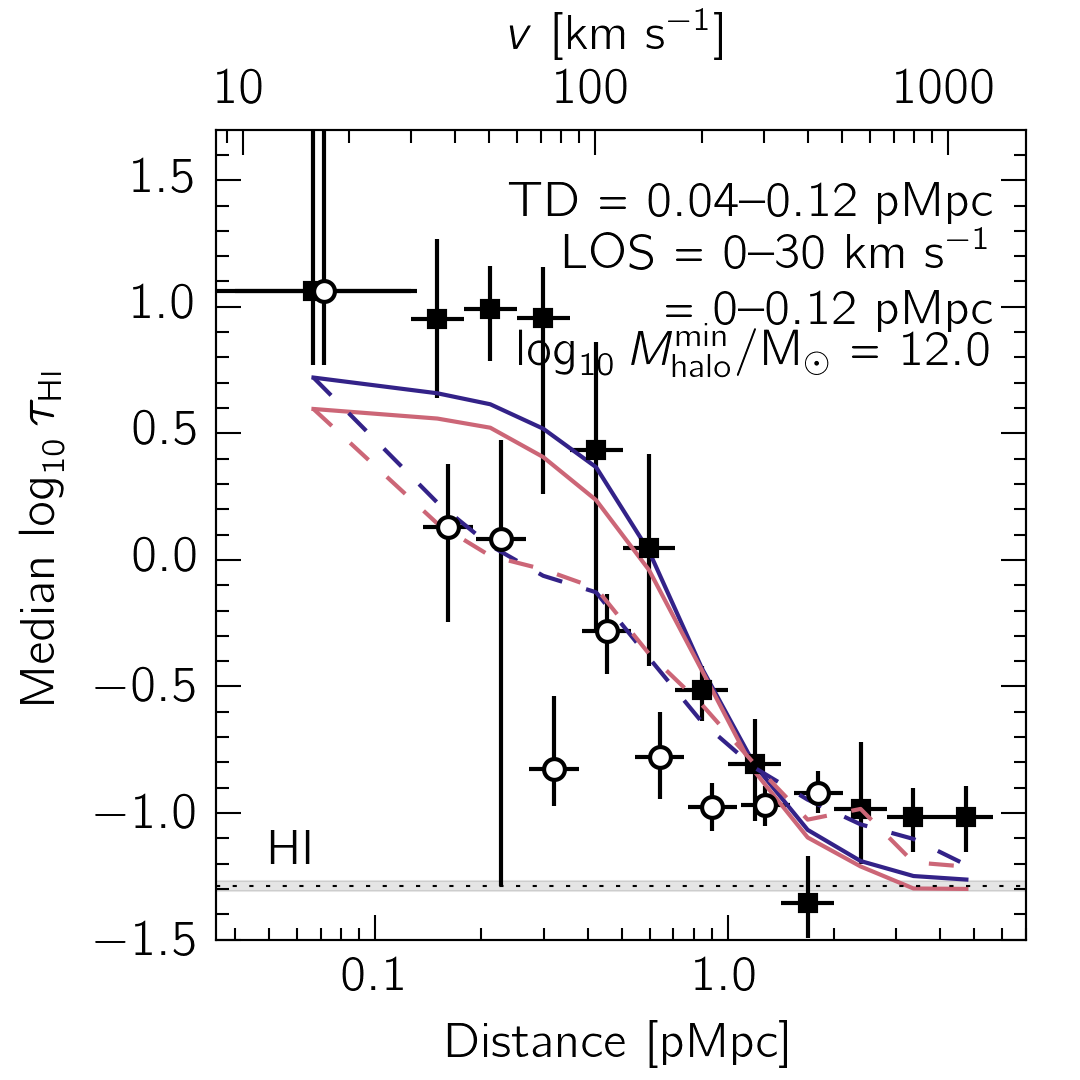} 
  \includegraphics[width=\wa]{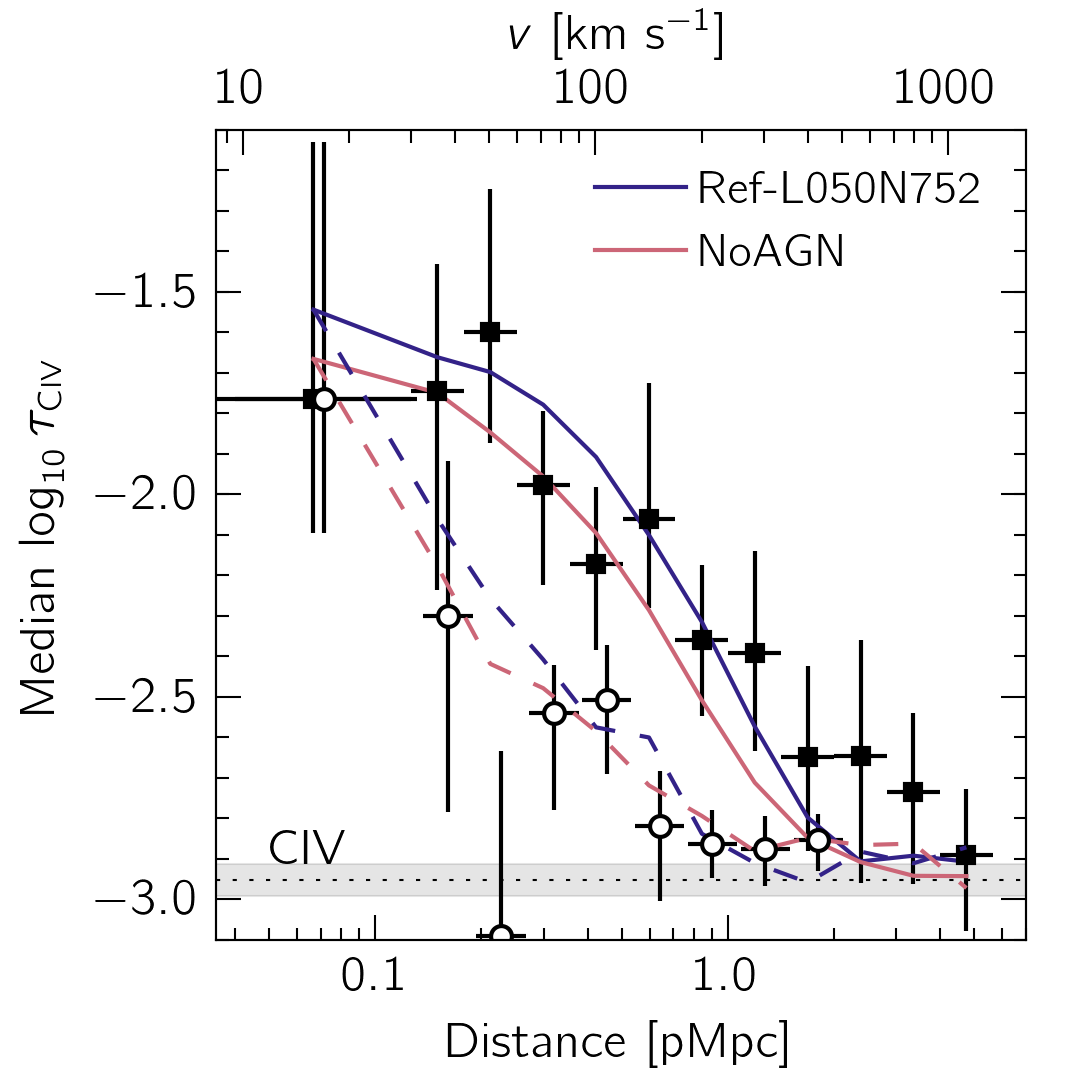} 
  \includegraphics[width=\wa]{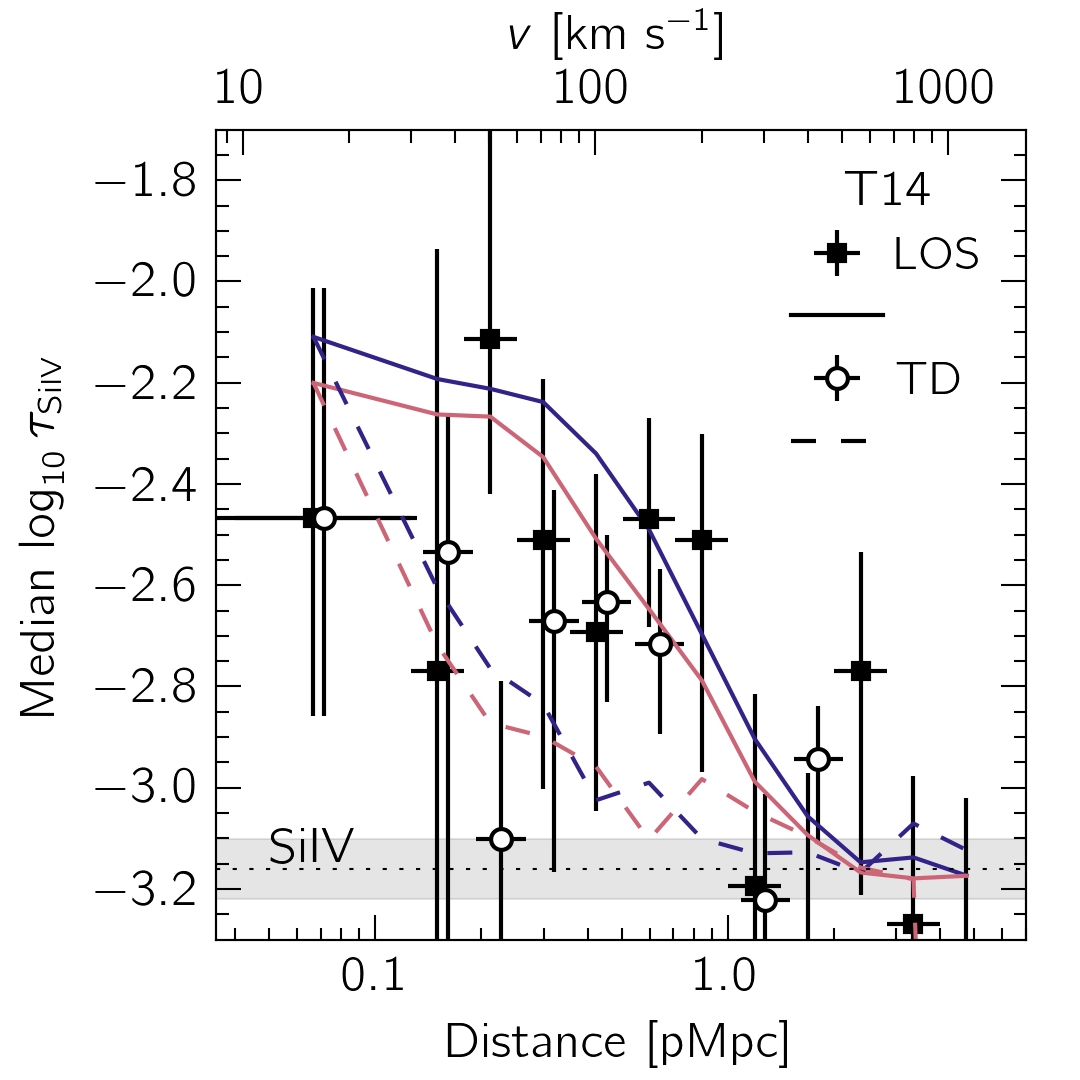} \\
   \caption{As Fig.~\ref{fig:zsd}, but for different AGN feedback prescriptions. 
   We examine both the Ref and NoAGN models using a 50~cMpc box, and find that turning 
    off AGN feedback does not have a significant impact on the optical depth profiles.}
 \label{fig:agn}
\end{figure*}

In \S~\ref{sec:variations}, we explored the effect that varying the stellar feedback
models had on velocities of the gas particles, and found that 
the ion mass-weighted radial velocities depend on halo mass (rather than stellar mass) and are
primarily infalling. In this section, we examine the impact that modifying the feedback
model has on the metal-line optical depths, and interpret the findings in the context of the 
results from \S~\ref{sec:variations}. 

Fig.~\ref{fig:wind} presents the outcome of varying the subgrid
stellar feedback for both a fixed minimum halo mass (top row) and stellar mass (bottom row).
We compare the median optical depths from
 Ref-L025N0376 to the WeakFB and StrongFB runs, which use respectively half
and twice as efficient stellar feedback as the reference model. Note that 
because these models use a 25~cMpc box, so at large transverse distances the 
results will not be converged with the size of the simulation volume. 

First examining the top row, we find that the optical depth profiles for \hone,
\cfour\ and \sifour\ do not depend strongly on the feedback model at fixed minimum halo mass.
While there are small differences between the models, these differences are not larger than
the errors on the data. 
This is consistent with the finding from \S~\ref{sec:variations} that the ion mass-weighted gas velocities 
do not change with feedback model for fixed minimum halo mass, as the gas probed is mainly infalling 
and has a radial velocity set by halo mass. 
Furthermore, the results for \hone\ are in agreement with \citet{rakic13}, who found that \hone\ optical depths around 
galaxies are set primarily by halo mass and are not strongly influenced by feedback model. 

Next, in the bottom row of Fig.~\ref{fig:wind} we show the same optical depth profiles, but 
using a fixed minimum stellar mass of $10^{9.5}$~\msol. In the case of \hone, we see an expected 
variation of the optical depth with subgrid model, where \hone\ optical depths increase
with feedback strength (or halo mass). On the other hand, for \cfour\ and \sifour\ we find
very little variation with halo mass. This may be due to two competing effects: halo mass 
and metallicty. Specifically, while we know the the infall velocities increase with halo mass
(stronger feedback), we have also found that at fixed stellar mass, \cfour- and \sifour-weighted
metallicity is higher with decreasing halo mass (weaker feedback), and the two effects my be cancelling out. 

Finally, in Fig.~\ref{fig:agn}  we compare
the median optical depths from the NoAGN run to the 50~cMpc
reference model, Ref-L050N752. We presents results for $\mhminm=10^{12.0}$~\msol,
as we do not expect the exclusion of AGN to have a strong effect below 
this halo mass. Overall, we do not find that the presence of AGN feedback has a
significant impact on the median optical depth profiles.

\end{document}

%% file: tables/med_od.tab
\begin{tabular}{lrllll}
\hline
Simulation  & $z$& $\Gamma_{\honem}$& $\log_{10} \taurndhonem$& $\log_{10} \taurndcfourm$& $\log_{10} \taurndsifourm$\\
& & [$10^{-12}$ s$^{-1}$]  & & &\\
\hline
\hline
T14 & $2.34$ & \ldots & $-1.29^{+0.02}_{-0.02}$ & $-2.95^{+0.04}_{-0.04}$ & $-3.16^{+0.06}_{-0.06}$ \\
   Ref-L100N1504  & $ 2.24$  & $ 0.88$ & $-1.43$ & $-3.60$ & $-4.37$ \\
   Ref-L050N0752  & $ 2.24$  & $ 0.90$ & $-1.43$ & $-3.60$ & $-4.19$ \\
   Ref-L025N0376  & $ 2.24$  & $ 1.04$ & $-1.38$ & $-3.65$ & $-4.37$ \\
   Ref-L025N0752  & $ 2.24$  & $ 0.98$ & $-1.40$ & $-3.64$ & $-4.31$ \\
 Recal-L025N0752  & $ 2.24$  & $ 0.99$ & $-1.39$ & $-3.63$ & $-4.36$ \\
           NoAGN  & $ 2.24$  & $ 0.91$ & $-1.42$ & $-3.59$ & $-4.19$ \\
          WeakFB  & $ 2.24$  & $ 1.00$ & $-1.39$ & $-3.57$ & $-4.29$ \\
        StrongFB  & $ 2.24$  & $ 1.06$ & $-1.37$ & $-3.85$ & $-4.59$ \\
   Ref-L100N1504  & $ 2.01$  & $ 0.61$ & $-1.56$ & $-3.56$ & $-4.33$ \\
   Ref-L100N1504  & $ 2.48$  & $ 1.27$ & $-1.28$ & $-3.64$ & $-4.41$ \\
\hline
\end{tabular}

%% file: tables/gal_props_b.tab
\begin{tabular}{lcrrrrrrrrrr}
\hline
Simulation  & $z$ & \multicolumn{5}{c}{Median $\log_{10} M_{\rm halo} / \msolm$} & \multicolumn{5}{c}{Number of galaxies} \\
$\log_{10} \mhminm / \msolm$ & & $10.5$ & $11.0$ & $11.5$ & $12.0$ & $12.5$ & $10.5$ & $11.0$ & $11.5$ & $12.0$ & $12.5$ \\
\hline
\hline
   Ref-L100N1504  & $ 2.24$  & 10.8 & 11.3 & 11.8 & 12.2 & 12.7 & 28112 & 9217 & 2729 &  689 &  111 \\
   Ref-L050N0752  & $ 2.24$  & 10.8 & 11.3 & 11.8 & 12.2 & 12.6 & 3775 & 1242 &  348 &   78 &    8 \\
   Ref-L025N0376  & $ 2.24$  & 10.8 & 11.3 & 11.7 & 12.1 & \ldots &  470 &  161 &   46 &   10 &    0 \\
   Ref-L025N0752  & $ 2.24$  & 10.8 & 11.3 & 11.7 & 12.1 & \ldots &  521 &  167 &   47 &   10 &    0 \\
 Recal-L025N0752  & $ 2.24$  & 10.8 & 11.3 & 11.7 & 12.1 & \ldots &  521 &  165 &   48 &   10 &    0 \\
           NoAGN  & $ 2.24$  & 10.8 & 11.3 & 11.7 & 12.2 & 12.6 & 3796 & 1236 &  349 &   80 &   11 \\
          WeakFB  & $ 2.24$  & 10.8 & 11.3 & 11.7 & 12.1 & \ldots &  500 &  162 &   48 &   10 &    0 \\
        StrongFB  & $ 2.24$  & 10.8 & 11.3 & 11.7 & 12.1 & \ldots &  460 &  156 &   47 &   10 &    0 \\
   Ref-L100N1504  & $ 2.01$  & 10.8 & 11.3 & 11.8 & 12.2 & 12.7 & 29129 & 9873 & 3047 &  814 &  161 \\
   Ref-L100N1504  & $ 2.48$  & 10.8 & 11.3 & 11.8 & 12.2 & 12.7 & 26771 & 8443 & 2402 &  540 &   87 \\
\hline
\end{tabular}

%% file: tables/zerr.tab
\begin{tabular}{@{}lllrrr}
\hline
Bin  & $r_{1}$  & $r_{2}$ & LR & NS &  MF  \\
\#	& (pMpc) & (pMpc) & & &  \\
\hline
\hline
1 & 0.04 & 0.13 & 0.23 & 0.15 & 0.62 \\ 
2 & 0.13 & 0.18 & 0.18 & 0.00 & 0.82 \\ 
3 & 0.18 & 0.25 & 0.36 & 0.09 & 0.55 \\ 
4 & 0.25 & 0.36 & 0.28 & 0.03 & 0.69 \\ 
5 & 0.36 & 0.50 & 0.34 & 0.07 & 0.59 \\ 
6 & 0.50 & 0.71 & 0.52 & 0.07 & 0.41 \\ 
7 & 0.71 & 1.00 & 0.48 & 0.03 & 0.49 \\ 
8 & 1.00 & 1.42 & 0.56 & 0.04 & 0.40 \\ 
9 & 1.42 & 2.00 & 0.71 & 0.05 & 0.24 \\ 
\hline
\end{tabular}

%% file: tables/gal_props_a.tab
\begin{tabular}{lcrrrrrrrrrr}
\hline
Simulation  & $z$ & \multicolumn{5}{c}{Median $\log_{10} M_{\ast} / \msolm$} & \multicolumn{5}{c}{Median SFR [$\msolm$ yr$^{-1}$]} \\
$\log_{10} \mhminm / \msolm$ & & $10.5$ & $11.0$ & $11.5$ & $12.0$ & $12.5$ & $10.5$ & $11.0$ & $11.5$ & $12.0$ & $12.5$ \\
\hline
\hline
   Ref-L100N1504  & $ 2.24$  &  8.3 &  9.1 &  9.8 & 10.4 & 10.7 &  0.2 &  1.3 &  6.5 & 15.3 & 19.4 \\
   Ref-L050N0752  & $ 2.24$  &  8.3 &  9.0 &  9.8 & 10.3 & 10.5 &  0.2 &  1.4 &  6.0 & 10.4 & 11.2 \\
   Ref-L025N0376  & $ 2.24$  &  8.3 &  9.1 &  9.8 & 10.2 & \ldots &  0.2 &  1.7 &  6.9 & 18.4 & \ldots \\
   Ref-L025N0752  & $ 2.24$  &  8.3 &  9.2 &  9.7 & 10.1 & \ldots &  0.3 &  2.1 &  4.6 &  9.1 & \ldots \\
 Recal-L025N0752  & $ 2.24$  &  8.2 &  9.1 &  9.7 & 10.1 & \ldots &  0.2 &  1.4 &  4.1 & 15.9 & \ldots \\
           NoAGN  & $ 2.24$  &  8.3 &  9.0 &  9.8 & 10.7 & 11.0 &  0.2 &  1.3 &  9.2 & 39.7 & 103.8 \\
          WeakFB  & $ 2.24$  &  8.8 &  9.5 & 10.1 & 10.3 & \ldots &  0.5 &  2.8 &  7.7 & 14.3 & \ldots \\
        StrongFB  & $ 2.24$  &  7.7 &  8.4 &  9.1 &  9.6 & \ldots &  0.0 &  0.3 &  1.3 &  5.8 & \ldots \\
   Ref-L100N1504  & $ 2.01$  &  8.3 &  9.1 &  9.9 & 10.4 & 10.7 &  0.2 &  1.3 &  6.0 & 13.8 & 18.1 \\
   Ref-L100N1504  & $ 2.48$  &  8.3 &  9.0 &  9.8 & 10.4 & 10.7 &  0.2 &  1.4 &  7.0 & 17.2 & 28.0 \\
\hline
\end{tabular}

%% file: tables/2dmap_chisq.tab
\begin{tabular}{lrrrrr}
\hline
$\log_{10} \mhminm / \msolm$ & $10.5$ & $11.0$ & $11.5$ & $12.0$ & $12.5$ \\
\hline
\hline
      \hone, full map                               & $ 0.002$  & $ 0.059$  & $ 0.033$  & $ 0.000$  & $ 0.000$  \\
      \hone, $\text{LOS} = 0.71\text{--}2.83$ pMpc & $ 0.003$  & $ 0.085$  & $ 0.340$  & $ 0.034$  & $ 0.000$  \\
     \cfour, full map                               & $ 0.063$  & $ 0.410$  & $ 0.326$  & $ 0.012$  & $ 0.000$  \\
    \sifour, full map                               & $ 0.001$  & $ 0.000$  & $ 0.049$  & $ 0.011$  & $ 0.000$  \\
\hline
\end{tabular}

%% file: tables/mass_cuts_chisq.tab
\begin{tabular}{rrrrrr}
\hline
$\log_{10} \mhminm / \msolm$ & $10.5$ & $11.0$ & $11.5$ & $12.0$ & $12.5$ \\
\hline
\hline
      \hone  & $ 0.003$  & $ 0.055$  & $ 0.133$  & $ 0.010$  & $ 0.000$  \\
     \cfour  & $ 0.379$  & $ 0.926$  & $ 0.938$  & $ 0.285$  & $ 0.003$  \\
    \sifour  & $ 0.023$  & $ 0.035$  & $ 0.099$  & $ 0.068$  & $ 0.030$  \\
\hline
\end{tabular}

%% file: tables/mass_zspacedist_chisq.tab
\begin{tabular}{rrrrrr}
\hline
$\log_{10} \mhminm / \msolm$ & $10.5$ & $11.0$ & $11.5$ & $12.0$ & $12.5$ \\
\hline
\hline
      \hone  & $ 0.003$  & $ 0.026$  & $ 0.109$  & $ 0.027$  & $ 0.000$  \\
     \cfour  & $ 0.100$  & $ 0.484$  & $ 0.835$  & $ 0.474$  & $ 0.074$  \\
    \sifour  & $ 0.219$  & $ 0.040$  & $ 0.316$  & $ 0.229$  & $ 0.143$  \\
\hline
\end{tabular}

%% file: tables/nopecvel_chisq.tab
\begin{tabular}{rrrrrrrrrrr}
\hline
 & \multicolumn{3}{c}{\hone} & \multicolumn{3}{c}{\cfour} & \multicolumn{3}{c}{\sifour} \\
$\log_{10} \mhminm / \msolm$ & $11.0$ & $11.5$ & $12.0$  & $11.0$ & $11.5$ & $12.0$  & $11.0$ & $11.5$ & $12.0$  \\
\hline
\hline
                  Ref  & $ 0.022$  & $ 0.117$  & $ 0.182$  & $ 0.147$  & $ 0.352$  & $ 0.392$  & $ 0.033$  & $ 0.389$  & $ 0.180$  \\
            No z err.  & $ 0.005$  & $ 0.037$  & $ 0.092$  & $ 0.063$  & $ 0.223$  & $ 0.333$  & $ 0.033$  & $ 0.433$  & $ 0.214$  \\
         No pec. vel.  & $ 0.021$  & $ 0.034$  & $ 0.043$  & $ 0.038$  & $ 0.046$  & $ 0.040$  & $ 0.040$  & $ 0.576$  & $ 0.246$  \\
\hline
\end{tabular}